\documentclass[apj,twocolumn]{openjournal}
\usepackage[utf8]{inputenc}
\usepackage{natbib,graphics,epsfig}
\usepackage{graphicx}
\usepackage{natbib}
\usepackage{amsmath}
\usepackage{color}

\usepackage{amsmath}
\usepackage{amssymb}
\usepackage{textcomp}
\usepackage{hyperref}
\usepackage{txfonts}
\usepackage{float}
\usepackage[T1]{fontenc}
\usepackage{aecompl}
  \definecolor{snow}{rgb}{1.000000,0.980392,0.980392}
  \definecolor{ghost white}{rgb}{0.972549,0.972549,1.000000}
  \definecolor{GhostWhite}{rgb}{0.972549,0.972549,1.000000}
  \definecolor{white smoke}{rgb}{0.960784,0.960784,0.960784}
  \definecolor{WhiteSmoke}{rgb}{0.960784,0.960784,0.960784}
  \definecolor{gainsboro}{rgb}{0.862745,0.862745,0.862745}
  \definecolor{floral white}{rgb}{1.000000,0.980392,0.941176}
  \definecolor{FloralWhite}{rgb}{1.000000,0.980392,0.941176}
  \definecolor{old lace}{rgb}{0.992157,0.960784,0.901961}
  \definecolor{OldLace}{rgb}{0.992157,0.960784,0.901961}
  \definecolor{linen}{rgb}{0.980392,0.941176,0.901961}
  \definecolor{antique white}{rgb}{0.980392,0.921569,0.843137}
  \definecolor{AntiqueWhite}{rgb}{0.980392,0.921569,0.843137}
  \definecolor{papaya whip}{rgb}{1.000000,0.937255,0.835294}
  \definecolor{PapayaWhip}{rgb}{1.000000,0.937255,0.835294}
  \definecolor{blanched almond}{rgb}{1.000000,0.921569,0.803922}
  \definecolor{BlanchedAlmond}{rgb}{1.000000,0.921569,0.803922}
  \definecolor{bisque}{rgb}{1.000000,0.894118,0.768627}
  \definecolor{peach puff}{rgb}{1.000000,0.854902,0.725490}
  \definecolor{PeachPuff}{rgb}{1.000000,0.854902,0.725490}
  \definecolor{navajo white}{rgb}{1.000000,0.870588,0.678431}
  \definecolor{NavajoWhite}{rgb}{1.000000,0.870588,0.678431}
  \definecolor{moccasin}{rgb}{1.000000,0.894118,0.709804}
  \definecolor{cornsilk}{rgb}{1.000000,0.972549,0.862745}
  \definecolor{ivory}{rgb}{1.000000,1.000000,0.941176}
  \definecolor{lemon chiffon}{rgb}{1.000000,0.980392,0.803922}
  \definecolor{LemonChiffon}{rgb}{1.000000,0.980392,0.803922}
  \definecolor{seashell}{rgb}{1.000000,0.960784,0.933333}
  \definecolor{honeydew}{rgb}{0.941176,1.000000,0.941176}
  \definecolor{mint cream}{rgb}{0.960784,1.000000,0.980392}
  \definecolor{MintCream}{rgb}{0.960784,1.000000,0.980392}
  \definecolor{azure}{rgb}{0.941176,1.000000,1.000000}
  \definecolor{alice blue}{rgb}{0.941176,0.972549,1.000000}
  \definecolor{AliceBlue}{rgb}{0.941176,0.972549,1.000000}
  \definecolor{lavender}{rgb}{0.901961,0.901961,0.980392}
  \definecolor{lavender blush}{rgb}{1.000000,0.941176,0.960784}
  \definecolor{LavenderBlush}{rgb}{1.000000,0.941176,0.960784}
  \definecolor{misty rose}{rgb}{1.000000,0.894118,0.882353}
  \definecolor{MistyRose}{rgb}{1.000000,0.894118,0.882353}
  \definecolor{white}{rgb}{1.000000,1.000000,1.000000}
  \definecolor{black}{rgb}{0.000000,0.000000,0.000000}
  \definecolor{dark slate gray}{rgb}{0.184314,0.309804,0.309804}
  \definecolor{DarkSlateGray}{rgb}{0.184314,0.309804,0.309804}
  \definecolor{dark slate grey}{rgb}{0.184314,0.309804,0.309804}
  \definecolor{DarkSlateGrey}{rgb}{0.184314,0.309804,0.309804}
  \definecolor{dim gray}{rgb}{0.411765,0.411765,0.411765}
  \definecolor{DimGray}{rgb}{0.411765,0.411765,0.411765}
  \definecolor{dim grey}{rgb}{0.411765,0.411765,0.411765}
  \definecolor{DimGrey}{rgb}{0.411765,0.411765,0.411765}
  \definecolor{slate gray}{rgb}{0.439216,0.501961,0.564706}
  \definecolor{SlateGray}{rgb}{0.439216,0.501961,0.564706}
  \definecolor{slate grey}{rgb}{0.439216,0.501961,0.564706}
  \definecolor{SlateGrey}{rgb}{0.439216,0.501961,0.564706}
  \definecolor{light slate gray}{rgb}{0.466667,0.533333,0.600000}
  \definecolor{LightSlateGray}{rgb}{0.466667,0.533333,0.600000}
  \definecolor{light slate grey}{rgb}{0.466667,0.533333,0.600000}
  \definecolor{LightSlateGrey}{rgb}{0.466667,0.533333,0.600000}
  \definecolor{gray}{rgb}{0.745098,0.745098,0.745098}
  \definecolor{grey}{rgb}{0.745098,0.745098,0.745098}
  \definecolor{light grey}{rgb}{0.827451,0.827451,0.827451}
  \definecolor{LightGrey}{rgb}{0.827451,0.827451,0.827451}
  \definecolor{light gray}{rgb}{0.827451,0.827451,0.827451}
  \definecolor{LightGray}{rgb}{0.827451,0.827451,0.827451}
  \definecolor{midnight blue}{rgb}{0.098039,0.098039,0.439216}
  \definecolor{MidnightBlue}{rgb}{0.098039,0.098039,0.439216}
  \definecolor{navy}{rgb}{0.000000,0.000000,0.501961}
  \definecolor{navy blue}{rgb}{0.000000,0.000000,0.501961}
  \definecolor{NavyBlue}{rgb}{0.000000,0.000000,0.501961}
  \definecolor{cornflower blue}{rgb}{0.392157,0.584314,0.929412}
  \definecolor{CornflowerBlue}{rgb}{0.392157,0.584314,0.929412}
  \definecolor{dark slate blue}{rgb}{0.282353,0.239216,0.545098}
  \definecolor{DarkSlateBlue}{rgb}{0.282353,0.239216,0.545098}
  \definecolor{slate blue}{rgb}{0.415686,0.352941,0.803922}
  \definecolor{SlateBlue}{rgb}{0.415686,0.352941,0.803922}
  \definecolor{medium slate blue}{rgb}{0.482353,0.407843,0.933333}
  \definecolor{MediumSlateBlue}{rgb}{0.482353,0.407843,0.933333}
  \definecolor{light slate blue}{rgb}{0.517647,0.439216,1.000000}
  \definecolor{LightSlateBlue}{rgb}{0.517647,0.439216,1.000000}
  \definecolor{medium blue}{rgb}{0.000000,0.000000,0.803922}
  \definecolor{MediumBlue}{rgb}{0.000000,0.000000,0.803922}
  \definecolor{royal blue}{rgb}{0.254902,0.411765,0.882353}
  \definecolor{RoyalBlue}{rgb}{0.254902,0.411765,0.882353}
  \definecolor{blue}{rgb}{0.000000,0.000000,1.000000}
  \definecolor{dodger blue}{rgb}{0.117647,0.564706,1.000000}
  \definecolor{DodgerBlue}{rgb}{0.117647,0.564706,1.000000}
  \definecolor{deep sky blue}{rgb}{0.000000,0.749020,1.000000}
  \definecolor{DeepSkyBlue}{rgb}{0.000000,0.749020,1.000000}
  \definecolor{sky blue}{rgb}{0.529412,0.807843,0.921569}
  \definecolor{SkyBlue}{rgb}{0.529412,0.807843,0.921569}
  \definecolor{light sky blue}{rgb}{0.529412,0.807843,0.980392}
  \definecolor{LightSkyBlue}{rgb}{0.529412,0.807843,0.980392}
  \definecolor{steel blue}{rgb}{0.274510,0.509804,0.705882}
  \definecolor{SteelBlue}{rgb}{0.274510,0.509804,0.705882}
  \definecolor{light steel blue}{rgb}{0.690196,0.768627,0.870588}
  \definecolor{LightSteelBlue}{rgb}{0.690196,0.768627,0.870588}
  \definecolor{light blue}{rgb}{0.678431,0.847059,0.901961}
  \definecolor{LightBlue}{rgb}{0.678431,0.847059,0.901961}
  \definecolor{powder blue}{rgb}{0.690196,0.878431,0.901961}
  \definecolor{PowderBlue}{rgb}{0.690196,0.878431,0.901961}
  \definecolor{pale turquoise}{rgb}{0.686275,0.933333,0.933333}
  \definecolor{PaleTurquoise}{rgb}{0.686275,0.933333,0.933333}
  \definecolor{dark turquoise}{rgb}{0.000000,0.807843,0.819608}
  \definecolor{DarkTurquoise}{rgb}{0.000000,0.807843,0.819608}
  \definecolor{medium turquoise}{rgb}{0.282353,0.819608,0.800000}
  \definecolor{MediumTurquoise}{rgb}{0.282353,0.819608,0.800000}
  \definecolor{turquoise}{rgb}{0.250980,0.878431,0.815686}
  \definecolor{cyan}{rgb}{0.000000,1.000000,1.000000}
  \definecolor{light cyan}{rgb}{0.878431,1.000000,1.000000}
  \definecolor{LightCyan}{rgb}{0.878431,1.000000,1.000000}
  \definecolor{cadet blue}{rgb}{0.372549,0.619608,0.627451}
  \definecolor{CadetBlue}{rgb}{0.372549,0.619608,0.627451}
  \definecolor{medium aquamarine}{rgb}{0.400000,0.803922,0.666667}
  \definecolor{MediumAquamarine}{rgb}{0.400000,0.803922,0.666667}
  \definecolor{aquamarine}{rgb}{0.498039,1.000000,0.831373}
  \definecolor{dark green}{rgb}{0.000000,0.392157,0.000000}
  \definecolor{DarkGreen}{rgb}{0.000000,0.392157,0.000000}
  \definecolor{dark olive green}{rgb}{0.333333,0.419608,0.184314}
  \definecolor{DarkOliveGreen}{rgb}{0.333333,0.419608,0.184314}
  \definecolor{dark sea green}{rgb}{0.560784,0.737255,0.560784}
  \definecolor{DarkSeaGreen}{rgb}{0.560784,0.737255,0.560784}
  \definecolor{sea green}{rgb}{0.180392,0.545098,0.341176}
  \definecolor{SeaGreen}{rgb}{0.180392,0.545098,0.341176}
  \definecolor{medium sea green}{rgb}{0.235294,0.701961,0.443137}
  \definecolor{MediumSeaGreen}{rgb}{0.235294,0.701961,0.443137}
  \definecolor{light sea green}{rgb}{0.125490,0.698039,0.666667}
  \definecolor{LightSeaGreen}{rgb}{0.125490,0.698039,0.666667}
  \definecolor{pale green}{rgb}{0.596078,0.984314,0.596078}
  \definecolor{PaleGreen}{rgb}{0.596078,0.984314,0.596078}
  \definecolor{spring green}{rgb}{0.000000,1.000000,0.498039}
  \definecolor{SpringGreen}{rgb}{0.000000,1.000000,0.498039}
  \definecolor{lawn green}{rgb}{0.486275,0.988235,0.000000}
  \definecolor{LawnGreen}{rgb}{0.486275,0.988235,0.000000}
  \definecolor{green}{rgb}{0.000000,1.000000,0.000000}
  \definecolor{chartreuse}{rgb}{0.498039,1.000000,0.000000}
  \definecolor{medium spring green}{rgb}{0.000000,0.980392,0.603922}
  \definecolor{MediumSpringGreen}{rgb}{0.000000,0.980392,0.603922}
  \definecolor{green yellow}{rgb}{0.678431,1.000000,0.184314}
  \definecolor{GreenYellow}{rgb}{0.678431,1.000000,0.184314}
  \definecolor{lime green}{rgb}{0.196078,0.803922,0.196078}
  \definecolor{LimeGreen}{rgb}{0.196078,0.803922,0.196078}
  \definecolor{yellow green}{rgb}{0.603922,0.803922,0.196078}
  \definecolor{YellowGreen}{rgb}{0.603922,0.803922,0.196078}
  \definecolor{forest green}{rgb}{0.133333,0.545098,0.133333}
  \definecolor{ForestGreen}{rgb}{0.133333,0.545098,0.133333}
  \definecolor{olive drab}{rgb}{0.419608,0.556863,0.137255}
  \definecolor{OliveDrab}{rgb}{0.419608,0.556863,0.137255}
  \definecolor{dark khaki}{rgb}{0.741176,0.717647,0.419608}
  \definecolor{DarkKhaki}{rgb}{0.741176,0.717647,0.419608}
  \definecolor{khaki}{rgb}{0.941176,0.901961,0.549020}
  \definecolor{pale goldenrod}{rgb}{0.933333,0.909804,0.666667}
  \definecolor{PaleGoldenrod}{rgb}{0.933333,0.909804,0.666667}
  \definecolor{light goldenrod yellow}{rgb}{0.980392,0.980392,0.823529}
  \definecolor{LightGoldenrodYellow}{rgb}{0.980392,0.980392,0.823529}
  \definecolor{light yellow}{rgb}{1.000000,1.000000,0.878431}
  \definecolor{LightYellow}{rgb}{1.000000,1.000000,0.878431}
  \definecolor{yellow}{rgb}{1.000000,1.000000,0.000000}
  \definecolor{gold}{rgb}{1.000000,0.843137,0.000000}
  \definecolor{light goldenrod}{rgb}{0.933333,0.866667,0.509804}
  \definecolor{LightGoldenrod}{rgb}{0.933333,0.866667,0.509804}
  \definecolor{goldenrod}{rgb}{0.854902,0.647059,0.125490}
  \definecolor{dark goldenrod}{rgb}{0.721569,0.525490,0.043137}
  \definecolor{DarkGoldenrod}{rgb}{0.721569,0.525490,0.043137}
  \definecolor{rosy brown}{rgb}{0.737255,0.560784,0.560784}
  \definecolor{RosyBrown}{rgb}{0.737255,0.560784,0.560784}
  \definecolor{indian red}{rgb}{0.803922,0.360784,0.360784}
  \definecolor{IndianRed}{rgb}{0.803922,0.360784,0.360784}
  \definecolor{saddle brown}{rgb}{0.545098,0.270588,0.074510}
  \definecolor{SaddleBrown}{rgb}{0.545098,0.270588,0.074510}
  \definecolor{sienna}{rgb}{0.627451,0.321569,0.176471}
  \definecolor{peru}{rgb}{0.803922,0.521569,0.247059}
  \definecolor{burlywood}{rgb}{0.870588,0.721569,0.529412}
  \definecolor{beige}{rgb}{0.960784,0.960784,0.862745}
  \definecolor{wheat}{rgb}{0.960784,0.870588,0.701961}
  \definecolor{sandy brown}{rgb}{0.956863,0.643137,0.376471}
  \definecolor{SandyBrown}{rgb}{0.956863,0.643137,0.376471}
  \definecolor{tan}{rgb}{0.823529,0.705882,0.549020}
  \definecolor{chocolate}{rgb}{0.823529,0.411765,0.117647}
  \definecolor{firebrick}{rgb}{0.698039,0.133333,0.133333}
  \definecolor{brown}{rgb}{0.647059,0.164706,0.164706}
  \definecolor{dark salmon}{rgb}{0.913725,0.588235,0.478431}
  \definecolor{DarkSalmon}{rgb}{0.913725,0.588235,0.478431}
  \definecolor{salmon}{rgb}{0.980392,0.501961,0.447059}
  \definecolor{light salmon}{rgb}{1.000000,0.627451,0.478431}
  \definecolor{LightSalmon}{rgb}{1.000000,0.627451,0.478431}
  \definecolor{orange}{rgb}{1.000000,0.647059,0.000000}
  \definecolor{dark orange}{rgb}{1.000000,0.549020,0.000000}
  \definecolor{DarkOrange}{rgb}{1.000000,0.549020,0.000000}
  \definecolor{coral}{rgb}{1.000000,0.498039,0.313726}
  \definecolor{light coral}{rgb}{0.941176,0.501961,0.501961}
  \definecolor{LightCoral}{rgb}{0.941176,0.501961,0.501961}
  \definecolor{tomato}{rgb}{1.000000,0.388235,0.278431}
  \definecolor{orange red}{rgb}{1.000000,0.270588,0.000000}
  \definecolor{OrangeRed}{rgb}{1.000000,0.270588,0.000000}
  \definecolor{red}{rgb}{1.000000,0.000000,0.000000}
  \definecolor{hot pink}{rgb}{1.000000,0.411765,0.705882}
  \definecolor{HotPink}{rgb}{1.000000,0.411765,0.705882}
  \definecolor{deep pink}{rgb}{1.000000,0.078431,0.576471}
  \definecolor{DeepPink}{rgb}{1.000000,0.078431,0.576471}
  \definecolor{pink}{rgb}{1.000000,0.752941,0.796078}
  \definecolor{light pink}{rgb}{1.000000,0.713726,0.756863}
  \definecolor{LightPink}{rgb}{1.000000,0.713726,0.756863}
  \definecolor{pale violet red}{rgb}{0.858824,0.439216,0.576471}
  \definecolor{PaleVioletRed}{rgb}{0.858824,0.439216,0.576471}
  \definecolor{maroon}{rgb}{0.690196,0.188235,0.376471}
  \definecolor{medium violet red}{rgb}{0.780392,0.082353,0.521569}
  \definecolor{MediumVioletRed}{rgb}{0.780392,0.082353,0.521569}
  \definecolor{violet red}{rgb}{0.815686,0.125490,0.564706}
  \definecolor{VioletRed}{rgb}{0.815686,0.125490,0.564706}
  \definecolor{magenta}{rgb}{1.000000,0.000000,1.000000}
  \definecolor{violet}{rgb}{0.933333,0.509804,0.933333}
  \definecolor{plum}{rgb}{0.866667,0.627451,0.866667}
  \definecolor{orchid}{rgb}{0.854902,0.439216,0.839216}
  \definecolor{medium orchid}{rgb}{0.729412,0.333333,0.827451}
  \definecolor{MediumOrchid}{rgb}{0.729412,0.333333,0.827451}
  \definecolor{dark orchid}{rgb}{0.600000,0.196078,0.800000}
  \definecolor{DarkOrchid}{rgb}{0.600000,0.196078,0.800000}
  \definecolor{dark violet}{rgb}{0.580392,0.000000,0.827451}
  \definecolor{DarkViolet}{rgb}{0.580392,0.000000,0.827451}
  \definecolor{blue violet}{rgb}{0.541176,0.168627,0.886275}
  \definecolor{BlueViolet}{rgb}{0.541176,0.168627,0.886275}
  \definecolor{purple}{rgb}{0.627451,0.125490,0.941176}
  \definecolor{medium purple}{rgb}{0.576471,0.439216,0.858824}
  \definecolor{MediumPurple}{rgb}{0.576471,0.439216,0.858824}
  \definecolor{thistle}{rgb}{0.847059,0.749020,0.847059}
  \definecolor{snow1}{rgb}{1.000000,0.980392,0.980392}
  \definecolor{snow2}{rgb}{0.933333,0.913725,0.913725}
  \definecolor{snow3}{rgb}{0.803922,0.788235,0.788235}
  \definecolor{snow4}{rgb}{0.545098,0.537255,0.537255}
  \definecolor{seashell1}{rgb}{1.000000,0.960784,0.933333}
  \definecolor{seashell2}{rgb}{0.933333,0.898039,0.870588}
  \definecolor{seashell3}{rgb}{0.803922,0.772549,0.749020}
  \definecolor{seashell4}{rgb}{0.545098,0.525490,0.509804}
  \definecolor{AntiqueWhite1}{rgb}{1.000000,0.937255,0.858824}
  \definecolor{AntiqueWhite2}{rgb}{0.933333,0.874510,0.800000}
  \definecolor{AntiqueWhite3}{rgb}{0.803922,0.752941,0.690196}
  \definecolor{AntiqueWhite4}{rgb}{0.545098,0.513726,0.470588}
  \definecolor{bisque1}{rgb}{1.000000,0.894118,0.768627}
  \definecolor{bisque2}{rgb}{0.933333,0.835294,0.717647}
  \definecolor{bisque3}{rgb}{0.803922,0.717647,0.619608}
  \definecolor{bisque4}{rgb}{0.545098,0.490196,0.419608}
  \definecolor{PeachPuff1}{rgb}{1.000000,0.854902,0.725490}
  \definecolor{PeachPuff2}{rgb}{0.933333,0.796078,0.678431}
  \definecolor{PeachPuff3}{rgb}{0.803922,0.686275,0.584314}
  \definecolor{PeachPuff4}{rgb}{0.545098,0.466667,0.396078}
  \definecolor{NavajoWhite1}{rgb}{1.000000,0.870588,0.678431}
  \definecolor{NavajoWhite2}{rgb}{0.933333,0.811765,0.631373}
  \definecolor{NavajoWhite3}{rgb}{0.803922,0.701961,0.545098}
  \definecolor{NavajoWhite4}{rgb}{0.545098,0.474510,0.368627}
  \definecolor{LemonChiffon1}{rgb}{1.000000,0.980392,0.803922}
  \definecolor{LemonChiffon2}{rgb}{0.933333,0.913725,0.749020}
  \definecolor{LemonChiffon3}{rgb}{0.803922,0.788235,0.647059}
  \definecolor{LemonChiffon4}{rgb}{0.545098,0.537255,0.439216}
  \definecolor{cornsilk1}{rgb}{1.000000,0.972549,0.862745}
  \definecolor{cornsilk2}{rgb}{0.933333,0.909804,0.803922}
  \definecolor{cornsilk3}{rgb}{0.803922,0.784314,0.694118}
  \definecolor{cornsilk4}{rgb}{0.545098,0.533333,0.470588}
  \definecolor{ivory1}{rgb}{1.000000,1.000000,0.941176}
  \definecolor{ivory2}{rgb}{0.933333,0.933333,0.878431}
  \definecolor{ivory3}{rgb}{0.803922,0.803922,0.756863}
  \definecolor{ivory4}{rgb}{0.545098,0.545098,0.513726}
  \definecolor{honeydew1}{rgb}{0.941176,1.000000,0.941176}
  \definecolor{honeydew2}{rgb}{0.878431,0.933333,0.878431}
  \definecolor{honeydew3}{rgb}{0.756863,0.803922,0.756863}
  \definecolor{honeydew4}{rgb}{0.513726,0.545098,0.513726}
  \definecolor{LavenderBlush1}{rgb}{1.000000,0.941176,0.960784}
  \definecolor{LavenderBlush2}{rgb}{0.933333,0.878431,0.898039}
  \definecolor{LavenderBlush3}{rgb}{0.803922,0.756863,0.772549}
  \definecolor{LavenderBlush4}{rgb}{0.545098,0.513726,0.525490}
  \definecolor{MistyRose1}{rgb}{1.000000,0.894118,0.882353}
  \definecolor{MistyRose2}{rgb}{0.933333,0.835294,0.823529}
  \definecolor{MistyRose3}{rgb}{0.803922,0.717647,0.709804}
  \definecolor{MistyRose4}{rgb}{0.545098,0.490196,0.482353}
  \definecolor{azure1}{rgb}{0.941176,1.000000,1.000000}
  \definecolor{azure2}{rgb}{0.878431,0.933333,0.933333}
  \definecolor{azure3}{rgb}{0.756863,0.803922,0.803922}
  \definecolor{azure4}{rgb}{0.513726,0.545098,0.545098}
  \definecolor{SlateBlue1}{rgb}{0.513726,0.435294,1.000000}
  \definecolor{SlateBlue2}{rgb}{0.478431,0.403922,0.933333}
  \definecolor{SlateBlue3}{rgb}{0.411765,0.349020,0.803922}
  \definecolor{SlateBlue4}{rgb}{0.278431,0.235294,0.545098}
  \definecolor{RoyalBlue1}{rgb}{0.282353,0.462745,1.000000}
  \definecolor{RoyalBlue2}{rgb}{0.262745,0.431373,0.933333}
  \definecolor{RoyalBlue3}{rgb}{0.227451,0.372549,0.803922}
  \definecolor{RoyalBlue4}{rgb}{0.152941,0.250980,0.545098}
  \definecolor{blue1}{rgb}{0.000000,0.000000,1.000000}
  \definecolor{blue2}{rgb}{0.000000,0.000000,0.933333}
  \definecolor{blue3}{rgb}{0.000000,0.000000,0.803922}
  \definecolor{blue4}{rgb}{0.000000,0.000000,0.545098}
  \definecolor{DodgerBlue1}{rgb}{0.117647,0.564706,1.000000}
  \definecolor{DodgerBlue2}{rgb}{0.109804,0.525490,0.933333}
  \definecolor{DodgerBlue3}{rgb}{0.094118,0.454902,0.803922}
  \definecolor{DodgerBlue4}{rgb}{0.062745,0.305882,0.545098}
  \definecolor{SteelBlue1}{rgb}{0.388235,0.721569,1.000000}
  \definecolor{SteelBlue2}{rgb}{0.360784,0.674510,0.933333}
  \definecolor{SteelBlue3}{rgb}{0.309804,0.580392,0.803922}
  \definecolor{SteelBlue4}{rgb}{0.211765,0.392157,0.545098}
  \definecolor{DeepSkyBlue1}{rgb}{0.000000,0.749020,1.000000}
  \definecolor{DeepSkyBlue2}{rgb}{0.000000,0.698039,0.933333}
  \definecolor{DeepSkyBlue3}{rgb}{0.000000,0.603922,0.803922}
  \definecolor{DeepSkyBlue4}{rgb}{0.000000,0.407843,0.545098}
  \definecolor{SkyBlue1}{rgb}{0.529412,0.807843,1.000000}
  \definecolor{SkyBlue2}{rgb}{0.494118,0.752941,0.933333}
  \definecolor{SkyBlue3}{rgb}{0.423529,0.650980,0.803922}
  \definecolor{SkyBlue4}{rgb}{0.290196,0.439216,0.545098}
  \definecolor{LightSkyBlue1}{rgb}{0.690196,0.886275,1.000000}
  \definecolor{LightSkyBlue2}{rgb}{0.643137,0.827451,0.933333}
  \definecolor{LightSkyBlue3}{rgb}{0.552941,0.713726,0.803922}
  \definecolor{LightSkyBlue4}{rgb}{0.376471,0.482353,0.545098}
  \definecolor{SlateGray1}{rgb}{0.776471,0.886275,1.000000}
  \definecolor{SlateGray2}{rgb}{0.725490,0.827451,0.933333}
  \definecolor{SlateGray3}{rgb}{0.623529,0.713726,0.803922}
  \definecolor{SlateGray4}{rgb}{0.423529,0.482353,0.545098}
  \definecolor{LightSteelBlue1}{rgb}{0.792157,0.882353,1.000000}
  \definecolor{LightSteelBlue2}{rgb}{0.737255,0.823529,0.933333}
  \definecolor{LightSteelBlue3}{rgb}{0.635294,0.709804,0.803922}
  \definecolor{LightSteelBlue4}{rgb}{0.431373,0.482353,0.545098}
  \definecolor{LightBlue1}{rgb}{0.749020,0.937255,1.000000}
  \definecolor{LightBlue2}{rgb}{0.698039,0.874510,0.933333}
  \definecolor{LightBlue3}{rgb}{0.603922,0.752941,0.803922}
  \definecolor{LightBlue4}{rgb}{0.407843,0.513726,0.545098}
  \definecolor{LightCyan1}{rgb}{0.878431,1.000000,1.000000}
  \definecolor{LightCyan2}{rgb}{0.819608,0.933333,0.933333}
  \definecolor{LightCyan3}{rgb}{0.705882,0.803922,0.803922}
  \definecolor{LightCyan4}{rgb}{0.478431,0.545098,0.545098}
  \definecolor{PaleTurquoise1}{rgb}{0.733333,1.000000,1.000000}
  \definecolor{PaleTurquoise2}{rgb}{0.682353,0.933333,0.933333}
  \definecolor{PaleTurquoise3}{rgb}{0.588235,0.803922,0.803922}
  \definecolor{PaleTurquoise4}{rgb}{0.400000,0.545098,0.545098}
  \definecolor{CadetBlue1}{rgb}{0.596078,0.960784,1.000000}
  \definecolor{CadetBlue2}{rgb}{0.556863,0.898039,0.933333}
  \definecolor{CadetBlue3}{rgb}{0.478431,0.772549,0.803922}
  \definecolor{CadetBlue4}{rgb}{0.325490,0.525490,0.545098}
  \definecolor{turquoise1}{rgb}{0.000000,0.960784,1.000000}
  \definecolor{turquoise2}{rgb}{0.000000,0.898039,0.933333}
  \definecolor{turquoise3}{rgb}{0.000000,0.772549,0.803922}
  \definecolor{turquoise4}{rgb}{0.000000,0.525490,0.545098}
  \definecolor{cyan1}{rgb}{0.000000,1.000000,1.000000}
  \definecolor{cyan2}{rgb}{0.000000,0.933333,0.933333}
  \definecolor{cyan3}{rgb}{0.000000,0.803922,0.803922}
  \definecolor{cyan4}{rgb}{0.000000,0.545098,0.545098}
  \definecolor{DarkSlateGray1}{rgb}{0.592157,1.000000,1.000000}
  \definecolor{DarkSlateGray2}{rgb}{0.552941,0.933333,0.933333}
  \definecolor{DarkSlateGray3}{rgb}{0.474510,0.803922,0.803922}
  \definecolor{DarkSlateGray4}{rgb}{0.321569,0.545098,0.545098}
  \definecolor{aquamarine1}{rgb}{0.498039,1.000000,0.831373}
  \definecolor{aquamarine2}{rgb}{0.462745,0.933333,0.776471}
  \definecolor{aquamarine3}{rgb}{0.400000,0.803922,0.666667}
  \definecolor{aquamarine4}{rgb}{0.270588,0.545098,0.454902}
  \definecolor{DarkSeaGreen1}{rgb}{0.756863,1.000000,0.756863}
  \definecolor{DarkSeaGreen2}{rgb}{0.705882,0.933333,0.705882}
  \definecolor{DarkSeaGreen3}{rgb}{0.607843,0.803922,0.607843}
  \definecolor{DarkSeaGreen4}{rgb}{0.411765,0.545098,0.411765}
  \definecolor{SeaGreen1}{rgb}{0.329412,1.000000,0.623529}
  \definecolor{SeaGreen2}{rgb}{0.305882,0.933333,0.580392}
  \definecolor{SeaGreen3}{rgb}{0.262745,0.803922,0.501961}
  \definecolor{SeaGreen4}{rgb}{0.180392,0.545098,0.341176}
  \definecolor{PaleGreen1}{rgb}{0.603922,1.000000,0.603922}
  \definecolor{PaleGreen2}{rgb}{0.564706,0.933333,0.564706}
  \definecolor{PaleGreen3}{rgb}{0.486275,0.803922,0.486275}
  \definecolor{PaleGreen4}{rgb}{0.329412,0.545098,0.329412}
  \definecolor{SpringGreen1}{rgb}{0.000000,1.000000,0.498039}
  \definecolor{SpringGreen2}{rgb}{0.000000,0.933333,0.462745}
  \definecolor{SpringGreen3}{rgb}{0.000000,0.803922,0.400000}
  \definecolor{SpringGreen4}{rgb}{0.000000,0.545098,0.270588}
  \definecolor{green1}{rgb}{0.000000,1.000000,0.000000}
  \definecolor{green2}{rgb}{0.000000,0.933333,0.000000}
  \definecolor{green3}{rgb}{0.000000,0.803922,0.000000}
  \definecolor{green4}{rgb}{0.000000,0.545098,0.000000}
  \definecolor{chartreuse1}{rgb}{0.498039,1.000000,0.000000}
  \definecolor{chartreuse2}{rgb}{0.462745,0.933333,0.000000}
  \definecolor{chartreuse3}{rgb}{0.400000,0.803922,0.000000}
  \definecolor{chartreuse4}{rgb}{0.270588,0.545098,0.000000}
  \definecolor{OliveDrab1}{rgb}{0.752941,1.000000,0.243137}
  \definecolor{OliveDrab2}{rgb}{0.701961,0.933333,0.227451}
  \definecolor{OliveDrab3}{rgb}{0.603922,0.803922,0.196078}
  \definecolor{OliveDrab4}{rgb}{0.411765,0.545098,0.133333}
  \definecolor{DarkOliveGreen1}{rgb}{0.792157,1.000000,0.439216}
  \definecolor{DarkOliveGreen2}{rgb}{0.737255,0.933333,0.407843}
  \definecolor{DarkOliveGreen3}{rgb}{0.635294,0.803922,0.352941}
  \definecolor{DarkOliveGreen4}{rgb}{0.431373,0.545098,0.239216}
  \definecolor{khaki1}{rgb}{1.000000,0.964706,0.560784}
  \definecolor{khaki2}{rgb}{0.933333,0.901961,0.521569}
  \definecolor{khaki3}{rgb}{0.803922,0.776471,0.450980}
  \definecolor{khaki4}{rgb}{0.545098,0.525490,0.305882}
  \definecolor{LightGoldenrod1}{rgb}{1.000000,0.925490,0.545098}
  \definecolor{LightGoldenrod2}{rgb}{0.933333,0.862745,0.509804}
  \definecolor{LightGoldenrod3}{rgb}{0.803922,0.745098,0.439216}
  \definecolor{LightGoldenrod4}{rgb}{0.545098,0.505882,0.298039}
  \definecolor{LightYellow1}{rgb}{1.000000,1.000000,0.878431}
  \definecolor{LightYellow2}{rgb}{0.933333,0.933333,0.819608}
  \definecolor{LightYellow3}{rgb}{0.803922,0.803922,0.705882}
  \definecolor{LightYellow4}{rgb}{0.545098,0.545098,0.478431}
  \definecolor{yellow1}{rgb}{1.000000,1.000000,0.000000}
  \definecolor{yellow2}{rgb}{0.933333,0.933333,0.000000}
  \definecolor{yellow3}{rgb}{0.803922,0.803922,0.000000}
  \definecolor{yellow4}{rgb}{0.545098,0.545098,0.000000}
  \definecolor{gold1}{rgb}{1.000000,0.843137,0.000000}
  \definecolor{gold2}{rgb}{0.933333,0.788235,0.000000}
  \definecolor{gold3}{rgb}{0.803922,0.678431,0.000000}
  \definecolor{gold4}{rgb}{0.545098,0.458824,0.000000}
  \definecolor{goldenrod1}{rgb}{1.000000,0.756863,0.145098}
  \definecolor{goldenrod2}{rgb}{0.933333,0.705882,0.133333}
  \definecolor{goldenrod3}{rgb}{0.803922,0.607843,0.113725}
  \definecolor{goldenrod4}{rgb}{0.545098,0.411765,0.078431}
  \definecolor{DarkGoldenrod1}{rgb}{1.000000,0.725490,0.058824}
  \definecolor{DarkGoldenrod2}{rgb}{0.933333,0.678431,0.054902}
  \definecolor{DarkGoldenrod3}{rgb}{0.803922,0.584314,0.047059}
  \definecolor{DarkGoldenrod4}{rgb}{0.545098,0.396078,0.031373}
  \definecolor{RosyBrown1}{rgb}{1.000000,0.756863,0.756863}
  \definecolor{RosyBrown2}{rgb}{0.933333,0.705882,0.705882}
  \definecolor{RosyBrown3}{rgb}{0.803922,0.607843,0.607843}
  \definecolor{RosyBrown4}{rgb}{0.545098,0.411765,0.411765}
  \definecolor{IndianRed1}{rgb}{1.000000,0.415686,0.415686}
  \definecolor{IndianRed2}{rgb}{0.933333,0.388235,0.388235}
  \definecolor{IndianRed3}{rgb}{0.803922,0.333333,0.333333}
  \definecolor{IndianRed4}{rgb}{0.545098,0.227451,0.227451}
  \definecolor{sienna1}{rgb}{1.000000,0.509804,0.278431}
  \definecolor{sienna2}{rgb}{0.933333,0.474510,0.258824}
  \definecolor{sienna3}{rgb}{0.803922,0.407843,0.223529}
  \definecolor{sienna4}{rgb}{0.545098,0.278431,0.149020}
  \definecolor{burlywood1}{rgb}{1.000000,0.827451,0.607843}
  \definecolor{burlywood2}{rgb}{0.933333,0.772549,0.568627}
  \definecolor{burlywood3}{rgb}{0.803922,0.666667,0.490196}
  \definecolor{burlywood4}{rgb}{0.545098,0.450980,0.333333}
  \definecolor{wheat1}{rgb}{1.000000,0.905882,0.729412}
  \definecolor{wheat2}{rgb}{0.933333,0.847059,0.682353}
  \definecolor{wheat3}{rgb}{0.803922,0.729412,0.588235}
  \definecolor{wheat4}{rgb}{0.545098,0.494118,0.400000}
  \definecolor{tan1}{rgb}{1.000000,0.647059,0.309804}
  \definecolor{tan2}{rgb}{0.933333,0.603922,0.286275}
  \definecolor{tan3}{rgb}{0.803922,0.521569,0.247059}
  \definecolor{tan4}{rgb}{0.545098,0.352941,0.168627}
  \definecolor{chocolate1}{rgb}{1.000000,0.498039,0.141176}
  \definecolor{chocolate2}{rgb}{0.933333,0.462745,0.129412}
  \definecolor{chocolate3}{rgb}{0.803922,0.400000,0.113725}
  \definecolor{chocolate4}{rgb}{0.545098,0.270588,0.074510}
  \definecolor{firebrick1}{rgb}{1.000000,0.188235,0.188235}
  \definecolor{firebrick2}{rgb}{0.933333,0.172549,0.172549}
  \definecolor{firebrick3}{rgb}{0.803922,0.149020,0.149020}
  \definecolor{firebrick4}{rgb}{0.545098,0.101961,0.101961}
  \definecolor{brown1}{rgb}{1.000000,0.250980,0.250980}
  \definecolor{brown2}{rgb}{0.933333,0.231373,0.231373}
  \definecolor{brown3}{rgb}{0.803922,0.200000,0.200000}
  \definecolor{brown4}{rgb}{0.545098,0.137255,0.137255}
  \definecolor{salmon1}{rgb}{1.000000,0.549020,0.411765}
  \definecolor{salmon2}{rgb}{0.933333,0.509804,0.384314}
  \definecolor{salmon3}{rgb}{0.803922,0.439216,0.329412}
  \definecolor{salmon4}{rgb}{0.545098,0.298039,0.223529}
  \definecolor{LightSalmon1}{rgb}{1.000000,0.627451,0.478431}
  \definecolor{LightSalmon2}{rgb}{0.933333,0.584314,0.447059}
  \definecolor{LightSalmon3}{rgb}{0.803922,0.505882,0.384314}
  \definecolor{LightSalmon4}{rgb}{0.545098,0.341176,0.258824}
  \definecolor{orange1}{rgb}{1.000000,0.647059,0.000000}
  \definecolor{orange2}{rgb}{0.933333,0.603922,0.000000}
  \definecolor{orange3}{rgb}{0.803922,0.521569,0.000000}
  \definecolor{orange4}{rgb}{0.545098,0.352941,0.000000}
  \definecolor{DarkOrange1}{rgb}{1.000000,0.498039,0.000000}
  \definecolor{DarkOrange2}{rgb}{0.933333,0.462745,0.000000}
  \definecolor{DarkOrange3}{rgb}{0.803922,0.400000,0.000000}
  \definecolor{DarkOrange4}{rgb}{0.545098,0.270588,0.000000}
  \definecolor{coral1}{rgb}{1.000000,0.447059,0.337255}
  \definecolor{coral2}{rgb}{0.933333,0.415686,0.313726}
  \definecolor{coral3}{rgb}{0.803922,0.356863,0.270588}
  \definecolor{coral4}{rgb}{0.545098,0.243137,0.184314}
  \definecolor{tomato1}{rgb}{1.000000,0.388235,0.278431}
  \definecolor{tomato2}{rgb}{0.933333,0.360784,0.258824}
  \definecolor{tomato3}{rgb}{0.803922,0.309804,0.223529}
  \definecolor{tomato4}{rgb}{0.545098,0.211765,0.149020}
  \definecolor{OrangeRed1}{rgb}{1.000000,0.270588,0.000000}
  \definecolor{OrangeRed2}{rgb}{0.933333,0.250980,0.000000}
  \definecolor{OrangeRed3}{rgb}{0.803922,0.215686,0.000000}
  \definecolor{OrangeRed4}{rgb}{0.545098,0.145098,0.000000}
  \definecolor{red1}{rgb}{1.000000,0.000000,0.000000}
  \definecolor{red2}{rgb}{0.933333,0.000000,0.000000}
  \definecolor{red3}{rgb}{0.803922,0.000000,0.000000}
  \definecolor{red4}{rgb}{0.545098,0.000000,0.000000}
  \definecolor{DeepPink1}{rgb}{1.000000,0.078431,0.576471}
  \definecolor{DeepPink2}{rgb}{0.933333,0.070588,0.537255}
  \definecolor{DeepPink3}{rgb}{0.803922,0.062745,0.462745}
  \definecolor{DeepPink4}{rgb}{0.545098,0.039216,0.313726}
  \definecolor{HotPink1}{rgb}{1.000000,0.431373,0.705882}
  \definecolor{HotPink2}{rgb}{0.933333,0.415686,0.654902}
  \definecolor{HotPink3}{rgb}{0.803922,0.376471,0.564706}
  \definecolor{HotPink4}{rgb}{0.545098,0.227451,0.384314}
  \definecolor{pink1}{rgb}{1.000000,0.709804,0.772549}
  \definecolor{pink2}{rgb}{0.933333,0.662745,0.721569}
  \definecolor{pink3}{rgb}{0.803922,0.568627,0.619608}
  \definecolor{pink4}{rgb}{0.545098,0.388235,0.423529}
  \definecolor{LightPink1}{rgb}{1.000000,0.682353,0.725490}
  \definecolor{LightPink2}{rgb}{0.933333,0.635294,0.678431}
  \definecolor{LightPink3}{rgb}{0.803922,0.549020,0.584314}
  \definecolor{LightPink4}{rgb}{0.545098,0.372549,0.396078}
  \definecolor{PaleVioletRed1}{rgb}{1.000000,0.509804,0.670588}
  \definecolor{PaleVioletRed2}{rgb}{0.933333,0.474510,0.623529}
  \definecolor{PaleVioletRed3}{rgb}{0.803922,0.407843,0.537255}
  \definecolor{PaleVioletRed4}{rgb}{0.545098,0.278431,0.364706}
  \definecolor{maroon1}{rgb}{1.000000,0.203922,0.701961}
  \definecolor{maroon2}{rgb}{0.933333,0.188235,0.654902}
  \definecolor{maroon3}{rgb}{0.803922,0.160784,0.564706}
  \definecolor{maroon4}{rgb}{0.545098,0.109804,0.384314}
  \definecolor{VioletRed1}{rgb}{1.000000,0.243137,0.588235}
  \definecolor{VioletRed2}{rgb}{0.933333,0.227451,0.549020}
  \definecolor{VioletRed3}{rgb}{0.803922,0.196078,0.470588}
  \definecolor{VioletRed4}{rgb}{0.545098,0.133333,0.321569}
  \definecolor{magenta1}{rgb}{1.000000,0.000000,1.000000}
  \definecolor{magenta2}{rgb}{0.933333,0.000000,0.933333}
  \definecolor{magenta3}{rgb}{0.803922,0.000000,0.803922}
  \definecolor{magenta4}{rgb}{0.545098,0.000000,0.545098}
  \definecolor{orchid1}{rgb}{1.000000,0.513726,0.980392}
  \definecolor{orchid2}{rgb}{0.933333,0.478431,0.913725}
  \definecolor{orchid3}{rgb}{0.803922,0.411765,0.788235}
  \definecolor{orchid4}{rgb}{0.545098,0.278431,0.537255}
  \definecolor{plum1}{rgb}{1.000000,0.733333,1.000000}
  \definecolor{plum2}{rgb}{0.933333,0.682353,0.933333}
  \definecolor{plum3}{rgb}{0.803922,0.588235,0.803922}
  \definecolor{plum4}{rgb}{0.545098,0.400000,0.545098}
  \definecolor{MediumOrchid1}{rgb}{0.878431,0.400000,1.000000}
  \definecolor{MediumOrchid2}{rgb}{0.819608,0.372549,0.933333}
  \definecolor{MediumOrchid3}{rgb}{0.705882,0.321569,0.803922}
  \definecolor{MediumOrchid4}{rgb}{0.478431,0.215686,0.545098}
  \definecolor{DarkOrchid1}{rgb}{0.749020,0.243137,1.000000}
  \definecolor{DarkOrchid2}{rgb}{0.698039,0.227451,0.933333}
  \definecolor{DarkOrchid3}{rgb}{0.603922,0.196078,0.803922}
  \definecolor{DarkOrchid4}{rgb}{0.407843,0.133333,0.545098}
  \definecolor{purple1}{rgb}{0.607843,0.188235,1.000000}
  \definecolor{purple2}{rgb}{0.568627,0.172549,0.933333}
  \definecolor{purple3}{rgb}{0.490196,0.149020,0.803922}
  \definecolor{purple4}{rgb}{0.333333,0.101961,0.545098}
  \definecolor{MediumPurple1}{rgb}{0.670588,0.509804,1.000000}
  \definecolor{MediumPurple2}{rgb}{0.623529,0.474510,0.933333}
  \definecolor{MediumPurple3}{rgb}{0.537255,0.407843,0.803922}
  \definecolor{MediumPurple4}{rgb}{0.364706,0.278431,0.545098}
  \definecolor{thistle1}{rgb}{1.000000,0.882353,1.000000}
  \definecolor{thistle2}{rgb}{0.933333,0.823529,0.933333}
  \definecolor{thistle3}{rgb}{0.803922,0.709804,0.803922}
  \definecolor{thistle4}{rgb}{0.545098,0.482353,0.545098}
  \definecolor{gray0}{rgb}{0.000000,0.000000,0.000000}
  \definecolor{grey0}{rgb}{0.000000,0.000000,0.000000}
  \definecolor{gray1}{rgb}{0.011765,0.011765,0.011765}
  \definecolor{grey1}{rgb}{0.011765,0.011765,0.011765}
  \definecolor{gray2}{rgb}{0.019608,0.019608,0.019608}
  \definecolor{grey2}{rgb}{0.019608,0.019608,0.019608}
  \definecolor{gray3}{rgb}{0.031373,0.031373,0.031373}
  \definecolor{grey3}{rgb}{0.031373,0.031373,0.031373}
  \definecolor{gray4}{rgb}{0.039216,0.039216,0.039216}
  \definecolor{grey4}{rgb}{0.039216,0.039216,0.039216}
  \definecolor{gray5}{rgb}{0.050980,0.050980,0.050980}
  \definecolor{grey5}{rgb}{0.050980,0.050980,0.050980}
  \definecolor{gray6}{rgb}{0.058824,0.058824,0.058824}
  \definecolor{grey6}{rgb}{0.058824,0.058824,0.058824}
  \definecolor{gray7}{rgb}{0.070588,0.070588,0.070588}
  \definecolor{grey7}{rgb}{0.070588,0.070588,0.070588}
  \definecolor{gray8}{rgb}{0.078431,0.078431,0.078431}
  \definecolor{grey8}{rgb}{0.078431,0.078431,0.078431}
  \definecolor{gray9}{rgb}{0.090196,0.090196,0.090196}
  \definecolor{grey9}{rgb}{0.090196,0.090196,0.090196}
  \definecolor{gray10}{rgb}{0.101961,0.101961,0.101961}
  \definecolor{grey10}{rgb}{0.101961,0.101961,0.101961}
  \definecolor{gray11}{rgb}{0.109804,0.109804,0.109804}
  \definecolor{grey11}{rgb}{0.109804,0.109804,0.109804}
  \definecolor{gray12}{rgb}{0.121569,0.121569,0.121569}
  \definecolor{grey12}{rgb}{0.121569,0.121569,0.121569}
  \definecolor{gray13}{rgb}{0.129412,0.129412,0.129412}
  \definecolor{grey13}{rgb}{0.129412,0.129412,0.129412}
  \definecolor{gray14}{rgb}{0.141176,0.141176,0.141176}
  \definecolor{grey14}{rgb}{0.141176,0.141176,0.141176}
  \definecolor{gray15}{rgb}{0.149020,0.149020,0.149020}
  \definecolor{grey15}{rgb}{0.149020,0.149020,0.149020}
  \definecolor{gray16}{rgb}{0.160784,0.160784,0.160784}
  \definecolor{grey16}{rgb}{0.160784,0.160784,0.160784}
  \definecolor{gray17}{rgb}{0.168627,0.168627,0.168627}
  \definecolor{grey17}{rgb}{0.168627,0.168627,0.168627}
  \definecolor{gray18}{rgb}{0.180392,0.180392,0.180392}
  \definecolor{grey18}{rgb}{0.180392,0.180392,0.180392}
  \definecolor{gray19}{rgb}{0.188235,0.188235,0.188235}
  \definecolor{grey19}{rgb}{0.188235,0.188235,0.188235}
  \definecolor{gray20}{rgb}{0.200000,0.200000,0.200000}
  \definecolor{grey20}{rgb}{0.200000,0.200000,0.200000}
  \definecolor{gray21}{rgb}{0.211765,0.211765,0.211765}
  \definecolor{grey21}{rgb}{0.211765,0.211765,0.211765}
  \definecolor{gray22}{rgb}{0.219608,0.219608,0.219608}
  \definecolor{grey22}{rgb}{0.219608,0.219608,0.219608}
  \definecolor{gray23}{rgb}{0.231373,0.231373,0.231373}
  \definecolor{grey23}{rgb}{0.231373,0.231373,0.231373}
  \definecolor{gray24}{rgb}{0.239216,0.239216,0.239216}
  \definecolor{grey24}{rgb}{0.239216,0.239216,0.239216}
  \definecolor{gray25}{rgb}{0.250980,0.250980,0.250980}
  \definecolor{grey25}{rgb}{0.250980,0.250980,0.250980}
  \definecolor{gray26}{rgb}{0.258824,0.258824,0.258824}
  \definecolor{grey26}{rgb}{0.258824,0.258824,0.258824}
  \definecolor{gray27}{rgb}{0.270588,0.270588,0.270588}
  \definecolor{grey27}{rgb}{0.270588,0.270588,0.270588}
  \definecolor{gray28}{rgb}{0.278431,0.278431,0.278431}
  \definecolor{grey28}{rgb}{0.278431,0.278431,0.278431}
  \definecolor{gray29}{rgb}{0.290196,0.290196,0.290196}
  \definecolor{grey29}{rgb}{0.290196,0.290196,0.290196}
  \definecolor{gray30}{rgb}{0.301961,0.301961,0.301961}
  \definecolor{grey30}{rgb}{0.301961,0.301961,0.301961}
  \definecolor{gray31}{rgb}{0.309804,0.309804,0.309804}
  \definecolor{grey31}{rgb}{0.309804,0.309804,0.309804}
  \definecolor{gray32}{rgb}{0.321569,0.321569,0.321569}
  \definecolor{grey32}{rgb}{0.321569,0.321569,0.321569}
  \definecolor{gray33}{rgb}{0.329412,0.329412,0.329412}
  \definecolor{grey33}{rgb}{0.329412,0.329412,0.329412}
  \definecolor{gray34}{rgb}{0.341176,0.341176,0.341176}
  \definecolor{grey34}{rgb}{0.341176,0.341176,0.341176}
  \definecolor{gray35}{rgb}{0.349020,0.349020,0.349020}
  \definecolor{grey35}{rgb}{0.349020,0.349020,0.349020}
  \definecolor{gray36}{rgb}{0.360784,0.360784,0.360784}
  \definecolor{grey36}{rgb}{0.360784,0.360784,0.360784}
  \definecolor{gray37}{rgb}{0.368627,0.368627,0.368627}
  \definecolor{grey37}{rgb}{0.368627,0.368627,0.368627}
  \definecolor{gray38}{rgb}{0.380392,0.380392,0.380392}
  \definecolor{grey38}{rgb}{0.380392,0.380392,0.380392}
  \definecolor{gray39}{rgb}{0.388235,0.388235,0.388235}
  \definecolor{grey39}{rgb}{0.388235,0.388235,0.388235}
  \definecolor{gray40}{rgb}{0.400000,0.400000,0.400000}
  \definecolor{grey40}{rgb}{0.400000,0.400000,0.400000}
  \definecolor{gray41}{rgb}{0.411765,0.411765,0.411765}
  \definecolor{grey41}{rgb}{0.411765,0.411765,0.411765}
  \definecolor{gray42}{rgb}{0.419608,0.419608,0.419608}
  \definecolor{grey42}{rgb}{0.419608,0.419608,0.419608}
  \definecolor{gray43}{rgb}{0.431373,0.431373,0.431373}
  \definecolor{grey43}{rgb}{0.431373,0.431373,0.431373}
  \definecolor{gray44}{rgb}{0.439216,0.439216,0.439216}
  \definecolor{grey44}{rgb}{0.439216,0.439216,0.439216}
  \definecolor{gray45}{rgb}{0.450980,0.450980,0.450980}
  \definecolor{grey45}{rgb}{0.450980,0.450980,0.450980}
  \definecolor{gray46}{rgb}{0.458824,0.458824,0.458824}
  \definecolor{grey46}{rgb}{0.458824,0.458824,0.458824}
  \definecolor{gray47}{rgb}{0.470588,0.470588,0.470588}
  \definecolor{grey47}{rgb}{0.470588,0.470588,0.470588}
  \definecolor{gray48}{rgb}{0.478431,0.478431,0.478431}
  \definecolor{grey48}{rgb}{0.478431,0.478431,0.478431}
  \definecolor{gray49}{rgb}{0.490196,0.490196,0.490196}
  \definecolor{grey49}{rgb}{0.490196,0.490196,0.490196}
  \definecolor{gray50}{rgb}{0.498039,0.498039,0.498039}
  \definecolor{grey50}{rgb}{0.498039,0.498039,0.498039}
  \definecolor{gray51}{rgb}{0.509804,0.509804,0.509804}
  \definecolor{grey51}{rgb}{0.509804,0.509804,0.509804}
  \definecolor{gray52}{rgb}{0.521569,0.521569,0.521569}
  \definecolor{grey52}{rgb}{0.521569,0.521569,0.521569}
  \definecolor{gray53}{rgb}{0.529412,0.529412,0.529412}
  \definecolor{grey53}{rgb}{0.529412,0.529412,0.529412}
  \definecolor{gray54}{rgb}{0.541176,0.541176,0.541176}
  \definecolor{grey54}{rgb}{0.541176,0.541176,0.541176}
  \definecolor{gray55}{rgb}{0.549020,0.549020,0.549020}
  \definecolor{grey55}{rgb}{0.549020,0.549020,0.549020}
  \definecolor{gray56}{rgb}{0.560784,0.560784,0.560784}
  \definecolor{grey56}{rgb}{0.560784,0.560784,0.560784}
  \definecolor{gray57}{rgb}{0.568627,0.568627,0.568627}
  \definecolor{grey57}{rgb}{0.568627,0.568627,0.568627}
  \definecolor{gray58}{rgb}{0.580392,0.580392,0.580392}
  \definecolor{grey58}{rgb}{0.580392,0.580392,0.580392}
  \definecolor{gray59}{rgb}{0.588235,0.588235,0.588235}
  \definecolor{grey59}{rgb}{0.588235,0.588235,0.588235}
  \definecolor{gray60}{rgb}{0.600000,0.600000,0.600000}
  \definecolor{grey60}{rgb}{0.600000,0.600000,0.600000}
  \definecolor{gray61}{rgb}{0.611765,0.611765,0.611765}
  \definecolor{grey61}{rgb}{0.611765,0.611765,0.611765}
  \definecolor{gray62}{rgb}{0.619608,0.619608,0.619608}
  \definecolor{grey62}{rgb}{0.619608,0.619608,0.619608}
  \definecolor{gray63}{rgb}{0.631373,0.631373,0.631373}
  \definecolor{grey63}{rgb}{0.631373,0.631373,0.631373}
  \definecolor{gray64}{rgb}{0.639216,0.639216,0.639216}
  \definecolor{grey64}{rgb}{0.639216,0.639216,0.639216}
  \definecolor{gray65}{rgb}{0.650980,0.650980,0.650980}
  \definecolor{grey65}{rgb}{0.650980,0.650980,0.650980}
  \definecolor{gray66}{rgb}{0.658824,0.658824,0.658824}
  \definecolor{grey66}{rgb}{0.658824,0.658824,0.658824}
  \definecolor{gray67}{rgb}{0.670588,0.670588,0.670588}
  \definecolor{grey67}{rgb}{0.670588,0.670588,0.670588}
  \definecolor{gray68}{rgb}{0.678431,0.678431,0.678431}
  \definecolor{grey68}{rgb}{0.678431,0.678431,0.678431}
  \definecolor{gray69}{rgb}{0.690196,0.690196,0.690196}
  \definecolor{grey69}{rgb}{0.690196,0.690196,0.690196}
  \definecolor{gray70}{rgb}{0.701961,0.701961,0.701961}
  \definecolor{grey70}{rgb}{0.701961,0.701961,0.701961}
  \definecolor{gray71}{rgb}{0.709804,0.709804,0.709804}
  \definecolor{grey71}{rgb}{0.709804,0.709804,0.709804}
  \definecolor{gray72}{rgb}{0.721569,0.721569,0.721569}
  \definecolor{grey72}{rgb}{0.721569,0.721569,0.721569}
  \definecolor{gray73}{rgb}{0.729412,0.729412,0.729412}
  \definecolor{grey73}{rgb}{0.729412,0.729412,0.729412}
  \definecolor{gray74}{rgb}{0.741176,0.741176,0.741176}
  \definecolor{grey74}{rgb}{0.741176,0.741176,0.741176}
  \definecolor{gray75}{rgb}{0.749020,0.749020,0.749020}
  \definecolor{grey75}{rgb}{0.749020,0.749020,0.749020}
  \definecolor{gray76}{rgb}{0.760784,0.760784,0.760784}
  \definecolor{grey76}{rgb}{0.760784,0.760784,0.760784}
  \definecolor{gray77}{rgb}{0.768627,0.768627,0.768627}
  \definecolor{grey77}{rgb}{0.768627,0.768627,0.768627}
  \definecolor{gray78}{rgb}{0.780392,0.780392,0.780392}
  \definecolor{grey78}{rgb}{0.780392,0.780392,0.780392}
  \definecolor{gray79}{rgb}{0.788235,0.788235,0.788235}
  \definecolor{grey79}{rgb}{0.788235,0.788235,0.788235}
  \definecolor{gray80}{rgb}{0.800000,0.800000,0.800000}
  \definecolor{grey80}{rgb}{0.800000,0.800000,0.800000}
  \definecolor{gray81}{rgb}{0.811765,0.811765,0.811765}
  \definecolor{grey81}{rgb}{0.811765,0.811765,0.811765}
  \definecolor{gray82}{rgb}{0.819608,0.819608,0.819608}
  \definecolor{grey82}{rgb}{0.819608,0.819608,0.819608}
  \definecolor{gray83}{rgb}{0.831373,0.831373,0.831373}
  \definecolor{grey83}{rgb}{0.831373,0.831373,0.831373}
  \definecolor{gray84}{rgb}{0.839216,0.839216,0.839216}
  \definecolor{grey84}{rgb}{0.839216,0.839216,0.839216}
  \definecolor{gray85}{rgb}{0.850980,0.850980,0.850980}
  \definecolor{grey85}{rgb}{0.850980,0.850980,0.850980}
  \definecolor{gray86}{rgb}{0.858824,0.858824,0.858824}
  \definecolor{grey86}{rgb}{0.858824,0.858824,0.858824}
  \definecolor{gray87}{rgb}{0.870588,0.870588,0.870588}
  \definecolor{grey87}{rgb}{0.870588,0.870588,0.870588}
  \definecolor{gray88}{rgb}{0.878431,0.878431,0.878431}
  \definecolor{grey88}{rgb}{0.878431,0.878431,0.878431}
  \definecolor{gray89}{rgb}{0.890196,0.890196,0.890196}
  \definecolor{grey89}{rgb}{0.890196,0.890196,0.890196}
  \definecolor{gray90}{rgb}{0.898039,0.898039,0.898039}
  \definecolor{grey90}{rgb}{0.898039,0.898039,0.898039}
  \definecolor{gray91}{rgb}{0.909804,0.909804,0.909804}
  \definecolor{grey91}{rgb}{0.909804,0.909804,0.909804}
  \definecolor{gray92}{rgb}{0.921569,0.921569,0.921569}
  \definecolor{grey92}{rgb}{0.921569,0.921569,0.921569}
  \definecolor{gray93}{rgb}{0.929412,0.929412,0.929412}
  \definecolor{grey93}{rgb}{0.929412,0.929412,0.929412}
  \definecolor{gray94}{rgb}{0.941176,0.941176,0.941176}
  \definecolor{grey94}{rgb}{0.941176,0.941176,0.941176}
  \definecolor{gray95}{rgb}{0.949020,0.949020,0.949020}
  \definecolor{grey95}{rgb}{0.949020,0.949020,0.949020}
  \definecolor{gray96}{rgb}{0.960784,0.960784,0.960784}
  \definecolor{grey96}{rgb}{0.960784,0.960784,0.960784}
  \definecolor{gray97}{rgb}{0.968627,0.968627,0.968627}
  \definecolor{grey97}{rgb}{0.968627,0.968627,0.968627}
  \definecolor{gray98}{rgb}{0.980392,0.980392,0.980392}
  \definecolor{grey98}{rgb}{0.980392,0.980392,0.980392}
  \definecolor{gray99}{rgb}{0.988235,0.988235,0.988235}
  \definecolor{grey99}{rgb}{0.988235,0.988235,0.988235}
  \definecolor{gray100}{rgb}{1.000000,1.000000,1.000000}
  \definecolor{grey100}{rgb}{1.000000,1.000000,1.000000}
  \definecolor{dark grey}{rgb}{0.662745,0.662745,0.662745}
  \definecolor{DarkGrey}{rgb}{0.662745,0.662745,0.662745}
  \definecolor{dark gray}{rgb}{0.662745,0.662745,0.662745}
  \definecolor{DarkGray}{rgb}{0.662745,0.662745,0.662745}
  \definecolor{dark blue}{rgb}{0.000000,0.000000,0.545098}
  \definecolor{DarkBlue}{rgb}{0.000000,0.000000,0.545098}
  \definecolor{dark cyan}{rgb}{0.000000,0.545098,0.545098}
  \definecolor{DarkCyan}{rgb}{0.000000,0.545098,0.545098}
  \definecolor{dark magenta}{rgb}{0.545098,0.000000,0.545098}
  \definecolor{DarkMagenta}{rgb}{0.545098,0.000000,0.545098}
  \definecolor{dark red}{rgb}{0.545098,0.000000,0.000000}
  \definecolor{DarkRed}{rgb}{0.545098,0.000000,0.000000}
  \definecolor{light green}{rgb}{0.564706,0.933333,0.564706}
  \definecolor{LightGreen}{rgb}{0.564706,0.933333,0.564706}

\DeclareRobustCommand{\VAN}[3]{#2}
\let\VANthebibliography\thebibliography
\def\thebibliography{\DeclareRobustCommand{\VAN}[3]{##3}\VANthebibliography}

\shorttitle{MBHs in Cluster Dwarfs}
\shortauthors{Tremmel et al.}

\begin{document}

\title[MBH Occupation in Cluster Galaxies]{An Enhanced Massive Black Hole Occupation Predicted in Cluster Dwarf Galaxies\vspace{-1.5cm}
}%\vspace{-300pt}
\author{M. ~Tremmel$^{1}$}
\author{A.~Ricarte$^{2,3}$}
\author{P.~Natarajan$^{2,4,5}$}
\author{J.~Bellovary$^{6,7}$}
\author{R.~Sharma$^8$}
\author{T.~R.~Quinn$^9$}

\affiliation{$^1$ School of Physics, University College Cork, Cork, Ireland}
\affiliation{$^2$ Black Hole Initiative, 20 Garden Street, Cambridge, MA 02138, USA}
\affiliation{$^3$ Center for Astrophysics -- Harvard \& Smithsonian, 60 Garden Street, Cambridge, MA 02138, USA}
\affiliation{$^4$ Department of Astronomy, Yale University, 52 Hillhouse Avenue New Haven, CT 06511, USA\\
$^5$Department of Physics, Yale University, P.O. Box 208121, New Haven, CT 06511, USA}
\affiliation{$^6$ Department of Physics, Queensborough Community College, City University of New York, 222-05 56th Ave, Bayside, NY, 11364}
\affiliation{$^7$ Department of Astrophysics, American Museum of Natural History, Central Park West at 79th Street, New York, NY 10024, USA}
\affiliation{$^8$Department of Physics \& Astronomy, Rutgers, The State University of New Jersey, 136 Frelinghuysen Road, Piscataway, NJ 08854, USA}
\affiliation{$^9$ Astronomy Department, University of Washington, Box 351580, Seattle, WA, 98195-1580
}
\email{Corresponding Author: mtremmel@ucc.ie}

% These dates will be filled out by the publisher
%\date{Accepted XXX. Received YYY; in original form ZZZ}

% Enter the current year, for the copyright statements etc.
%\pubyear{2022}

% Don't change these lines
%\label{firstpage}
%\maketitle

\begin{abstract}
The occupation fraction of massive black holes (MBHs) in low mass galaxies offers interesting insights into initial black hole seeding mechanisms and their mass assembly history, though disentangling these two effects remains challenging. Using the {\sc Romulus} cosmological simulations we examine the impact of environment on the occupation fraction of MBHs in low mass galaxies. Unlike most modern cosmological simulations, {\sc Romulus} seeds MBHs based on local gas properties, selecting dense ($n>3$ cm$^{-3}$, 15 times the threshold for star formation), pristine ($Z<3\times10^{-4}Z_{\odot}$), and rapidly collapsing regions in the early Universe as sites to host MBHs without assuming anything about MBH occupation as a function of galaxy stellar mass, or halo mass, {\it a priori}. The simulations predict that dwarf galaxies with M$_{\star}<10^9$ M$_{\odot}$ in cluster environments are $\sim2$ times more likely to host a MBH compared to those in the field. The predicted occupation fractions are remarkably consistent with those of nuclear star clusters. Across cluster and field environments, dwarf galaxies with earlier formation times are more likely to host a MBH. Thus, while the MBH occupation function is similar between cluster and field environments at high redshift ($z>3$), a difference arises as late-forming dwarfs -- which do not exist in the cluster environment -- begin to dominate in the field and pull the MBH occupation fraction down for low mass galaxies. Additionally, prior to in-fall some cluster dwarfs are similar to progenitors of massive, isolated galaxies, indicating that they might have grown to higher masses had they not been impeded by the cluster environment. While the population of MBHs in dwarf galaxies is already widely understood to be important for understanding MBH formation, this work demonstrates that environmental dependence is important to consider as future observations search for low mass black holes in dwarf galaxies.
\end{abstract}

%\begin{keywords}
%galaxies:dwarf -- black hole physics -- galaxies:clusters
%\end{keywords}

%%%%%%%%%%%%%%%%%%%%%%%%%%%%%%%%%%%%%%%%%%%%%%%%%%

%%%%%%%%%%%%%%%%% BODY OF PAPER %%%%%%%%%%%%%%%%%%

\section{Introduction}

The origin and evolution of massive black holes (MBHs) of mass $>10^5$ M$_{\odot}$, which are ubiquitously found in the centers of massive galaxies \citep{tremaine_etal02,kormendy95,kormendy2013}, remains an important open question. There are different models for how the seeds of MBHs form in the early Universe \citep{volonteri10,natarajan14} which predict different formation rates and initial masses. Seeds formed through the stellar evolution of population III stars would have initial masses on the order of $\sim100$ M$_{\odot}$ and would exist in virtually every galaxy, with a fraction capable of attaining much larger masses through, e.g. super-Eddington accretion events \citep[][]{volonteriRees2005, talPN2014, inayoshi16, sessano23}. Other formation models require more specific conditions at high redshift, such as very high densities and a lack of both metals and molecular gas, but can produce seed black holes of mass $10^4-10^6$ M$_{\odot}$ \citep{natarajan11,talPN2014}. The collapse of a dense star cluster can produce a very massive star through runaway stellar mergers that then forms a massive black hole \citep{devecchi2009,davies11}. Alternatively, if fragmentation into a star cluster is prevented, gas can collapse directly into a very massive (quasi-) star type object, which then forms a massive black hole \citep{LN2006BH,LN2007BH}. The origins of MBHs are notoriously difficult to constrain observationally, as the early seeding epochs are currently inaccessible (though JWST is expected to change that soon); and typical detection methods are best only at finding very massive and/or rapidly growing MBHs, which effectively have their initial conditions mostly erased \citep{volonteri08,volonteri09}. JWST and next-generation telescopes, such as Euclid, have the potential to observe some of the earliest phases of MBH growth \citep{sesana15,natarajan17,pacucci19} and could shine new light into their formation physics and environments. Electromagnetic observations will be complemented by new gravitational wave detectors like LISA (Laser Interferometer Space Antenna), with the ability to detect merging, low-mass black holes out to $z>20$ \citep{colpi19wp, lisa_astro}. Constraining the MBH population at high redshift will help to constrain the relative efficiency of different seeding mechanisms \citep{sesana13,sesana15,ricarte18} while providing new insight into the mechanisms driving their growth and co-evolution with their host (proto-)galaxies.

However, clues to MBH formation exist more locally as well. Dwarf galaxies (with stellar mass M$_{\star} < 10^{10}$ M$_{\odot}$) may host black holes that have not grown significantly, neither through accretion nor mergers, and therefore may maintain their connection to their initial conditions \citep{volonteriMSIGMA2009,vanwassenhove10, greene12,ricarte18, regan23b}. There is a growing sample of MBHs detected in dwarf galaxies using a variety of methods and wavelengths, typically observing MBHs as low luminosity active galactic nuclei \citep[AGN; e.g.][]{reines11,reines12,reines20,baldassare15, baldassare16, mezcua18b, nguyen19, woo19, birchall20, molina21, latimer21, cann21, burke22}. Observations like these have been used to estimate the true occupation fraction - the fraction of galaxies hosting any MBH, regardless of whether it is detected as an AGN \citep{greene12, miller15, nguyen19, burke22, askar22}. While there is evidence that the occupation fraction does not dramatically change down to stellar masses as low as $10^{9}$ M$_{\odot}$ \citep{miller15,baldassare20,burke22}, mapping detected AGN fractions to the underlying population of MBHs in low mass galaxies remains difficult and highly uncertain. JWST and next-generation telescopes like Vera Rubin will also prove useful in expanding our view of MBHs in dwarf galaxies \citep{baldassare18, cann21, burke22}.

Cosmological simulations, which self-consistently model the collapse and merger history of dark matter halos with the baryonic evolution (gas accretion, star formaiton, MBH growth) of galaxies, have proven to be an invaluable tool to study the co-evolution of MBHs and galaxies \citep[e.g.][]{diMatteo2008,dimatteo17,okamoto2008_BHFB,IllustrisBH15,rosasGuevara16,dubois16,nelson19,blank19,habouzit21,koudmani21,koudmani22,ni22}. While important work has been done to study the evolution of MBHs in dwarf galaxies using large-scale simulations \citep[e.g.][]{haidaar22}, difficulties arise due to a wide range of model assumptions which often have strong effects on the results. Even the most modern large-scale simulations lack the resolution to correctly model galaxies much lower than $10^9$ M$_{\odot}$ in stellar mass. Many smaller-scale simulations that can attain very high resolution, such as TNG50 \citep{nelson19} or FABLE \citep{henden18}, utilize simplistic prescriptions for seeding MBHs whereby all galaxies residing in dark matter halos above a certain mass are seeded with a MBH at their centers. While this may allow for predictions of the accretion rates and luminosities of MBHs in low mass galaxies (i.e the active fraction), the underlying occupation fraction of MBHs is an explicitly assumed prior. This type of prescription will mean that for many low mass galaxies black holes are seeded at rather late times instead of at high redshift like most theoretical models (although recent works by \cite{natarajan21} and \cite{mayer23} have noted that MBH formation could continue throughout cosmic time and may not be limited to metal poor regions). In addition, when considering satellite or backsplash galaxies, simplistic schemes to force MBHs to the centers of galaxies can generate unrealistic numerical effects and artificially impact the occupation fraction in some environments \citep{borrow22}.

Recent high-resolution simulations that resolve dwarf galaxies while also incorporating more predictive models for MBH formation, such as the {\sc Romulus} simulations \citep{tremmel17,tremmel19}, the Obelisk Simulations \citep{obelisk21}, the New Horizon simulations \citep{dubois21, beckmann23}, and the MARVELous and DC Justice League simulations \citep{bellovary19, bellovary21, munshi21,applebaum21} are valuable tools in predicting MBH evolution and occupation fraction within low mass galaxies. In this Paper we utilize the {\sc Romulus} suite of simulations to examine the environmental dependence of MBH occupation in dwarf galaxies. {\sc Romulus} forms MBHs from dense, pristine gas in the early Universe ($z > 5$) without any {\it a priori} assumptions about which halos should or should not host a MBH. The dynamics of MBHs is followed realistically down to sub-kpc separations, so they are not always forced to the centers of galaxies \citep{tremmel15}. The physics models governing MBH growth, star formation, and supernovae feedback that have been incorporated into {\sc Romulus} have been optimized to broadly reproduce observed scaling relations across nearly five orders of magnitude in halo mass \citep{tremmel17,tremmel19,ricarte19, sharma20}.

MBHs in field dwarf galaxies have been extensively studied in {\sc Romulus}, which produces a realistic population of dwarf galaxy AGN consistent with observations \citep{sharma22} that can also affect their evolution \citep{sharma20, sharma22b}, something which is seen in other simulations as well \citep{koudmani21,koudmani22}. We expand on previous analysis in this work by examining how the dwarf galaxy MBH occupation fraction evolves with both time and environment. To do this we compare results from {\sc Romulus25}, a 25 Mpc-per-side uniform volume simulation, with {\sc RomulusC}, a zoom-in simulation of a $10^{14}$ M$_{\odot}$ galaxy cluster. Between these two simulations, we have several hundred simulated dwarf galaxies with stellar masses in the range $10^7-10^{10}$ M$_{\odot}$ that are resolved with at least 200 baryonic resolution elements (and $10^4$ for dark matter).

In Section 2 we provide a brief overview of the {\sc Romulus} simulations and the relevant physics models incorporated in them. In Section 3 we present our results of the dependence of occupation fraction on environment and its evolution with time. In Section 4 we explore the origins of the enhanced MBH occupation fraction in $z=0$ cluster dwarf galaxies that we find. Section 5 discusses our results and Section 6 provides a summary of our conclusions.

\section{The Romulus Simulations}

In this section we briefly describe the relevant properties of the simulations. For a more detailed discussion, including how the parameters were chosen, we point the reader to \citet{tremmel17, tremmel19}.

\subsection{Overview of the Romulus Simulations}

The {\sc Romulus} Simulations \citep{tremmel17,tremmel19} are a suite of cosmological hydrodynamic simulations that includes a 25 Mpc-per-side uniform volume simulation ({\sc Romulus25}) and a zoom-in simulation of a $10^{14}$ M$_{\odot}$ galaxy cluster ({\sc RomulusC}). With a dark matter mass resolution of $3.4 \times 10^5$ M$_{\odot}$, both {\sc Romulus} simulations are able to resolve halos as small as $\sim3\times10^9$ M$_{\odot}$ with more than 10,000 particles. With typical gas and star particle masses of $2.1 \times 10^5$ and $6 \times 10^4$ M$_{\odot}$ respectively, and a spline softening of 350 pc (equivalent to 250 pc plummer softening), the two simulations are also able to resolve the the baryonic structure of dwarf galaxies as small as $\sim10^7$ M$_{\odot}$ in stellar mass with hundreds of resolution elements. Both simulations are run with the same cosmology \citep[$\Omega_0 = 0.3086, \Lambda = 0.6914, h= 0.6777, \sigma_8 = 0.8288$;][]{planck16} and physics.

The {\sc Romulus} simulations were run using the N-body+Smoothed particle hydrodynamics code, {\sc ChaNGa} \citep{changa15}, which incorporates standard physics modules previously used in GASOLINE \citep{wadsley04,wadsley08,wadsley17}. This includes a cosmic UV background \citep{HM12} with self-shielding \citep{pontzen08}, star formation with `blastwave' supernovae feedback \citep{Stinson06}, low temperature metal cooling \citep{eris11}, and thermal and metal diffusion \citep{shen10, governato15}. It also includes recent improvements, such as an SPH force calculation that uses the geometric mean density \citep{ritchie01,changa15,governato15}, an updated turbulent diffusion implementation \citep{wadsley17}, and a time-dependent artificial viscosity and time-step adjustment system \citep{saitoh09, wadsley17}.

\subsection{Star Formation, Supernovae Feedback and Gas Cooling}

Star formation occurs within dense (n$ >0.2$ cm$^{-3}$), cold (T$ < 10^4$ K) gas. Each gas particle that meets these criteria is allowed to form star particles on a characteristic timescale of $10^6$ years with the following probability:

\begin{equation}
p =\frac{m_{gas}}{m_{star}}\left(1 - e^{-c_{\star}(10^6/t_{\mathrm{dyn}})}\right),
\end{equation}

\noindent where $m_\mathrm{star} = 0.3 m_\mathrm{gas}$, $c_{\star} = 0.15$, and $t_\mathrm{dyn}$ is the dynamical time of the gas particle. Energy from supernovae couples thermally to nearby gas with an efficiency of 75\%. Supernova feedback uses the `blastwave' implementation \citep{Stinson06}, where gas cooling is shutoff for a period of time to avoid numerical overcooling. This implementation of feedback and star formation produces dwarf galaxies that lie on the observed stellar mass-halo mass relations \citep{tremmel17} and includes a realistic population of `ultra-diffuse' galaxies \citep{tremmel20,wright21,jvn22}. 

An important limitation of {\sc Romulus} is the lack of high temperature metal-line cooling. This can affect the accretion history of gas onto massive galaxies \citep{vandevoort11}. This choice, discussed in more detail in \citet{tremmel19}, is motivated by a variety of previous simulations showing that the inclusion of metal-line cooling in simulations without adding molecular hydrogen physics and more detailed star formation prescriptions results in unrealistic dwarf galaxies \citep{christensen14b}. Despite being among the highest resolution simulations of its class, even {\sc Romulus} is unable to resolve the multiphase ISM so it is not possible to include these more detailed physical processes while maintaining realistic low mass galaxies. It has been shown that {\sc RomulusC}, our most massive halo using these input physics, maintains a realistic intracluster medium \citep{tremmel19}. Because this paper is focused on dwarf galaxies and on the formation of MBHs at high redshift from very metal-poor gas (see below), this choice does not impact our results.

\subsection{Black Hole Accretion and Feedback}

Massive black holes are allowed to grow by accreting nearby gas via a modified Bondi-Hoyle formalism \citep{bondihoyle44} that accounts for angular momentum support and includes a density-dependent boost factor \citep{BoothBH2009} that is meant to account for unresolved, multiphase gas:

\begin{equation}
    \dot{M}_\bullet = \left( \frac{n}{n_*} \right)^\beta \begin{cases}
    \frac{(GM_\bullet)^2\rho}{(v^2_\mathrm{bulk} + c^2_s)^{3/2}} & \text{if $v_\mathrm{bulk} > v_\theta$} \\ 
    \frac{(GM_\bullet)^2\rho c_s}{(v^2_\theta + c^2_s)^{2}} & \text{if $v_\mathrm{bulk} < v_\theta$}, \\ 
    \end{cases}
\end{equation}

\noindent where $G$ is the gravitational constant; $n_*$ is the star formation threshold (0.2 cm$^{-3}$; see previous section); $v_\mathrm{bulk}$ is the bulk velocity of the gas; $v_\theta$ is its rotational velocity; $c_s$ is its sound speed; $\rho$ is its mass density; and $\beta$ is set to 2. The radiative efficiency ($\epsilon_r$) of accreting MBHs is assumed to be 0.1 and all accretion is capped at the Eddington rate. 

Growing MBHs produce feedback at a rate 
\begin{eqnarray}
\dot{E}_{BH} = \epsilon_r \epsilon_f \dot{M}_\bullet c^2,
\label{eqn3}
\end{eqnarray}
where $\epsilon_f$ is set to 0.02 and $c$ is the speed of light. This energy is distributed instantaneously as thermal energy to the surrounding 32 nearest gas particles. Feedback from MBHs has been shown to successfully regulate the star formation in simulated massive galaxies in {\sc Romulus} \citep{tremmel17,tremmel19,chadayammuri21}. However, too many massive galaxies remain star forming and disk-dominated in {\sc Romulus} at $z=0$ \citep{jung22}, indicating that stronger modes of feedback are still needed for higher mass galaxies. On the other hand, at the lower mass end, feedback from MBHs has also been shown to influence the evolution of dwarf galaxies in {\sc Romulus} \citep{sharma20,sharma22,sharma22b} and has been shown to over-quench star formation at low masses, indicating that feedback may be too efficient at low masses. Because this paper focuses on the formation of black holes and less on the detailed evolution of their host galaxies, this should not affect our results. The majority of low mass galaxies in {\sc RomulusC} are quenched by the cluster environment and not by internal processes \citep{tremmel19}.

\subsection{Black Hole Dynamics}

Unlike most cosmological simulations that include MBHs, {\sc Romulus} does not force black holes to the centers of galaxies. Rather, they are allowed to move realistically within their host galaxies. This is done through the implementation of a sub-grid routine to account for unresolved dynamical friction \citep{tremmel15}. This estimates and applies the necessary force each MBH would feel due to dynamical friction by integrating the Chandrasekhar formula \citep{chandrasekhar43} out to the gravitational softening length, $\epsilon_g$, during each black hole timestep.

\begin{equation}
    \boldsymbol{a}_\mathrm{DF} = -4 G^2 M_\bullet \rho(<v_\bullet) \ln \Lambda \frac{\boldsymbol{v}_\bullet}{v_\bullet^3},
\end{equation}

Here $v_\bullet$ is the velocity of the black hole relative to nearby star and dark matter particles, $\rho(<v_\bullet)$ is the local density of star and dark matter particles that are moving slower than the black hole relative the background particles, and $\ln \Lambda$ is the Coulomb logarithm and equal to $\ln (\epsilon_g/r_{90})$. Here $r_{90}$ is the $90^\circ$ deflection radius based on the black hole's mass and local relative velocity. As shown in \citet{tremmel15}, this method is able to produce realistically decaying orbits. The result of this model is that MBHs take non-negligible time to reach the halo center after galaxy mergers \citep{tremmel18}, and sometimes their orbits fail to decay and they end up as `wandering', off-center MBHs \citep{tremmel18b,bellovary21,ricarte21, ricarte21b}. Two MBHs are allowed to merge when they are within 700 pc ($2\times\epsilon_g$) and  mutually bound to one another. 

The technique used here to model dynamical friction is similar to that employed in a growing number of other simulations, including Magneticum \citep{Hirschmann14}, Obelisk \citep{obelisk21,pfister19} and {\sc Astrid} \citep{chen22,ni22}. The high resolution of {\sc Romulus} allows for the dynamical evolution of MBHs to be accurately tracked down to sub-kpc scales with dynamical friction applied only locally, sampling densities of stars and dark matter near each MBH (distances $<350$ pc). Other simulations like Horizon-AGN \citep{dubois16} have models that account only for dynamical friction from gas \citep{ostriker99} while {\sc Romulus} only accounts for dynamical friction from stars and dark matter (i.e. collision-less particles). While it is possible that gas may dominate the density of a galaxy at times, it is difficult to fully account for the effects of torques due to unresolved structure and turbulence within the ISM which can be at least as important \citep{roskar15,bortolas22,lescaudron22}.

\begin{figure}
\centering
\includegraphics[trim=20mm 10mm 40mm 15mm, clip, width=85mm]{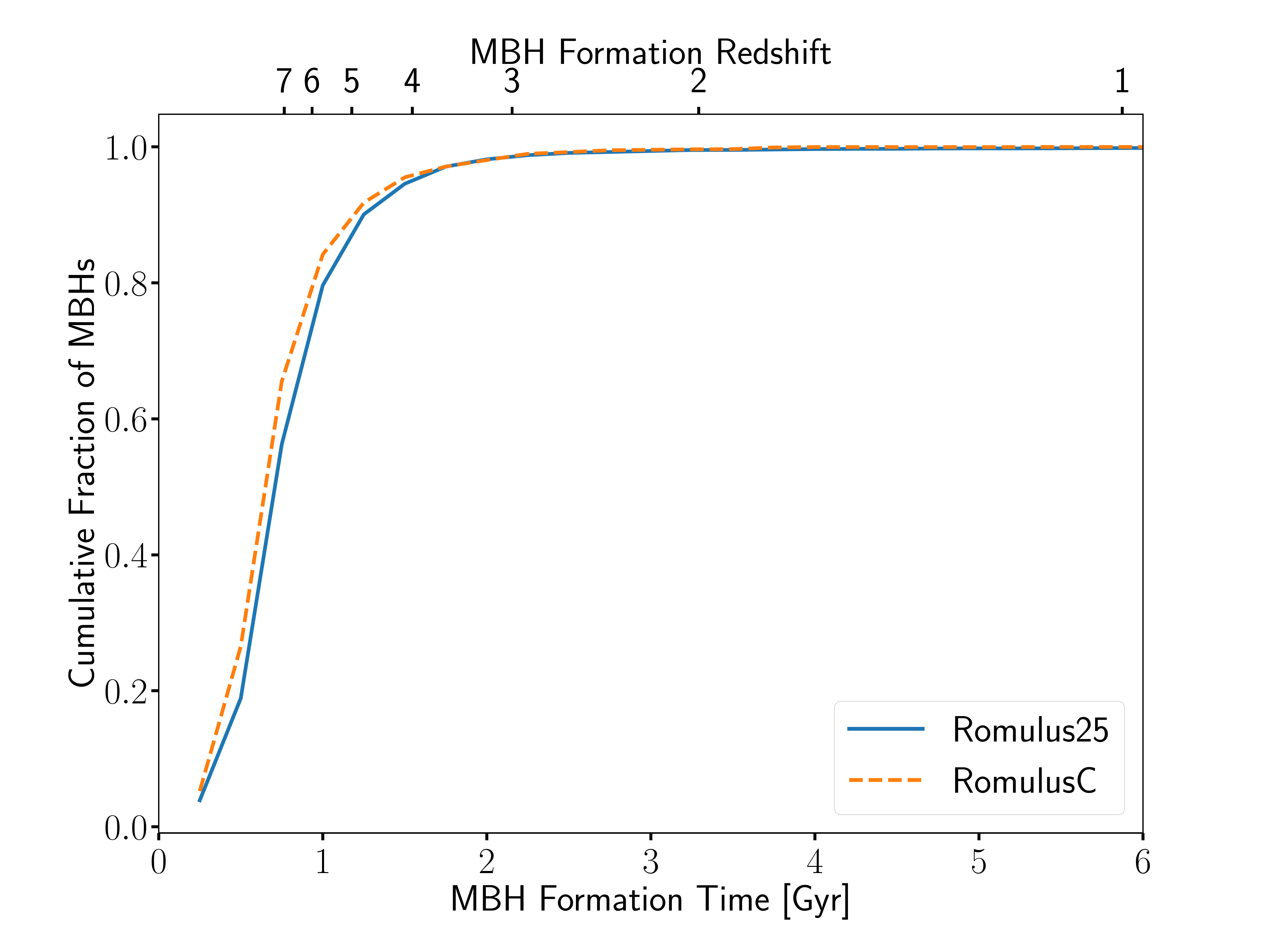}
\caption{{\sc Distribution of Black Hole Formation Times}. The cumulative distribution of formation times for MBHs in {\sc Romulus25} and {\sc RomulusC}. Despite sampling different environments, both simulations seed MBHs at very similar times.}
\label{bhformtime}
\end{figure}

\subsection{Black Hole Seeding}

A critical aspect of the {\sc Romulus} simulations to this work is the prescription for MBH seeding. Like most modern, large-scale cosmological simulations, the actual formation sites for MBH seeds are unresolved in {\sc Romulus}. To approximate the regions most likely to create an early, massive black hole, we form MBH seeds in gas that would otherwise be forming stars (see above) but with the following additional criteria:

\begin{itemize}
    \item Metallicity less than $3 \times 10^{-4} \ Z_\odot$.
    \item Particle density is greater than $3 \ cm^{-3} = 15n_{\star}$.
    \item Temperature is between $9500 \ \mathrm{K}$ and $10000 \ \mathrm{K}$.
\end{itemize}

\noindent This prescription will naturally pick out regions in the very early Universe that have yet to be polluted with metals from star formation, yet are still able to collapse to high densities (15 times greater than the threshold for star formation) on timescales shorter than that for star formation (i.e. the gas is able to reach densities well beyond the star formation threshold before forming stars) and without cooling much below $10^4$ K (densities are growing faster than the gas can effectively cool). While these criteria are most directly reminiscent of the direct collapse formation scenario \citep[gas collapsing into a single massive object;][]{LN2006BH,LN2007BH, regan17, wise19, regan20}, or the collapse of a dense star cluster \citep[runaway stellar collisions produce a single massive object;][]{devecchi2009,davies11}, they are also consistent with regions where a low mass black hole seed (e.g. from a population III star) may be able to grow rapidly in a very short amount of time thanks to the presence of of dense, rapidly collapsing gas \citep[e.g.][]{volonteriRees2005, talPN2014, inayoshi16, sessano23}.

Importantly, these adopted criteria result in an epoch of MBH seeding that is roughly 90\% complete by $z = 5$ and makes no assumptions as to which halos should (or should not) host a MBH \citep{tremmel17}. This is in stark contrast to the methods more typically utilized in cosmological simulations, where MBHs are seeded based on a halo mass threshold (typically a few $\times 10^{10}$ M$_{\odot}$), which preferentially prevent MBHs from populating low mass galaxies early on, and will result in a prolonged period of MBH seeding. By predicting the existence of MBHs based only on local gas properties, {\sc Romulus} can therefore \textit{predict} the occupation of MBHs in galaxies in a way that even simulations of similar resolution often cannot. Some cosmological simulations have started to employ a similar MBH formation prescription. For example, the New Horizon simulations have shown the importance of both resolving low mass galaxies and allowing them to be seeded with MBHs in studying MBH mergers and potential gravitational wave sources \citep{volonteri20}. However, it is important to note that this method would not capture formation processes that occur at later times in more metal-enriched gas \citep[e.g.][]{regan20b,natarajan21, mayer23}.

In Figure \ref{bhformtime} we compare the distribution of black hole formation times in the field ( {\sc Romulus25}) with that of the galaxy cluster ({\sc RomulusC}), confirming that they are virtually indistinguishable in our simulations. This means that the initial seeding of MBHs at high redshift, which only depends on local gas properties, is not impacted directly by the environment. This will be an important fact to keep in mind as we delve deeper into the differences inferred in the MBH population between the two environments.

\subsection{Halo Selection and Galaxy Properties}

Halo finding is performed with the Amiga Halo Finder \citep{knollmann09}, which uses a spherical top-hat collapse technique to define the virial radius (R$_{vir}$) and mass (M$_{vir}$) of each halo and sub-halo. It also assigns all of the baryonic content belonging to each dark matter halo. In our analysis we use R$_{200}$ which is the radius enclosing a mean density 200 times the critical density of the Universe at a given redshift. We only include halos with M$_{vir}>3\times10^9$ M$_{\odot}$, such that each halo included in our analysis has at least $\sim 10,000$ particles. An exception is made for our analysis of {\sc RomulusC} where we include halos down to $3\times10^8$ M$_{\odot}$ to account for the fact that many halos have experienced tidal stripping of the outer regions of their dark matter halos in the dense cluster environment. As discussed in \citet{tremmel20}, while the total halo mass is affected by the cluster environment, dwarf galaxies often keep most of their stellar mass intact. This choice to lower the halo mass threshold is to avoid discounting galaxies that would otherwise have been previously considered well-resolved prior to falling into the cluster. We have confirmed that our results are not sensitive to this choice.

The position of each galaxy/halo is calculated using the shrinking spheres approach \citep{power03}, which reliably extracts the centers of the central galaxies within each halo. As {\sc RomulusC} is a zoom-in simulation, it is surrounded by a region of low resolution dark matter. Galaxies on the outskirts may be contaminated by high mass (low resolution) dark matter particles. We avoid including such galaxies in our analysis by requiring each galaxy to have less than $5\%$ of its dark matter particles be contaminated by low resolution elements. This cut removes only 3\% of galaxies within 2 Mpc ($\sim 2 $R$_{200}$) of the cluster center, the vast majority of which are beyond the virial radius.

When quoting the stellar mass of each galaxy, we estimate what the observed stellar mass would be from typical techniques. \citet{munshi13} found that typical observational techniques result in a systematic underestimate of galaxy stellar mass. Based on those results, we apply a correction factor of 0.6 to the `raw' stellar masses (the summed mass of all star particles associated with a given halo), a conservative estimate given the results of \citet{munshi13}. \citet{leja19} also find a similar discrepancy when comparing advanced spectroscopic techniques to estimating stellar masses with common photometric techniques. This choice affects only the stellar masses we present, but since all galaxies are treated equally in this analysis this has no effect on the substance of our results comparing different simulation regions.

Throughout this work we also classify some galaxies as `isolated'. We base our definition of isolated on the results of \citet{geha12} that show the onset of environmental effects for dwarf galaxies (M$_{\star}<10^{10}$ M$_{\odot}$) occur when they are approximately 1.5 Mpc away from a massive galaxy. The same environmental effects are seen on the quenched fraction of galaxies in {\sc Romulus} \citep{sharma22b}. Isolated galaxies in {\sc Romulus25} are therefore defined so that they are not within R$_{200}$ of any more massive halo. Additionally, for low mass galaxies with M$_{\star}<10^{10}$ M$_{\odot}$, they must be further than 1.5 Mpc from any galaxy with stellar mass greater than $2.5\times10^{10}$ M$_{\odot}$. 

%\begin{figure}
%    \centering
%   \includegraphics[trim=35mm 10mm 50mm 40mm, clip, width=80mm]{plots/occ_frac_cen_any_compare.png}
%    \caption{{\sc The effect of excluding off-center MBHs for the galaxy occupation fraction.} As in Figure~\ref{occ_frac_all} blue points represent field galaxies ({\sc Romulus25}) while the orange is for cluster galaxies ({\sc RomulusC}). Open points are the same as in Figure~\ref{occ_frac_all} and include any MBH. The filled points are the fraction of galaxies hosting specifically a central (within 1 kpc of galaxy center) MBH. The exclusion of off-center MBHs does not strongly influence the overall occupation fraction, nor does it affect the enhancement we see among simulated cluster dwarf galaxies.}
%    \label{of_any_cen_nsc}
%\end{figure}

\section{The Predicted occupation fraction of MBHs in Romulus}

\begin{figure*}
\centering
\includegraphics[trim=35mm 70mm 50mm 100mm, clip, width=135mm]{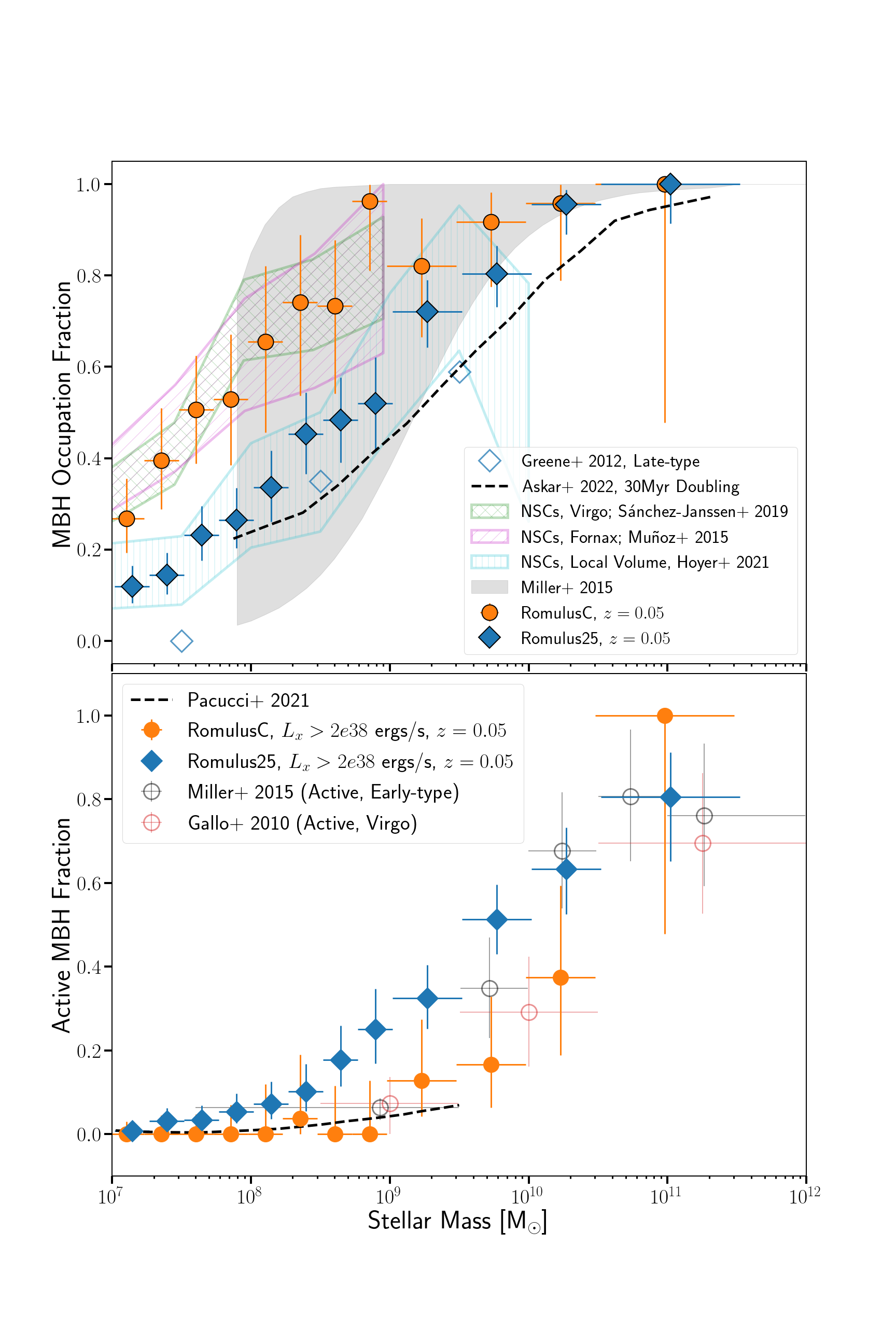}
\caption{{\sc The excess of mostly quiescent massive black holes in cluster dwarf galaxies}. {\it Top:} The fraction of galaxies hosting at least one MBH as a function of stellar mass for galaxies in {\sc Romulus25} (blue) and {\sc RomulusC} (orange). Error bars represent 95\% binomial 
confidence intervals \citep{cameron11}. The grey region and open diamonds are both estimates derived from observations in \citet{miller15} and \citet{greene12} respectively. The black dashed line is the predicted occupation of MBHs above $10^5$ M$_{\odot}$ from \citet{askar22} assuming a 30 Myr mass doubling time for MBHs. The three hatched regions are occupation fractions of nuclear star clusters (NSCs) observed in dwarf galaxies in Virgo \citep[green;][]{sanchezJanssen19}, Fornax \citep[magenta;][]{munos15}, and the local volume \citep[cyan;][]{hoyer2021}. For the local volume data, the hatched region represents 95\% binomial confidence intervals, which were calculated using the total number of observed galaxies in each bin. While the field population in {\sc Romulus25} has occupation fractions roughly consistent with observationally derived estimates, {\sc RomulusC} has an enhanced occupation fraction at low masses (M$_{\star} < 10^9 $M$_{\odot}$). The occupation fraction in the {\sc RomulusC} simulation is remarkably consistent with that in the observed NSCs within galaxy clusters, while the occupation fraction in field galaxies ({\sc Romulus25}) is very consistent with local volume NSCs.  {\it Bottom:} The fraction of galaxies hosting a central ($D<1$ kpc), luminous (L$_{x} > 2\times10^{38} \ \mathrm{erg} \; \mathrm{s}^{-1}$) MBH as a function of stellar mass. Also shown are two observational data sets examining AGN in cluster and early-type galaxies with a similarly low luminosity threshold \citep{gallo10, miller15}. Also plotted in the dashed black line is the active fraction estimated from \citet{pacucci21}. The cluster dwarf population (orange), which are preferentially quenched galaxies, matches well with the observations and are noticeably lower than the occupation of luminous MBHs in field dwarfs (blue). Despite having a higher underlying MBH occupation fraction, cluster dwarfs are less likely to host low luminosity AGN compared to the field by $z=0$.}
\label{occ_frac_all}
\end{figure*}

The top panel of Figure~\ref{occ_frac_all} shows the fraction of galaxies hosting at least one MBH in both the galaxy cluster ({\sc RomulusC}) and in the field ({\sc Romulus25}). For {\sc Romulus25}, only galaxies within halos of mass $>3\times10^9$ M$_{\odot}$ are included. For {\sc RomulusC}, only galaxies in halos of mass $>3\times10^8$ M$_{\odot}$ and within $2R_{200}$ of cluster center are included (see previous section for more justification of this choice, though it does not affect our overall results). %This choice was made in order to also include halos that experienced significant stripping of the outskirts of their dark matter halos in the cluster environment but were otherwise similar to low mass field galaxies. As discussed in \citet{tremmel20}, cluster dwarf galaxies have the outskirts of their dark matter halos tidally stripped, but it is much rarer for galaxies to have their stellar masses significantly affected. We confirm that this choice does not influence our conclusions. 
The results are compared with multiple observationally derived estimates for the underlying occupation fraction \citep{greene12,miller15, askar22} and the observed occupation fraction of nuclear star clusters (NSCs) in dwarf galaxies within the Virgo and Fornax clusters \citep{sanchezJanssen19, munos15}, as well as the Local Group \citep{hoyer2021}. Note that observational estimates for the underlying MBH occupation fraction are highly uncertain and rely on inference from active galactic nuclei. In the context of this work, these are useful benchmarks to compare our results while more detailed comparisons are not very useful. The error bars in occupation fraction in this and all following figures represent 95\% binomial confidence intervals \citep{cameron11}. %In recent work. \textbf{[Baldassare 2022]} show that a significant number of nuclear star clusters show evidence of accreting MBHs.

The occupation fractions in the field derived from {\sc Romulus25} are consistent with estimates derived from observations and match quite well with observed NSC occupation fractions in the local volume \citep{hoyer2021}. %They are also consistent with recent results from \citep{baldassare20} that find that the occupation fraction of MBHs does not evolve significantly in galaxies down to stellar masses of $10^9$ M$_{\odot}$, as we only see a factor of two decline in MBH occupation at $M_{\star}\sim10^9$ M$_{odot}$ compared to the most massive galaxies in the simulation. The most striking feature, however, is that there is
We find a significantly elevated occupation fraction among cluster dwarf galaxies compared to field dwarfs with M$_{\star} < 10^9$ M$_{\odot}$. Interestingly, the occupation fraction in these low mass cluster galaxies matches remarkably well with the occupation fraction of nuclear star clusters in Fornax and Virgo, which also show enhanced occupation compared to lower density environments \citep{sanchezJanssen19, munos15}.

The bottom panel of Figure~\ref{occ_frac_all} shows the occupation fraction of central (within 1 kpc of the halo center) black holes that would be observed as active galactic nuclei at $z = 0.05$ as a function of stellar mass. We limit the spatial offset to mimic the fact that observers often search for X-ray sources associated with galactic centers specifically. We find that cluster dwarf galaxies are less likely to host X-ray luminous AGN compared to the field. As discussed in \citet{tremmel19} this is because the majority of cluster dwarf galaxies have had their gas supply removed due to ram pressure stripping, resulting in very low MBH growth. We note that it is possible that ram pressure can also cause gas to compress to the center and, in some cases, momentarily activate black hole growth \citep{poggianti17}. We see this effect in {\sc RomulusC} \citep{ricarte20} but this process is transient and only common among the more massive in-falling galaxy population. Low mass galaxies quickly lose their gas and both star formation and black hole accretion are shut off.

To estimate the X-ray luminosity of growing MBHs in the simulation, we first estimated the bolometric luminosity. To do this we followed previous work \citep{churazov05, habouzit22, sharma22} and implemented a two-mode model that accounts for radiatively inefficient accretion flows during low Eddington ratio ($f_\mathrm{Edd}$) accretion:

\begin{equation}
	    L_{\rm bol} = \begin{cases}
    	    \epsilon_r \dot{M}_{\rm BH} c^2, \quad & f_{\rm Edd} \geq 0.1\\
    	    10 f_{\rm Edd} \epsilon_r \dot{M}_{\rm BH} c^2, \quad & f_{\rm Edd} < 0.1.
	    \end{cases}
     \label{eqn5}
\end{equation}

\noindent This assumes the same constant radiative efficiency ($\epsilon_r=0.1$) used in the simulation to calculate Eddington ratio, applying a scale factor to estimate an `effective' radiative efficiency for low, radiatively inefficient accretion rates ($f_{Edd}<0.1$). This is purely a post-processing calculation to estimate the total escaping luminosity from each black hole. In the simulation a constant $\epsilon_r =0.1$ was used to calculate both Eddington ratio and feedback (see equation \ref{eqn3}). While this estimate of bolometric luminosity doesn't directly contradict any of the physics implemented in the simulation, it does change how feedback relates to (observed) luminosity, as the original feedback model rationale was that some constant fraction of emitted light couples to nearby gas as thermal energy, which would no longer be true when using equation \ref{eqn5}.

\begin{figure*}
\centering
\includegraphics[trim=40mm 25mm 30mm 40mm, clip, width=180mm]{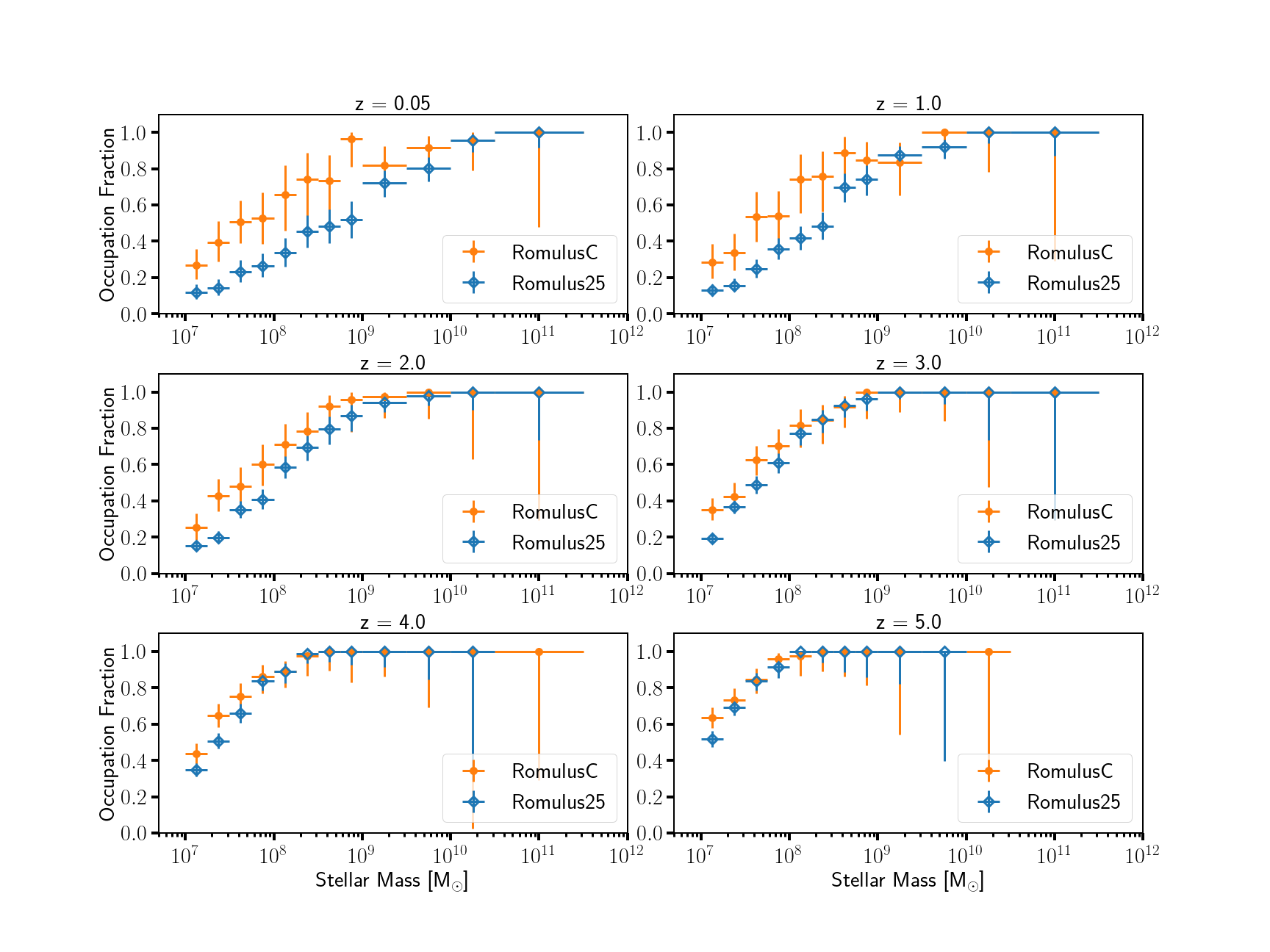}
\caption{{\sc Comparing MBH Occupation Fraction Across Cosmic Time}. The occupation fraction of MBHs as a function of galaxy stellar mass at different redshifts. Blue points show the results for galaxies in the field ({\sc Romulus25}) and the orange points show galaxies within 2 Mpc of the (proto-)cluster center at each redshift. At $z = 5$, when the vast majority ($>95\%$) of MBHs have formed in both simulations, the occupation fractions look very similar. The differences in the two distributions becomes more dramatic with cosmic time, particularly after $z \sim 3$. The large error bars are due to small number statistics at the highest mass bins at earlier times. %Error bars are binomial uncertainties \citep{cameron11}. %These values represent the fraction of galaxies that host any MBH, not limited to central MBHs.% do not limit the black hole distance from their host centers
%(PN: is this what you mean?). 
%At higher redshift, there are more wandering MBHs \citep[see][for more details]{ricarte21a} that makes the results noisier were we to select for only those within 1 kpc of galaxy center. However, the results are quantitatively the same regardless of this selection criteria.
}
\label{occ_frac_redshift_comp}
\end{figure*}

%The rate at which energy is transferred to nearby gas from a growing MBH in the simulation depends only on accretion rate while the (observed) luminosity estimated for a growing MBH via equation \ref{eqn5} depends also on the Eddington ratio. There is nothing inherent to the model used in the simulation that requires that feedback be directly proportional to observed luminosity, although the idea that some constant portion of the escaping luminosity becomes thermally coupled to gas would no longer be true in this scenario.
%applies an eddington ratio-dependent scale factor to the luminosity relative to the constant radiative efficiency ($\epsilon_r=0.1$) assumed in the simulation. In the simulation this radiative efficiency was used to calculate the eddington ratio, which would not be affected by this new model, as well as feedback
%, $\epsilon_r=0.1$, as in the simulation. This value is also used to estimate the Eddington ratio ($f_{Edd} = \epsilon_r \dot{M}_{\rm BH} c^2 / L_\mathrm{Edd}$). %The simulations themselves, like most other cosmological simulations with black holes, assume a constant radiative efficiency when calculating black hole feedback and enforcing the Eddington limit. 

After calculating the bolometric luminosity, we apply the bolometric correction from \citet{shen20} to estimate the 0.5-10 keV X-ray luminosity of each MBH. The rationale and application of this approach is discussed further in \citet{sharma22} where it is shown that calculating luminosity using equation \ref{eqn5} brings the simulated dwarf AGN fraction in {\sc Romulus} much closer to observed values. Note that here we are not making any additional cuts based on the estimated X-ray luminosity of stars and gas in the galaxy, as is done in \citet{sharma22}. Including this would likely bring down the active fraction in the field and have less of an effect for the cluster galaxies, bringing the two potentially closer together. Regardless, our results would remain the same in that the enhanced occupation fraction of MBHs in the cluster is not expressed in the active fraction, which only includes MBHs that are likely to be detectable with observations.

MBHs in the {\sc Romulus} simulations are allowed to be off-center and indeed often are \citep{tremmel18,tremmel18b,ricarte21, ricarte21b}. Such off-center black holes may even be preferentially more common in dwarf galaxies, which host lower mass black holes that experience more inefficient dynamical friction \citep{bellovary21}. Observations searching for MBHs in low mass galaxies often focus on their centers and this would artificially lower the observed active fraction. However, we confirm for both sides of Figure~\ref{occ_frac_all} that the choice to include/exclude off-center ($D>1$ kpc) MBHs makes very little difference in our predicted occupation fractions. Placing more strict criteria for hosting a MBH will decrease the overall fraction of galaxies but the effect is minor and of similar magnitude across environments. For luminous MBHs, it is difficult for non-central MBHs to accrete enough material to become luminous so the inclusion of off-center MBHs also has little effect here. Importantly, the decision to include/exclude off-center MBHs does not affect our main prediction: an enhanced occupation fraction of MBHs in low mass cluster galaxies.

%In Figure~\ref{occ_frac_all} we chose to examine only central ($D<1$ kpc from galaxy/halo center) MBHs because this is where observations typically focus when looking for AGN or kinematic signatures of MBHs. We confirm, however, that were we to remove this restriction and instead study the fraction of galaxies that host a MBH anywhere within it that our results are unchanged. While the occupation fractions predicted in both the field and cluster environments increase slightly by including systems previously excluded (those with only off-center MBHs) the difference is small and the relative enhancement in cluster dwarfs remains the same.

\subsection{Evolution with Redshift}

In Figure \ref{occ_frac_redshift_comp} we show the MBH occupation fraction as a function of stellar mass at six snapshots at different redshifts. We only examine this out to $z = 5$ because before this time many MBHs are still actively being seeded while, by $z = 5$, the vast majority ($\sim90\%$) of MBHs have formed (see Figure~\ref{bhformtime}). From Figure \ref{occ_frac_redshift_comp} we can see that the occupation fraction in the different environments begins to look very similar beyond $z\sim3$. At $z = 5$ the occupation of MBHs as a function of stellar mass is nearly identical between field and cluster environments. Therefore, both the formation times (Figure~\ref{bhformtime}) and host halos at high redshift (Figure~ \ref{occ_frac_redshift_comp}; lower right panel) are similar between {\sc RomulusC} and {\sc Romulus25}, indicating that the presence and location of dense, metal-free gas is very similar between environments at early epochs.

In Figure \ref{occ_frac_evol} we compare the occupation fraction within each environment at different redshifts. In the field (right), the occupation fraction decreases steadily with redshift, particularly at lower masses. For the cluster (left) the occupation fraction ceases to decrease significantly past $z\sim3$. It is this difference in evolution that results in the difference seen at $z=0.05$. The occupation fraction in the cluster at $z=0.05$ is more similar to the field at $z=3$, when the evolution stopped for cluster galaxies. In contrast, the decline in MBH occupation fraction continues for field dwarfs throughout cosmic time. As we will discuss further in Section 4, this decline seen in the field is driven by late-forming dwarf galaxies which do not exist in the cluster environment.

\subsection{Dependence on Cluster-centric Distance and Halo Mass}

\begin{figure*}
\centering
\includegraphics[trim=15mm 10mm 40mm 5mm, clip, width=85mm]{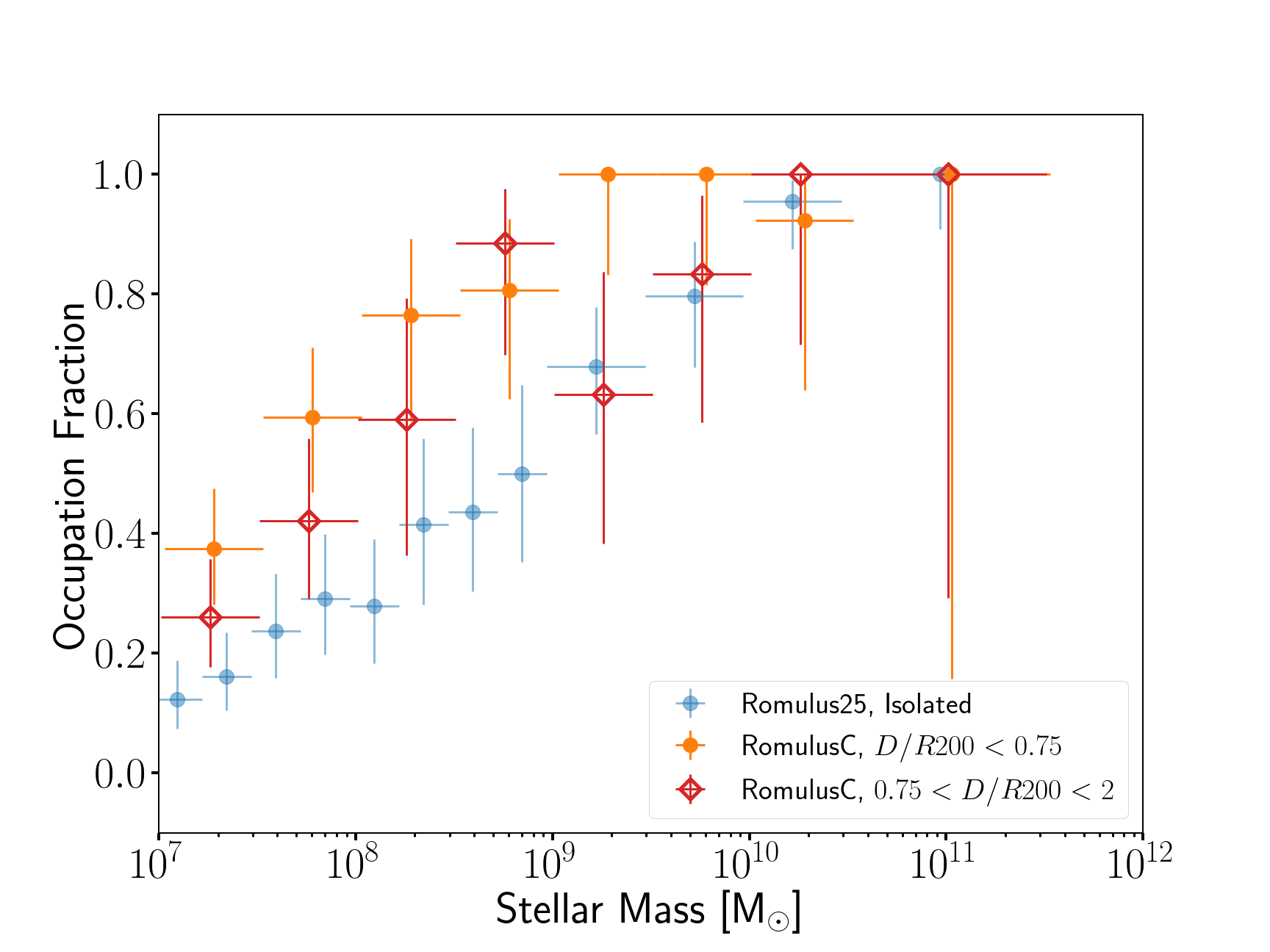}
\includegraphics[trim=15mm 10mm 40mm 5mm, clip, width=85mm]{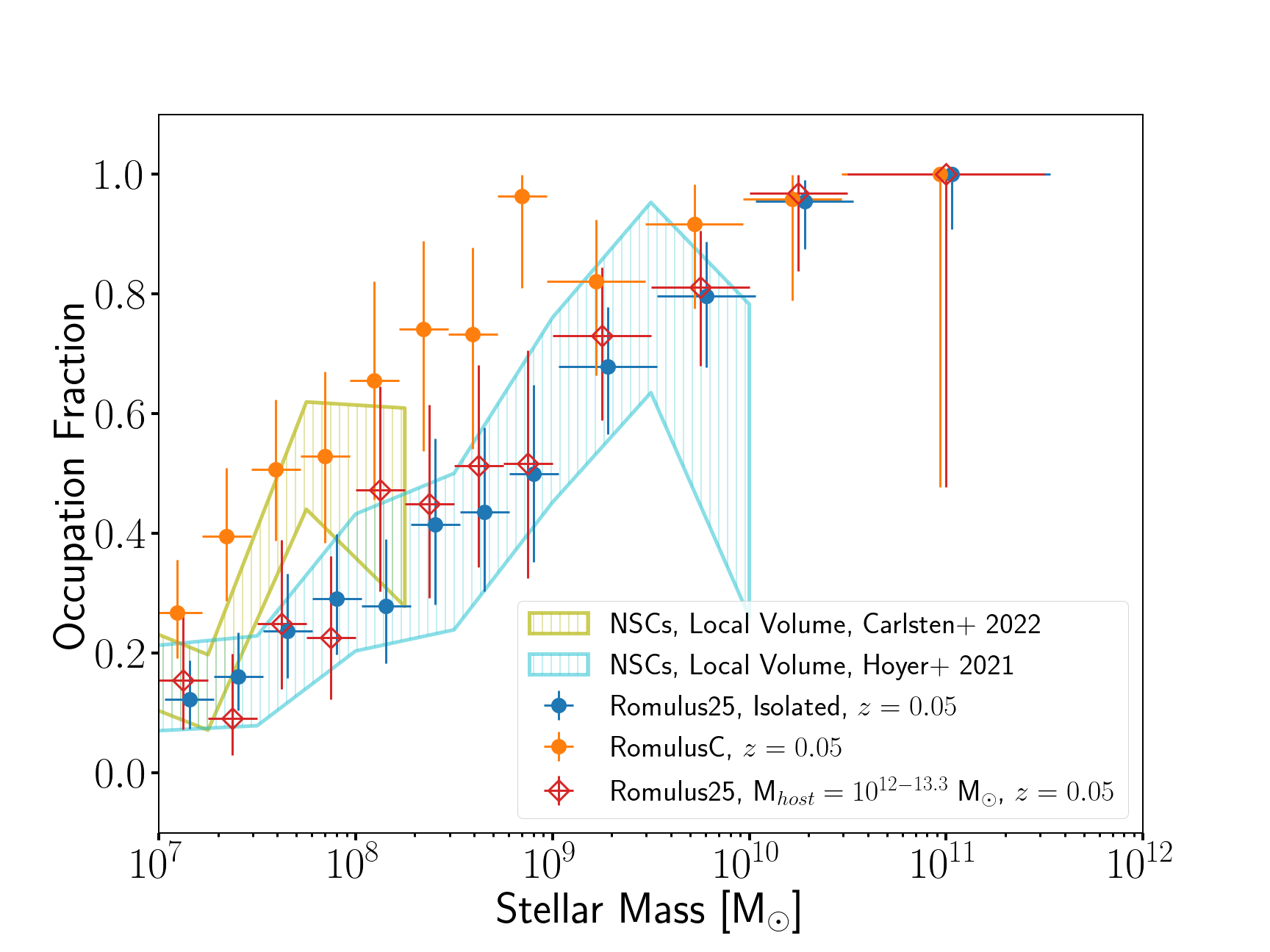}
\caption{{\sc The Occupation Fraction As A Function of Environment}. {\it Left:} The occupation fraction of MBHs in galaxies in {\sc RomulusC} at different cluster-centric distances. The orange points represent galaxies within 0.75 R$_{200}$ of cluster center while the red points are for galaxies in the cluster outskirts ($0.75 < \mathrm{D}/\mathrm{R}_{200} < 2$). The blue points are for isolated galaxies in {\sc Romulus25}. Although a subtle effect, there is evidence that galaxies in cluster outskirts do have less MBHs, though the low mass galaxies below 10$^{9}$ M$_{\odot}$ are still enhanced compared to isolated galaxies. {\it Right:} The occupation fraction of galaxies in {\sc Romulus25} in different environments. The red points show the occupation fraction of galaxies that are within 2R$_{200}$ of a halo of mass between $10^{12}$ and $10^{13.3}$ M$_{\odot}$. There is little difference between the occupation fraction at these intermediate masses and that for isolated dwarf galaxies (blue points). We also show observational results for local volume dwarfs from \citet{carlsten22} and \citet{hoyer2021} as hatched regions.}
\label{of_dist}
\end{figure*}

We can explore in more detail the extent to which the occupation fraction is dependent on environment. In the left-hand plot in Figure~\ref{of_dist} we find evidence that the occupation fraction evolves with cluster-centric distance. As one looks more toward the cluster outskirts ($0.75<$D$<2$ R$_{200}$) the MBH occupation fraction in galaxies does decline, although it remains systematically higher than that in isolated galaxies at stellar masses below $10^9$ M$_{\odot}$. This implies that while we might expect a gradual evolution from isolated to cluster galaxies, the influence of the environment persists even in the outskirts of clusters.

We can use {\sc Romulus25} to see if there are enhancements in the occupation fraction in less dense environments, such as low mass groups. The right-hand plot in Figure~\ref{of_dist} plots the occupation fraction of isolated galaxies in blue, cluster galaxies in orange, and galaxies within 2R$_{200}$ of a more massive halo with virial mass between $10^{12}$ - $10^{13.3}$ M$_{\odot}$ in red (i.e. the most massive halos in the 25 Mpc volume). We find no evidence that dwarf galaxies near these larger galaxies have any systematic enhancements to their MBH occupation. We compare again to the observed nuclear star cluster occupation in dwarfs from \citet{hoyer2021}, as well as \citet{carlsten22}, and find that the MBH occupation in dwarf galaxies associated with lower mass halos match well with observed NSCs in the local volume.

%While the MBH occupation in intermediate mass environments broadly matches these results we do not see an uptick at M$_{\star}\sim10^{7.5}$ as is seen with NSCs.

We confirm that these results are not sensitive to the specific host halo mass range used, though we are limited by the small volume of {\sc Romulus25}. These results imply that the enhancement in occupation fraction we see in the {\sc Romulus} simulations is isolated to very massive halos (massive groups and above; M$_{vir} > 10^{13.3}$ M$_{\odot}$). Lower mass halos assemble their mass earlier, meaning that low mass galaxies in-fall at higher redshift when the halo has a shorter characteristic dynamical time. These halos are more likely to become disrupted before $z=0$ and contribute to the population of wandering black holes in massive halos \citep{tremmel18, tremmel18b,ricarte21,ricarte21b}. It may also be that these lower density environments have a MBH occupation fraction enhancement only at galaxy masses that are currently unresolved in {\sc Romulus}.

%Figure~\ref{occ_frac_redshift_comp} compares the occupation fraction for {\sc RomulusC} and {\sc Romulus25} galaxies at different redshifts, including all SMBHs. The {\sc RomulusC} galaxies are always selected to be within 2 Mpc of the (proto-)cluster center at all redshifts. The occupation fraction changes with cosmic time in both populations, but at $z = 5$, just after the majority of SMBHs have finished being seeded in both simulations, the two populations look very similar. The difference in evolution among the two environments can be seen more clearly in Figure~\ref{occ_frac_evol}. In the field there is substantial evolution throughout cosmic time, while for cluster galaxies the evolution in the occupation fraction stalls around $z = 2-3$. This stalling is the origin of the difference in occupation fraction seen at $z = 0$ in Figures~\ref{occ_frac_all} and~\ref{of_any_cen_nsc}.

\section{Exploring the Origin of the Environmental Dependence of Dwarf Galaxy MBH Occupation}

\begin{figure*}
\centering
\includegraphics[trim=100mm 0mm 100mm 20mm, clip, width=180mm]{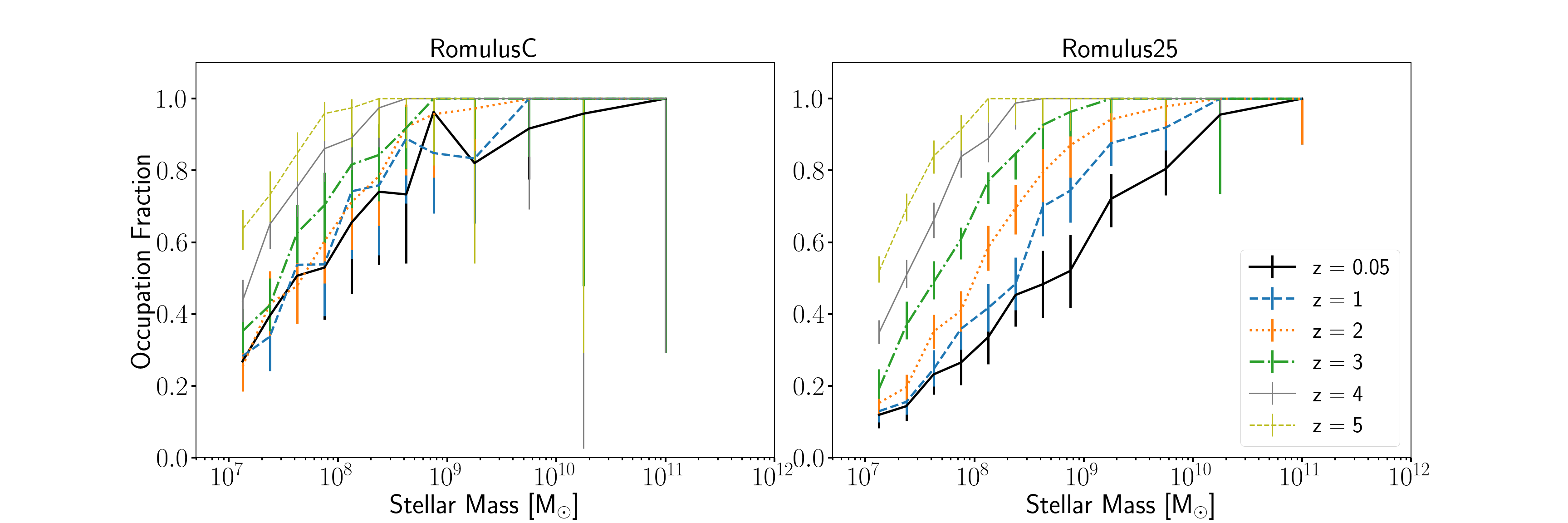}
\caption{{\sc The Evolution of the MBH Occupation Fraction Stalls in the Cluster Environment}. Similar to Figure ~\ref{occ_frac_redshift_comp} here we show the occupation fraction for galaxies as a function of stellar mass at six different redshifts for cluster galaxies (left) and field galaxies (right). While the occupation fraction in field dwarf galaxies decreases steadily through time, the evolution is slower in the cluster environment and plateaus at $z=2-3$.}
\label{occ_frac_evol}
\end{figure*}

The evolution of galaxies in dense environments like galaxy clusters is different compared with galaxies in the field. For cluster member galaxies, eventually, their ability to accrete new mass or form new stars will be shut off by the cluster environment \citep{mistani16}. However, even before galaxies become bound to the cluster, or in-fall to within R$_{200}$, they are growing within an over-dense region. This means that the galaxies that exist within a cluster environment at a given stellar (or halo) mass will have very different evolutionary histories compared with similar mass galaxies in the field \citep{gao05,gao07,croton07, bk09}. Cluster dwarf galaxies are likely to have earlier formation times because they must assemble their material before the cluster environment shuts down their growth. It may also be true that cluster dwarfs would have grown to much larger masses if their mass assembly was allowed to continue unabated by their environment \citep{mistani16}. Finally, cluster galaxies may also have their mass removed through tidal stripping as they interact with the cluster potential. In some cases, this stripping can unbind the majority of their stars, leaving behind only the dense, central stellar component (such as a NSC), which is then detected as an ultra-compact dwarf \citep{drinkwater03,bekki03b,seth14,voggel16}. In the following sections we evaluate the ability of each of these evolutionary scenarios in explaining the unique MBH occupation fraction in the simulated cluster ({\sc RomulusC}) environment compared to the field ({\sc Romulus25}).

\subsection{Cluster dwarfs as stripped remnants of more massive galaxies}

The enhanced dwarf galaxy MBH occupation fraction in {\sc RomulusC} could be explained if low mass host galaxies in the cluster were once much more massive. If a more massive galaxy were to become tidally stripped, it could lose significant stellar mass while retaining its central MBH. If this is true for a large number of low mass galaxies in {\sc RomulusC}, then it would make sense that their MBH occupation fractions would resemble that of more massive galaxies (i.e. their original mass prior to tidal stripping). However, as discussed in \citet{tremmel20}, while significant dark matter mass is lost due to interacting with the cluster potential, the majority of galaxies have not lost much stellar mass. While much of the dark matter stripping occurs on larger scales, stellar mass is confined within the galaxies themselves. Tidal stripping of this much more compact component requires closer, more intense interactions with the center cluster potential that are likely to result in the complete disruption of the galaxy, as far as the simulation and halo finder are concerned. It is important to note that some of this disruption could be artificial due to the limited resolution of the simulation \citep{vdBosch18b}. It is also important to note that {\sc Romulus} cannot resolve extremely compact structures within galaxies that would be able to best survive significant tidal interactions, such as NSCs. %Such structures left behind from more massive galaxies are thought to form the observed ultra-compact dwarf galaxies in groups and clusters \citep[e.g.][]{seth14,voggel19}.

The analysis from \citet{tremmel20} was done by tracing halos back in time to compare their $z=0$ stellar mass with their maximum stellar mass. This could induce a bias where the halos that pass closest to cluster center are more likely to have time-steps where they are missed by the halo finder, a well known issue \citep[e.g.][]{knebe11,onions12,joshi16}. This would cause us to be unable to fully trace their evolution through time and these potentially heavily stripped galaxies would be ignored. Focusing on MBH hosts, we can instead trace the MBHs themselves through time without relying on halo finding and compare the final host stellar mass with the maximum host stellar mass, excluding any intervening steps where they are temporarily taken to be hosted by the main cluster halo (i.e. times where the halo finder fails to extract their host sub-halo). 

Focusing only on MBHs hosted in cluster dwarf galaxies (M$_{\star}<10^{10}$M$_{\odot}$) at $z=0.05$, we find that the median MBH host galaxy has only experienced a net loss of $\sim20\%$ of its stellar mass as it interacts with the cluster environment. Only one fifth of the MBH hosts have seen their stellar mass decrease by more than a factor of $\sim3$. Looking at figure \ref{occ_frac_all}, a typical stellar mass loss of a factor of $\sim3-5$ is needed to bring the occupation fractions in line with the field. In {\sc RomulusC} such extreme mass loss is too rare to fully explain the overabundance of MBHs in low mass galaxies.

\subsection{Cluster Dwarfs as Failed Massive Galaxies}

Galaxies in cluster environments will eventually stop accreting new material, as both dark matter and gas will flow onto the primary halo instead. The lack of replenishing gas supply combined with ram pressure removing the ISM will eventually slow down or completely quench new star formation in the galaxy. However, it is possible that, were these galaxies allowed to continue to grow unimpeded, they would be more massive at $z=0$. If the progenitor galaxies to our cluster dwarfs are more similar to progenitors of massive field galaxies than they are to those of field dwarfs, this could explain the difference in MBH occupation fraction. In other words, it may be that cluster dwarfs actually represent progenitors to more massive field galaxies that were instead `frozen' in their growth by their environment. In this scenario it is not required the galaxies lose stellar mass, just that they fail to reach the same masses as their field counterparts.

In order to explore this scenario, we trace our cluster galaxies back in time, finding the redshift, stellar mass, halo mass, and concentration\footnote{We define concentration here as v$_{max}$/v$_{200}$, where v$_{max}$ is the maximum circular velocity and v$_{200}$ is the circular velocity at R$_{200}$.} at the time they reach maximum virial mass (t$_{max}$). We exclude all galaxies that fail to trace backward to a time prior to falling into the cluster. These progenitors are then matched to galaxies at that same redshift in {\sc Romulus25} which are major progenitors to isolated $z=0.05$ galaxies, requiring the stellar and halo masses be within 0.2 dex and the difference in concentration be less than 0.2. For each cluster galaxy, we require they match with at least 4 field galaxies with this criteria. We then recalculate the occupation fractions using for each cluster galaxy the median $z=0.05$ stellar mass among the matched isolated galaxies. The idea is that this should approximate the stellar mass they would have attained were they allowed to continue to grow.

%The median stellar mass among these matched, isolated $z=0$ galaxies are then used in place of each cluster galaxy's `true' stellar mass and the occupation fractions are re-calculated. In other words. we use {\sc Romulus25} to estimate the stellar mass each cluster galaxy would attain if they never fell into the cluster. We require that each cluster galaxy be matched with at least 4 such isolated galaxies to be considered for this analysis, though we find that our results are insensitive to this choice.

\begin{figure}
\centering
\includegraphics[trim=15mm 10mm 10mm 30mm, clip, width=90mm]{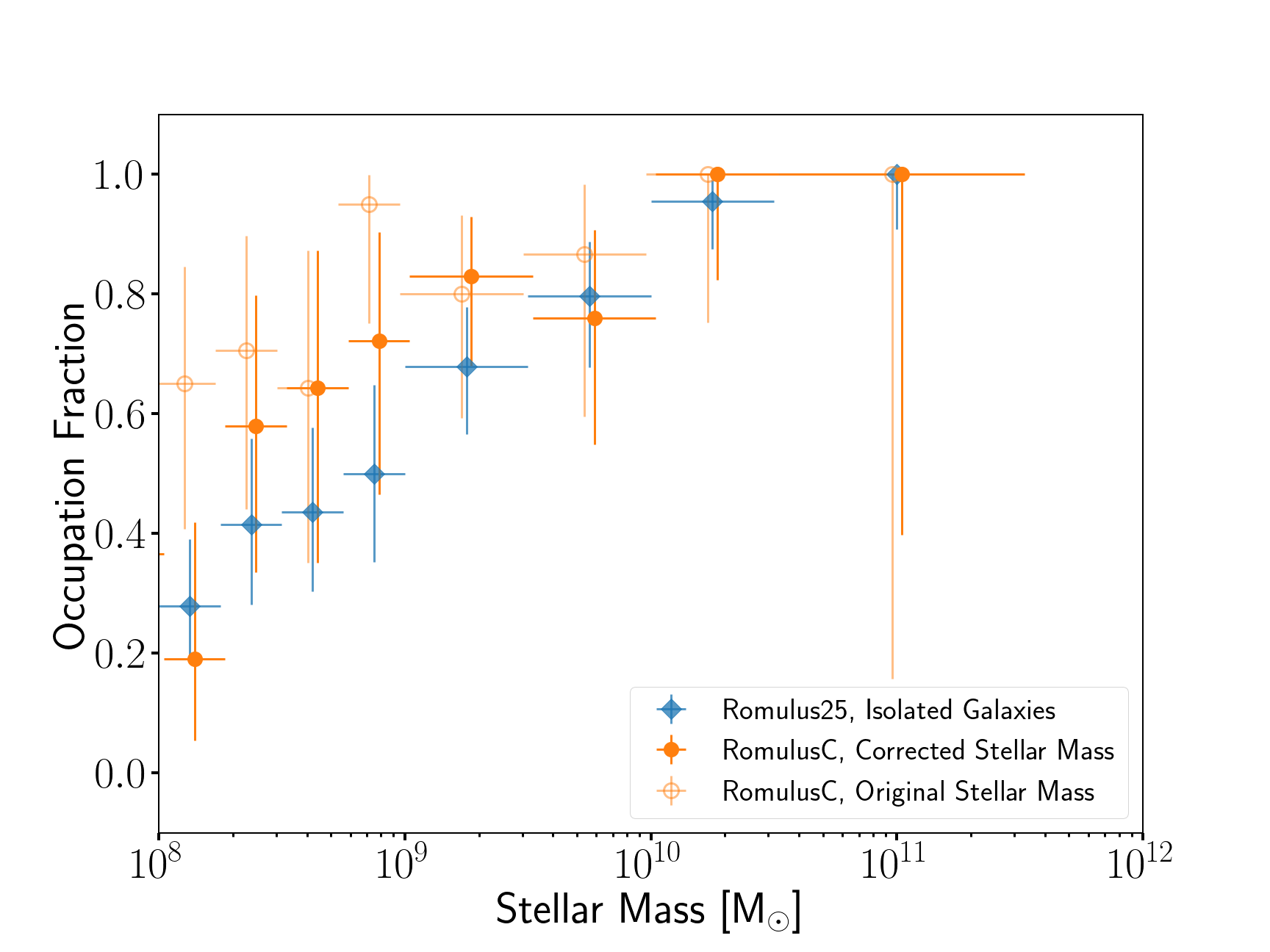}
\caption{{\sc The Occupation Fraction for Galaxies Matched with Isolated Galaxies}. Here we test the extent to which the difference in occupation fraction can be explained by cluster dwarf galaxies being `failed' larger galaxies, i.e. galaxies that, were they not in a cluster environment, would have grown to be more massive. In blue we plot the occupation fraction for isolated galaxies in {\sc Romulus25} at $z=0$. The solid orange points show the occupation fraction as a function of galaxy mass after each cluster galaxy, where its mass has been corrected to estimate what it might have attained had the galaxy not fallen into a cluster. We do this by matching each cluster galaxy with a $z=0$ isolated galaxy based on its stellar and halo masses at $t_{50,halo}$ (the time when the host dark matter halo reaches 50\% of its maximum mass; see text for details). The open, light orange points are these same galaxies but with their original final masses. Note that we do not include the lowest mass galaxies in this analysis because too many of them fail to be traced back in time successfully (see text for details). While this matching process does alleviate some of the differences, the `corrected' occupation fraction for cluster dwarfs remains systematically higher than isolated dwarf galaxies. This implies that stunted mass growth is only a part of the explanation for the different occupation fractions.}
\label{of_matched}
\end{figure}

Figure~\ref{of_matched} shows the results of this analysis, plotting the occupation fraction using both the original $z=0$ stellar masses  (open orange points) and the stellar masses calculated by matching the progenitors to isolated galaxies (solid orange points). We are only able to do this analysis with galaxies with original stellar masses above $10^8$ M$_{\odot}$. At lower masses too many galaxies become excluded because we fail to trace them back before in-fall into the cluster due to the halo finder failing to identify them at some point. We confirm that these results are not sensitive to our specific matching criteria, specifically which combination of halo mass, stellar mass, and concentration were used, as well as the number of matches required for the analysis.

%\begin{figure*}
%\centering
%\includegraphics[trim=40mm 10mm 10mm 0mm, clip, width=180mm]{plots/t50_dist_halo_dwarfs.png}
%\includegraphics[trim=40mm 10mm 10mm 0mm, clip, width=180mm]{plots/t50_dist_stars_dwarfs.png}
%\caption{{\sc Halo and Galaxy Formation Timescales}. \textit{Top:} Cumulative distributions for $t_{50,halo}$, the time at which each galaxy's dark matter halo reaches 50\% of its (maximum) mass, for dwarf galaxies within two stellar mass bins. For isolated galaxies in {\sc Romulus25} (blue, black) we use each halo's final $z=0$ mass and for cluster galaxies in {\sc RomulusC} (orange) we use each galaxy's maximum halo mass when performing this calculation. \textit{Bottom:} Cumulative distributions for $t_{50,stars}$, the time at which each galaxy forms half of its final stellar mass (again, using the maximum value for {\sc RomulusC} galaxies). The solid blue and orange lines show the disributions for galaxies hosting a MBH and the dashed lines are for galaxies without MBH. In isolated galaxies, MBH hosts have earlier formation times in terms of both their stellar mass and their halo masses. Cluster dwarf galaxies on average form significantly earlier than isolated dwarf galaxies with respect to both stellar and halo mass. This demonstrates that cluster dwarf galaxies have experienced significantly different evolutionary histories compared to typical isolated galaxies.}
%\label{t50_dist}
%\end{figure*}

If the majority of low mass MBH host galaxies are more like progenitors to massive, isolated galaxies, the `corrected' occupation fraction technique would see many MBHs hosts move to larger (corrected) stellar masses, decreasing the occupation fraction at low mass and bringing the results more in-line with the field. While we see in Figure~\ref{of_matched} that this matching technique does result in lower occupation fractions at low masses, the results are still systematically higher than the field. This implies that while some MBH host galaxies may be considered `failed' massive galaxies (i.e. they could have grown larger had their mass accretion not been shut down by the cluster environment) this scenario fails to fully explain the discrepancy in occupation fraction between environments. Of course, because some galaxies had to be excluded due to bad tracking, the error bars are larger and the difference between the corrected and original occupation fractions for individual mass bins are often only marginally significant.

\subsection{An Overabundance of Early-forming Dwarfs in Clusters}

%Figure \ref{t50_dist} plots the cumulative distribution of formation times (t$_{50}$, or the time to accumulate 50\% of their stellar/halo mass) of galaxies with and without black holes in terms of their stellar mass (bottom) and total halo mass (top).
As discussed in \citet{sharma20}, dwarf galaxies in {\sc Romulus25} hosting black holes, especially those that are more massive, tend to have earlier formation times for both their stars and their overall halo mass. While feedback from black holes could influence the assembly history of stars, potentially quenching star formation even in low mass galaxies \citep{sharma22b, koudmani21}, the fact that this trend is also seen in dark matter halo assembly indicates that it is likely something more fundamental. In cluster environments, the assembly of galaxies and the dark matter halos in which they reside is stopped by the cluster environment, such that a galaxy of a given mass in the cluster must have assembled that mass prior to in-fall. This is an expected result of assembly bias in the formation of dark matter halos \citep[e.g.][]{gao05, gao07, croton07, bk09} and is seen in other cosmological hydrodynamic simulations \citep[e.g.][]{mistani16, chavesMontero16}.

%We can see that in Figure~\ref{t50_dist} by comparing the solid blue (with black holes) to the dashed blue (without black holes) lines. The figure shows results for two bins of dwarf galaxy stellar mass. In both mass bins, galaxies with MBHs have preferentially earlier formation times. We can examine the same properties but for cluster dwarf galaxies (orange lines). We see the same trend where galaxies hosting black holes have earlier formation times, but the more significant difference is between environments, where the typical formation time of cluster dwarf galaxies is significantly earlier than for isolated dwarf galaxies in {\sc Romulus25}. This is an expected result of assembly bias in the formation of dark matter halos \citep[e.g.][]{gao05, gao07, croton07, bk09} and is seen in other cosmological hydrodynamic simulations \citep[e.g.][]{mistani16, chavesMontero16}.

%something which is seen in other simulations \citep[e.g.][]{mistani16} and is the result of assembly bias \citep{gao05}

\begin{figure*}
\centering
\includegraphics[trim=40mm 0mm 0mm 0mm, clip, width=180mm]{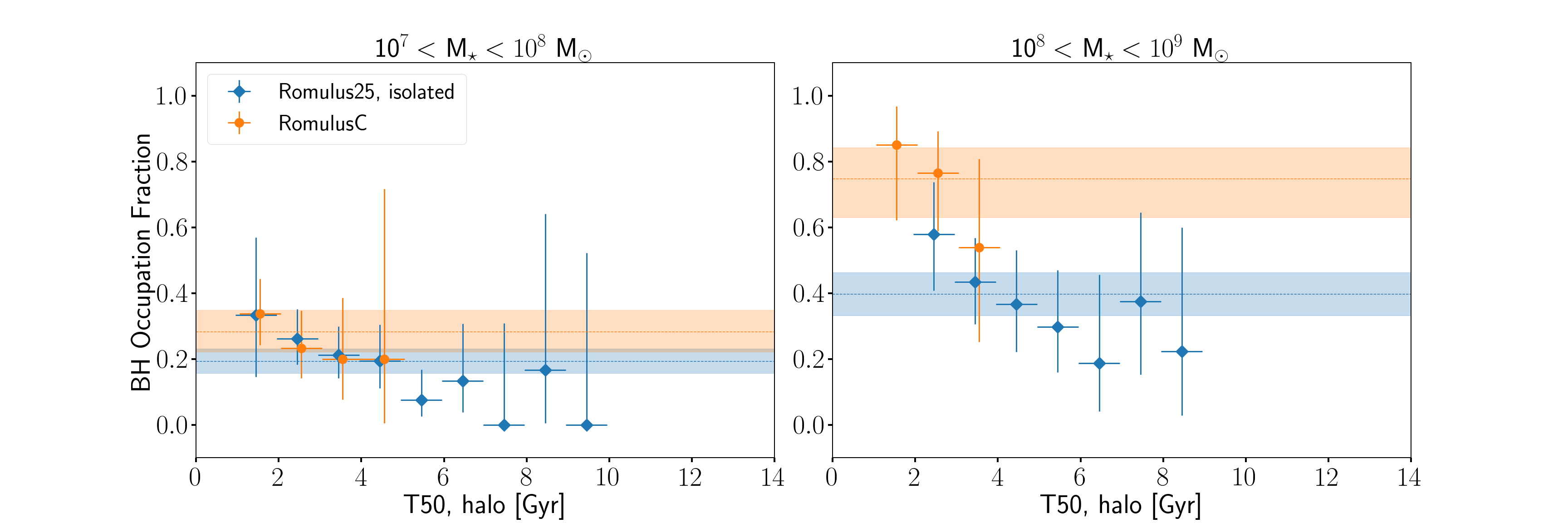}
\includegraphics[trim=40mm 0mm 0mm 0mm, clip, width=180mm]{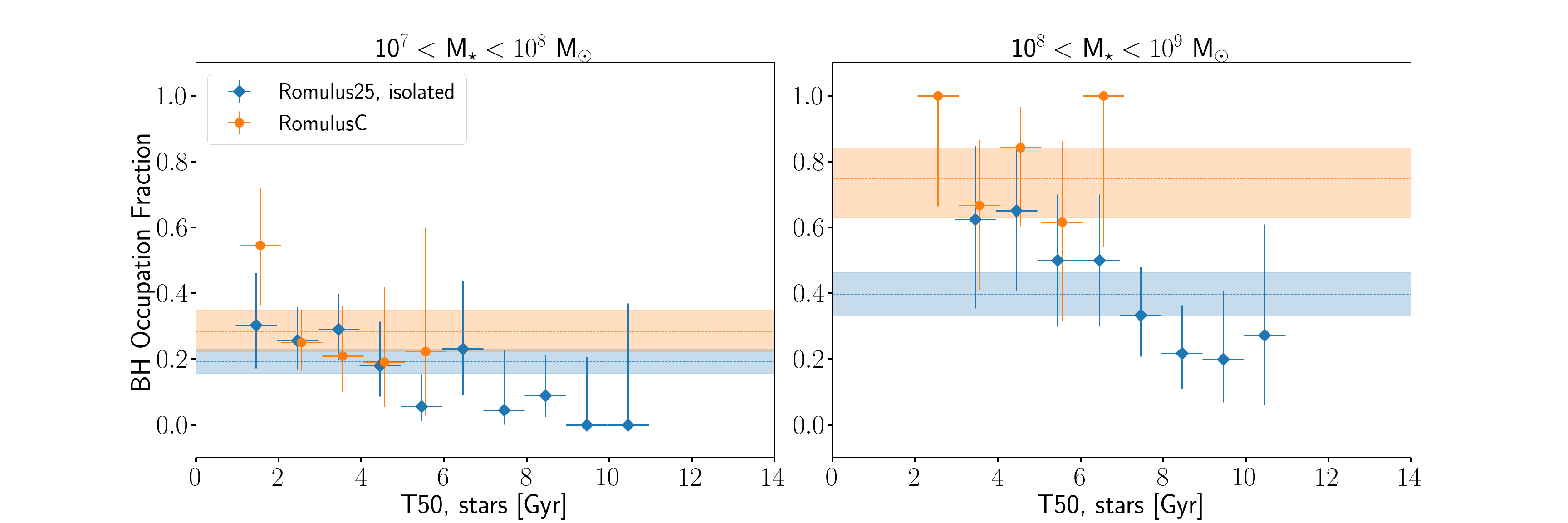}
\caption{{\sc Early Forming Dwarfs are More Likely to Host MBHs}. The occupation fraction of MBHs as a function of $t_{50,halo}$ (top) and $t_{50,stars}$ (bottom) for dwarf galaxies in two stellar mass bins. The bands represent the total occupation fraction in each mass bin (only for galaxies that can be traced adequately far back in time; see text for details). The relationship between formation time and MBH occupation fraction is similar for cluster dwarfs (orange) and isolated dwarfs (blue), though galaxies in the higher mass bin are still biased slightly high when controlling for either formation timescale. Dwarf galaxies that form earlier are more likely to host a MBH. The important difference between the two simulations, therefore, is that dwarf galaxies within cluster environments form earlier than those that are more isolated. }
\label{of_vs_t50}
\end{figure*}

Figure ~\ref{of_vs_t50} plots the occupation fraction as a function of formation time for the same two stellar mass bins of dwarf galaxies. The top panels show the halo formation time (time to accumulate 50\% of the maximum halo mass) and the bottom panels show the stellar formation time (time to form 50\% of the maximum stellar mass). The orange and blue bands show the average occupation fraction for successfully tracked galaxies in each mass bin. Note that the average values are calculated only for the galaxies that have been successfully traced back through time and included in the calculation of the individual data points shown here.

In both the field and cluster environments, dwarf galaxies with earlier formation times are more likely to host MBHs. This is true when examining either stars or halo mass. The MBH occupation fraction for cluster dwarfs is similar to field dwarfs when controlling for formation time, though this connection is better illustrated by halo mass when considering higher mass dwarfs. In the field, dwarfs of a given mass are allowed to form throughout cosmic time, but those that accumulate their mass later are less likely to host MBHs. This lack of late-forming dwarf galaxies in the cluster is what causes the evolution of the occupation fraction to stop evolving after $z\sim3$ (see Figures~\ref{occ_frac_redshift_comp} and~\ref{occ_frac_evol}). In the field, `new' dwarf galaxies grow at later times, filing those lower mass bins with galaxies that lack MBHs. While this is a function of the specific seeding criteria we use, these results show that regions of very dense, rapidly collapsing, pristine gas are more likely to exist in the progenitors to early-forming dwarf galaxies compared to late-forming dwarfs. In late-forming dwarfs, such early phases of collapse occur too slowly, allowing for the formation of stars and metal enrichment before the required high densities (far beyond the threshold for star formation in the simulation) are reached (if they ever are). In this scenario, the role of environment is more to stop the formation of late-forming dwarf galaxies, rather than influence on the formation sites of MBHs. As can be seen in Figures~\ref{bhformtime}, ~\ref{occ_frac_redshift_comp}, and~\ref{occ_frac_evol}, the time and host halos of MBH seeding is very similar between the two environments.

%decrease rapidly at low masses over time for field galaxies, as while `new' dwarf galaxies are continuously forming, however without black holes (PN: does this not completely reflect the seeding prescription adopted in Romulus - of pristine gas?). In the cluster, there are no late-forming dwarf galaxies so the occupation fraction becomes less diluted and indeed `freezes out' around $z=2$ (see Figures~\ref{occ_frac_evol} and~\ref{occ_frac_redshift_comp}).

Once again, we face the problem discussed in the previous section whereby low mass (M$_{\star} < 10^8$ M$_{\odot}$) galaxies with MBHs are more likely to be excluded because they fail to be tracked back in time successfully. These galaxies form earlier and fall into the cluster sooner so they are more likely to have passed closer to the cluster center and also be missed by the halo finder. However, there is still a significant difference in the mean occupation fraction in each mass bin relative to the field (comparing the orange and blue bands). %Likely, the bias arises because the earliest forming dwarf galaxies also fell into the cluster the earlier and have been more likely to experience a closer approach and be missed by the halo finder. As we see in figure ~\ref{of_vs_t50}, these earliest forming dwarfs are also the most likely to host a MBH.

Controlling for formation time results in a better match between the two populations of galaxies when looking at halo mass, rather than stellar mass. This makes sense, as many additional factors may play a role in the star formation history of a galaxy, including the presence of feedback from a MBH \citep{sharma20,sharma22b}. The more massive dwarfs in clusters still appear biased high relative to isolated galaxies with similar formation times. This may indicate that a combination of effects are needed to fully explain this enhanced MBH occupation population, i.e. some dwarfs could have assembled into more massive galaxies were they isolated (see previous section) combined with a lack of late-forming dwarf galaxies in clusters.

\section{Discussion}

Observational constraints on the underlying MBH occupation fraction in low mass galaxies are uncertain because it is difficult to detect the low mass, low luminosity black holes. Indeed while the {\sc Romulus} simulation has been shown to reproduce observed samples of dwarf galaxy AGN fractions and luminosities (with specific assumptions made to convert between black hole accretion rate and X-ray luminosity) the simulation predicts a large population of MBHs that would go undetected by even the most sensitive modern X-ray surveys \citep{sharma22}. Still, as observations improve, evidence increasingly points to a significant number of MBHs in low mass galaxies \citep{nguyen18, baldassare20,burke22}. Much work is still needed from the observational side, but upcoming time domain surveys from the Vera Rubin Telescope may offer hope for dramatically increasing the completeness of the observed MBH popuation through their intrinsic variability \citep{baldassare18,burke22} as well as tidal disruption events \citep{bricmanGomboc2020}. JWST may also be a powerful tool for detecting AGN with low X-ray luminosities \citep{cann21}. As observations continue to get better at detecting MBHs in low mass galaxies, predictions like the ones made in this work will be crucial in understanding and contextualising them with respect to MBH formation models. 

The challenge for simulations is, as always, resolution which comes at the cost of the size and statistical sample of the data. Large-scale simulations like {\sc Romulus} reach a middle ground by being large enough to have many galaxies while also capable of resolving dwarf galaxies. Still, smaller volumes means a lack of environment diversity with only a handful of groups and a single low-mass galaxy cluster. While newer, large-volume simulations are becoming better in terms of both resolution and the black hole physics they implement \citep[e.g.][]{dubois21, obelisk21, ni22} it remains a challenging balance. Even at the resolution of {\sc Romulus} and these other state-of-the-art simulations, the ISM remains largely unresolved, requiring relatively simple prescriptions for black hole formation (see below for further discussion). Zoom-in simulations are another viable path forward, allowing for more detailed formation prescriptions \citep[e.g.][]{dunn18} and more detailed analysis of both black hole dynamics and the internal structure of dwarf galaxies \citep{bellovary19, bellovary21}. Very high resolution simulations targeting the high redshift Universe have also proven useful tools for examining the physics of MBH formation \citep{wise19,regan17,regan20,regan23}.

The seeding algorithm for MBHs implemented in {\sc Romulus} is more predictive than many previous large-scale cosmological simulations because it seeds MBHs based on local gas properties (density, temperature, metalicity) without making any {\it a priori} assumptions on which halos/galaxies should host a MBH. As discussed below, our model is still simplistic because of limited resolution, but the simulations still have significant predictive power.
%simple because of limited resolution, it is still predictive. 
In particular, our results demonstrate that the gas properties of galaxies at $z>5$ is connected to their formation history and, therefore, so may be their likelihood of hosting a MBH. Higher resolution simulations will be able to further test this prediction, as will future observations. The environmental dependence of MBH occupation that we predict here should be considered a potential way to differentiate between MBH formation mechanisms. For example, an observed environmental dependence of MBH occupation would support the theory that the primary formation channel of MBHs occurs in the early ($z>5$) Universe from metal poor gas. However, a lack of observed environmental dependence, based on our results here, would indicate that other formation channels which have MBHs grow at later times and from more metal polluted gas \citep[e.g.][]{ regan20b,natarajan21, mayer23} likely dominate.

\subsection{The Effect of MBH Formation Model}

The most important caveat to these results is that they will naturally rely on our choice of MBH formation criteria. The main concern is whether our choices directly influence our results, which is not the case here. The criteria in {\sc Romulus} are common sense requirements given any of the leading MBH formation models and the requirement that each MBH should be able to grow to large masses in a relatively short amount of time. In practice, our criteria will pick out gas that is collapsing to very high densities on a timescale shorter than the typical star formation timescale (assumed to be $10^6$ yr) and faster than it can effectively cool. The additional criteria that this gas must be (nearly) pristine means that such locations must form prior to or simultaneously with the very first stars forming in the (proto-)galaxy. Despite the simplicity of the model, the connection between the high-redshift properties of (proto-)galaxies and their future assembly history and environment remains a prediction of the model, rather than a direct consequence of our choice in criteria. Still, we do not attempt to test different criteria and it is very possible that this would influence our results. For example, softening the metallicity or density requirements would make MBHs much more common likely wash out any environmental dependence. 

It should be noted that this model is only capable of capturing MBH formation channels that take place in pristine (or near-pristine) gas. This is primarily what results in an early formation epoch, as most gas becomes polluted as stars form in the simulation. However, it may be possible to grow MBHs at later times in metal enriched gas, either growing a low mass seed within star clusters \citep{natarajan21} or in massive merger events \citep{mayer23}. The formation model implemented in {\sc Romulus} would not capture such channels.

The fact that our theoretical results produce active fractions consistent with observations \citep[see Figure~\ref{occ_frac_all} and][]{sharma22} indicates that our model parameters are reasonable, at least. However, \citet{sharma22b} find that AGN feedback in dwarf galaxies is the primary cause for over-quenching low mass galaxies. This could indicate that our occupation fractions are too high in the field, though this is just as likely an issue with overly efficient MBH accretion and/or feedback. In any case, a more strict formation criteria would decrease occupation fractions of MBHs and could potentially increase the divide between environments even further.

\subsection{Halo Finder Limitations}

Our analysis has been limited by our ability to extract halos in consecutive timesteps. The difficulty of extracting substructure close to the centers of dense structures like clusters is a well-known issue with halo finding routines. While this may result in an artificial lack of low mass galaxies deep within the cluster, this should not effect our overall results on the occupation fraction of cluster dwarf galaxies. In fact, galaxies closer to the center of the cluster are likely to have fallen in earlier and are therefore more likely to have hosted a MBH, so including more central dwarfs could increase our cluster occupation fractions further. 

More important is the effect on our ability to trace halos backward in time. This requires that a given halo is detected in all timesteps while it is in the cluster, which may not happen if it passes close to the center at any point. Given the wide mass range we examine and the fact that we are able to successfully trace back the majority of even the smallest galaxies, these missed galaxies should not affect our conclusions. In fact, we should preferentially miss the earliest forming dwarf galaxies that fall into the cluster first, which would only strengthen the effect that we see already.

\subsection{Effect of Resolution}

An important numerical affect caused by limited resolution is the artificial disruption of dark matter halos (and galaxies). As discussed in detail in \citet{vdBosch18b}, limited particle count and gravity resolution results in substructure becoming artificially disrupted. In the simulation, most galaxies that do experience significant tidal stripping of their stars are very soon completely disrupted while, in reality, it is possible they should survive. This would mean that we underestimate the portion of MBHs that exist in significantly stripped galaxies. This would likely further increase the difference in occupation fraction we already see in the cluster, as we would have an additional population of dwarf galaxies hosting MBHs that we currently do not resolve. It would also mean that tidal stripping is more important than we currently predict to the overall MBH population in cluster dwarfs \citep[see][for more discussion on this population of MBHs based on observations]{voggel19}.

Similarly, limited resolution means that {\sc Romulus} will not resolve dense stellar structures, such as nuclear star clusters, at the centers of galaxies. These structures are more resilient to tidal effects, so even if the disruption of subhalos were all real it is possible that some of these structures would survive around the MBHs as ultra-compact dwarf galaxies \citep{drinkwater03, bekki03b,seth14}. Similar to artificial disruption, the effect of this would be that there is a more significant population of tidally stripped remnants that host MBHs. It would also further increase the predicted environmental dependence of the occupation fraction by, once again, adding a new population of dwarf galaxies hosting MBHs to our sample. Further, this would create a population of galaxies with significantly overmassive black holes. As discussed in \citet{ricarte19}, the accretion histories of MBHs in {\sc RomulusC} dwarf galaxies is very similar to that of field galaxies, and so galaxies in both environments exist on the same stellar mass-black hole mass relation. This might not remain the case if the number of artificially disrupted galaxies/nuclear star clusters is accounted for, but we leave this question to future work more focused on explaining observations of MBHs in ultra-compact dwarf galaxies \citep[e.g.][]{seth14,afanasiev18,voggel19}.

\subsection{Connection to Nuclear Star Clusters}

The MBH occupation fractions we predict with our relatively simple seed formation model matches remarkably well with observations of NSCs in cluster dwarf galaxies \citep{munos15,sanchezJanssen19}, as well as local group dwarfs \citep{hoyer2021, carlsten22}. 
There is reason to think that the formation of NSCs and MBHs could be connected. It may be that MBHs are seeded as a result of the evolution of a dense nuclear star cluster \citep{devecchi2009, davies11, kroupa20}. More broadly, the environment likely to form/grow a MBH (very dense, pristine gas) is also the site of very dense, early star formation that seeds an initial NSC \citep[note that NSCs often include stars with  variety of ages and metalicities, indicated extended star formation histories][]{seth06,carson15,kacharov18}. Some work suggests that NSCs and MBHs form from entirely separate mechanisms \citep[e.g.][]{scott13} or that NSCs may grow mostly from mergers of other star clusters, rather than in-situ formation \citep{antonini15, fahrion20}. In reality, it is likely that a combination of mechanisms are occurring \citep[e.g.][]{fahrion21, fahrion22}.

%The core collapse of dense NSCs can result in runaway stellar collisions that produce a supermassive star which then evolves into a MBH of $\sim10^3$ M$_{\odot}$ \citep{devecchi2009}. The collapse of a central, dense star cluster due to rapid gas accretion can result in mergers between stellar mass black holes that can result in a MBH as large as $10^5$ M$_{\odot}$ or more \citep{davies11}. Even in models where MBHs form from a single, supermassive star, or even the remnants of Population III stars, the environment in which these objects are able to grow to high masses are also likely locations for dense, early star formation. 

%The exact formation mechanism for NSCs remains uncertain and some work suggests that NSCs and MBHs form from entirely separate mechanisms \citep[e.g.][]{scott13} or that NSCs may grow mostly from mergers of other clusters, rather than in-situ formation \citep{antonini15}. In reality, it is likely that a combination of the two mechanisms are occurring \citep[e.g.][]{fahrion21, fahrion22}. 

Many NSCs co-exist with MBHs and their masses scale with the mass and properties of their host galaxies in similar ways \citep{wehner06,ferrarese06b, seth08, georgiev16, nguyen18, nguyen19, fahrion22b}. The potential well of a NSC can help low mass MBHs grow \citep{natarajan21,askar22}. Conversely, feedback from MBHs, as well as their dynamical interactions with nearby stars, can hinder the growth of NSCs and even disrupt them completely \citep{antonini13, antonini15, sanchezJanssen19}. The presence of MBHs within dense, nuclear regions are thought to explain the presence of supermassive black holes in ultra-compact dwarf galaxies typically found in groups and clusters \citep{seth14,afanasiev18,voggel19}.

%gas accrete more efficiently onto low mass MBHs, potentially representing a separate channel for intermediate mass black hole formation \citep{natarajan21,askar22}. On the other hand, feedback from MBHs, as well as their dynamical interaction with nearby stars, can hinder the growth and even completely disrupt NSCs \citep{antonini13, antonini15}. This effect is often invoked to explain the lack of NSCs in massive galaxies hosting the largest supermassive black holes \citep[e.g.][]{sanchezJanssen19}. 

%The presence of MBHs within dense, nuclear regions may also explain the presence of supermassive black holes within ultra-compact dwarf galaxies found in group and cluster environments \citep{seth14,afanasiev18,voggel19}. Given the unusually large masses of the black holes relative to their host galaxies, these systems are thought to be the dense central regions of a once more massive galaxy that had the rest of its mass tidally stripped by the host halo. Importantly, because we cannot resolve ultra-compact dwarf galaxies, none of the dwarf galaxies presented here are stripped remnants of more massive galaxies. As discussed in \citet{tremmel20}, in {\sc RomulusC} galaxies lose at most half of their stellar mass between in-fall and $z=0$, meaning that dwarf galaxies at $z=0$ were always dwarf galaxies.

The formation criteria for MBHs in {\sc Romulus} is relatively agnostic regarding the exact physics as we are far from resolving the formation process itself. Rather, it is meant to encapsulate the type of environment where one may expect a MBH to form and grow rapidly in the early Universe, i.e. where cold, low metalicity gas is collapsing on timescales much shorter than the star formation timecale. Such conditions are required for all formation channels of MBHs. Even starting as Population III remnants, the seeds would need a lot of dense gas nearby to quickly grow large enough to become $10^5$ M$_{\odot}$ in a short amount of time \citep[see][for further discussion on this]{volonteri12}. While we are far from being able to resolve NSCs in the simulation, the connection to observed NSC occupation makes sense regardless of the details of NSC formation and suggests that NSCs might serve as incubators for the formation and growth of MBH seeds over cosmic time as noted by \citet{natarajan21} . An in-situ formation channel would require similar properties to MBH formation (very dense gas in the early Universe) but a formation channel dominated by globular cluster mergers \citep[e.g.][]{antonini12, antonini15, fahrion20} would also fit with a connection to MBHs. {\sc Romulus} predicts that the environment required for MBH formation is more likely in galaxies (and dark matter halos) that form earlier. Such galaxies will be likely to form more globular clusters that will then have more time to sink and merge and form a NSC. Observations have shown an overabundance of globular clusters in dwarf ellipticals residing in galaxy cluster environments \citep[e.g.][]{miller98,miller07,jordan07,peng08,sanchezjanssen12} and the cause of this can be attributed to their earlier formation times \citep{mistani16, carleton21}.

The fact that our results match so well with observed NSC populations across environments \citep{munos15,sanchezJanssen19, hoyer2021, carlsten22} while producing realistic black hole occupation fractions in the field  (though observational constraints are murky at best) supports the notion that MBH formation could be connected with NSC formation. While we cannot directly resolve the formation of NSCs, our results indicate that their presence, like that of MBHs, may be connected not only to the properties of gas at high redshift, but to the overall formation history of galaxies and therefore, indirectly, their environment.

%Unfortunately, we are far from able to resolve NSCs with {\sc Romulus} as that would require more than 100 times better resolution, something that is difficult even for zoom-in simulations, so it is difficult to say what the exact connection would be. It may be that NSCs and MBHs form/grow at similar times and from similar gas reservoirs, which our MBH formation model picks out. It could also be that the same galaxies that form MBHs at very high-z are also the ones that have the right conditions to form NSCs later from, e.g., globular cluster mergers.

\section{Conclusions}

We use the {\sc Romulus} simulations to predict an enhanced MBH occupation fraction in cluster dwarf galaxies ($M_{\star}<10^9$M$_{\odot}$). The {\sc Romulus} simulations are unique in their ability to both resolve low mass galaxies and implement a model for MBH seeding that relies only on local gas properties (density, temperature, metallicity) rather than requiring any \textit{a priori} assumptions about which halos should or should not host a MBH. Despite forming black holes at similar times and within similar galaxy masses, we find that the cosmic evolution of the MBH occupation fraction in galaxies is halted at $z\sim3$ in the cluster environment relative to the field. This `freezing out' of the occupation fraction results in a factor of $\sim2$ enhancement to the fraction of dwarf galaxies that host MBHs in clusters relative to the field at $z=0.05$.

We investigate the cause of the enhancement in more detail and find that it can likely be explained by a combination of two mechanisms: 

\begin{enumerate}
    \item Early formation times of dwarf galaxies in cluster environments makes them more likely to host a MBH. When controlling for formation time, cluster and field galaxies have similar MBH occupation fractions, but late-forming dwarf galaxies that do not host MBHs dillute the field population and pull down the MBH occupation fraction, while these systems do not exist in clusters.
    
    \item Some cluster dwarf galaxies may be `failed' massive galaxies. Were they allowed to grow in the field, unimpeded by the cluster environment, they would likely have attained much higher masses. 
\end{enumerate}

\noindent We do not find evidence that many MBH host galaxies experience tidal stripping of their stars. However, future work will examine whether some MBHs in the simulation may have unresolved, compact stellar structures around them. 

The enhanced MBH occupation fraction in the cluster simulation appears to fall with increased cluster-centric distance, but it remains in place out to $2R_{200}$ for galaxies with $M_{\star}<10^9$M$_{\odot}$. While we do not attempt to model the detailed physics of MBH formation in these simulations, the connection between galaxy assembly history, environment, and the properties of gas at high redshift are important predictions. The presence of quickly collapsing, high density regions of pristine gas are likely formation sites of MBH seeds. While such environments are common in the progenitors of massive galaxies, we predict that only early forming dwarf galaxies are typically able to produce such environments. Such dwarf galaxies make up a much higher fraction of the overall population in dense environments like galaxy clusters. Our findings have important consequences for the origin and evolution of host galaxy-MBH co-evolution. 

Finally, we find that the predicted MBH occupation fraction in the cluster is remarkably consistent with the observed occupation fraction of nuclear star clusters in Virgo and Fornax, while the field occupation fraction is similar to that of dwarfs in the local volume. So, the dense, pristine gas at $z>5$ that the simulation attributes to MBH formation may also be connected to the formation of nuclear star clusters, either directly (they form from similar gas at similar times) or indirectly (e.g. connected to the earlier formation times associated with the host galaxies).

These results show how MBH occupation may be not just a function of galaxy mass but also environment, and that there are many connections between the high redshift properties of galaxies and their overall assembly history. We note that {\sc RomulusC} is a relatively low mass cluster and more massive ones may have even more enhancement in their occupation fractions. These results also indicate that we must be cautious in how we utilize AGN observations to constrain the underlying MBH occupation fraction, as the connection is heavily dependent on environment and, more generally, the age of the galaxy (i.e. its formation time). Further work and higher resolution simulations will be needed to better understand how the environmental dependence of the underlying MBH occupation fraction may be inferred from observations. More broadly, these results demonstrate how simulations with more predictive models for MBH physics are crucial to our understanding of MBH formation and evolution. %Simulations like {\sc Romulus} with predictive models for MBH seeding will be needed to inform and interpret observations of low mass MBHs in dwarf galaxies with current and future facilities.

\section*{Acknowledgements}

This work used the {\sc pynbody} \citep{pynbody} and {\sc Tangos} \citep{tangos} software packages. MT was supported by an NSF Astronomy and Astrophysics Postdoctoral Fellowship under award AST-2001810. PN gratefully acknowledges support from the Black Hole Initiative at Harvard University, which is funded by grants from the John Templeton Foundation and the Gordon and Betty Moore Foundation. The authors thank Frank van den Bosch, Marta Volonteri, Mark Kennedy, Paul Callanan, and Katja Fahrion for useful discussions and comments.

%%%%%%%%%%%%%%%%%%%%%%%%%%%%%%%%%%%%%%%%%%%%%%%%%%
\section*{Data Availability}

The {\sc Tangos} database for both {\sc Romulus} simulations are available upon email request. Recreaction of all of the figures in this paper with processed data (derived using {\sc Tangos} databases) can be done using the jupyter notebook, code, and data files that are available at \url{https://github.com/mtremmel/tremmel2023_mbh_occFrac}
 
%The inclusion of a Data Availability Statement is a requirement for articles published in MNRAS. Data Availability Statements provide a standardised format for readers to understand the availability of data underlying the research results described in the article. The statement may refer to original data generated in the course of the study or to third-party data analysed in the article. The statement should describe and provide means of access, where possible, by linking to the data or providing the required accession numbers for the relevant databases or DOIs.



\bibliographystyle{mnras}
\bibliography{bibref_mjt.bib} % if your bibtex file is called example.bib

\begin{thebibliography}{}
\makeatletter
\relax
\def\mn@urlcharsother{\let\do\@makeother \do\$\do\&\do\#\do\^\do\_\do\%\do\~}
\def\mn@doi{\begingroup\mn@urlcharsother \@ifnextchar [ {\mn@doi@}
  {\mn@doi@[]}}
\def\mn@doi@[#1]#2{\def\@tempa{#1}\ifx\@tempa\@empty \href
  {http://dx.doi.org/#2} {doi:#2}\else \href {http://dx.doi.org/#2} {#1}\fi
  \endgroup}
\def\mn@eprint#1#2{\mn@eprint@#1:#2::\@nil}
\def\mn@eprint@arXiv#1{\href {http://arxiv.org/abs/#1} {{\tt arXiv:#1}}}
\def\mn@eprint@dblp#1{\href {http://dblp.uni-trier.de/rec/bibtex/#1.xml}
  {dblp:#1}}
\def\mn@eprint@#1:#2:#3:#4\@nil{\def\@tempa {#1}\def\@tempb {#2}\def\@tempc
  {#3}\ifx \@tempc \@empty \let \@tempc \@tempb \let \@tempb \@tempa \fi \ifx
  \@tempb \@empty \def\@tempb {arXiv}\fi \@ifundefined
  {mn@eprint@\@tempb}{\@tempb:\@tempc}{\expandafter \expandafter \csname
  mn@eprint@\@tempb\endcsname \expandafter{\@tempc}}}

\bibitem[\protect\citeauthoryear{{Afanasiev} et~al.,}{{Afanasiev}
  et~al.}{2018}]{afanasiev18}
{Afanasiev} A.~V.,  et~al., 2018, \mn@doi [\mnras] {10.1093/mnras/sty913},
  \href {https://ui.adsabs.harvard.edu/abs/2018MNRAS.477.4856A} {477, 4856}

\bibitem[\protect\citeauthoryear{{Alexander} \& {Natarajan}}{{Alexander} \&
  {Natarajan}}{2014}]{talPN2014}
{Alexander} T.,  {Natarajan} P.,  2014, \mn@doi [Science]
  {10.1126/science.1251053}, \href
  {https://ui.adsabs.harvard.edu/abs/2014Sci...345.1330A} {345, 1330}

\bibitem[\protect\citeauthoryear{{Amaro-Seoane} et~al.,}{{Amaro-Seoane}
  et~al.}{2023}]{lisa_astro}
{Amaro-Seoane} P.,  et~al., 2023, \mn@doi [Living Reviews in Relativity]
  {10.1007/s41114-022-00041-y}, \href
  {https://ui.adsabs.harvard.edu/abs/2023LRR....26....2A} {26, 2}

\bibitem[\protect\citeauthoryear{{Antonini}}{{Antonini}}{2013}]{antonini13}
{Antonini} F.,  2013, \mn@doi [\apj] {10.1088/0004-637X/763/1/62}, \href
  {https://ui.adsabs.harvard.edu/abs/2013ApJ...763...62A} {763, 62}

\bibitem[\protect\citeauthoryear{{Antonini} \& {Merritt}}{{Antonini} \&
  {Merritt}}{2012}]{antonini12}
{Antonini} F.,  {Merritt} D.,  2012, \mn@doi [\apj]
  {10.1088/0004-637X/745/1/83}, \href
  {http://adsabs.harvard.edu/abs/2012ApJ...745...83A} {745, 83}

\bibitem[\protect\citeauthoryear{{Antonini}, {Barausse}  \& {Silk}}{{Antonini}
  et~al.}{2015}]{antonini15}
{Antonini} F.,  {Barausse} E.,   {Silk} J.,  2015, \mn@doi [\apj]
  {10.1088/0004-637X/812/1/72}, \href
  {http://adsabs.harvard.edu/abs/2015ApJ...812...72A} {812, 72}

\bibitem[\protect\citeauthoryear{{Applebaum}, {Brooks}, {Christensen},
  {Munshi}, {Quinn}, {Shen}  \& {Tremmel}}{{Applebaum}
  et~al.}{2021}]{applebaum21}
{Applebaum} E.,  {Brooks} A.~M.,  {Christensen} C.~R.,  {Munshi} F.,  {Quinn}
  T.~R.,  {Shen} S.,   {Tremmel} M.,  2021, \mn@doi [\apj]
  {10.3847/1538-4357/abcafa}, \href
  {https://ui.adsabs.harvard.edu/abs/2021ApJ...906...96A} {906, 96}

\bibitem[\protect\citeauthoryear{{Askar}, {Davies}  \& {Church}}{{Askar}
  et~al.}{2022}]{askar22}
{Askar} A.,  {Davies} M.~B.,   {Church} R.~P.,  2022, \mn@doi [\mnras]
  {10.1093/mnras/stab3741}, \href
  {https://ui.adsabs.harvard.edu/abs/2022MNRAS.511.2631A} {511, 2631}

\bibitem[\protect\citeauthoryear{{Baldassare}, {Reines}, {Gallo}  \&
  {Greene}}{{Baldassare} et~al.}{2015}]{baldassare15}
{Baldassare} V.~F.,  {Reines} A.~E.,  {Gallo} E.,   {Greene} J.~E.,  2015,
  \mn@doi [\apjl] {10.1088/2041-8205/809/1/L14}, \href
  {https://ui.adsabs.harvard.edu/abs/2015ApJ...809L..14B} {809, L14}

\bibitem[\protect\citeauthoryear{{Baldassare} et~al.,}{{Baldassare}
  et~al.}{2016}]{baldassare16}
{Baldassare} V.~F.,  et~al., 2016, \mn@doi [\apj] {10.3847/0004-637X/829/1/57},
  \href {http://adsabs.harvard.edu/abs/2016ApJ...829...57B} {829, 57}

\bibitem[\protect\citeauthoryear{{Baldassare}, {Geha}  \&
  {Greene}}{{Baldassare} et~al.}{2018}]{baldassare18}
{Baldassare} V.~F.,  {Geha} M.,   {Greene} J.,  2018, preprint, \href
  {https://ui.adsabs.harvard.edu/#abs/2018arXiv180809578B} {p.
  arXiv:1808.09578} (\mn@eprint {arXiv} {1808.09578})

\bibitem[\protect\citeauthoryear{{Baldassare}, {Geha}  \&
  {Greene}}{{Baldassare} et~al.}{2020}]{baldassare20}
{Baldassare} V.~F.,  {Geha} M.,   {Greene} J.,  2020, \mn@doi [\apj]
  {10.3847/1538-4357/ab8936}, \href
  {https://ui.adsabs.harvard.edu/abs/2020ApJ...896...10B} {896, 10}

\bibitem[\protect\citeauthoryear{{Beckmann} et~al.,}{{Beckmann}
  et~al.}{2023}]{beckmann23}
{Beckmann} R.~S.,  et~al., 2023, \mn@doi [\mnras] {10.1093/mnras/stad1544},
  \href {https://ui.adsabs.harvard.edu/abs/2023MNRAS.tmp.1558B} {}

\bibitem[\protect\citeauthoryear{{Bekki}, {Couch}, {Drinkwater}  \&
  {Shioya}}{{Bekki} et~al.}{2003}]{bekki03b}
{Bekki} K.,  {Couch} W.~J.,  {Drinkwater} M.~J.,   {Shioya} Y.,  2003, \mn@doi
  [\mnras] {10.1046/j.1365-8711.2003.06916.x}, \href
  {https://ui.adsabs.harvard.edu/abs/2003MNRAS.344..399B} {344, 399}

\bibitem[\protect\citeauthoryear{{Bellovary} et~al.,}{{Bellovary}
  et~al.}{2019}]{bellovary19}
{Bellovary} J.,  et~al., 2019, \baas, \href
  {https://ui.adsabs.harvard.edu/abs/2019BAAS...51c.175B} {51, 175}

\bibitem[\protect\citeauthoryear{{Bellovary} et~al.,}{{Bellovary}
  et~al.}{2021}]{bellovary21}
{Bellovary} J.~M.,  et~al., 2021, \mn@doi [\mnras] {10.1093/mnras/stab1665},
  \href {https://ui.adsabs.harvard.edu/abs/2021MNRAS.505.5129B} {505, 5129}

\bibitem[\protect\citeauthoryear{{Birchall}, {Watson}  \& {Aird}}{{Birchall}
  et~al.}{2020}]{birchall20}
{Birchall} K.~L.,  {Watson} M.~G.,   {Aird} J.,  2020, \mn@doi [\mnras]
  {10.1093/mnras/staa040}, \href
  {https://ui.adsabs.harvard.edu/abs/2020MNRAS.492.2268B} {492, 2268}

\bibitem[\protect\citeauthoryear{{Blank}, {Macci{\`o}}, {Dutton}  \&
  {Obreja}}{{Blank} et~al.}{2019}]{blank19}
{Blank} M.,  {Macci{\`o}} A.~V.,  {Dutton} A.~A.,   {Obreja} A.,  2019, \mn@doi
  [\mnras] {10.1093/mnras/stz1688}, \href
  {https://ui.adsabs.harvard.edu/abs/2019MNRAS.487.5476B} {487, 5476}

\bibitem[\protect\citeauthoryear{{Bondi} \& {Hoyle}}{{Bondi} \&
  {Hoyle}}{1944}]{bondihoyle44}
{Bondi} H.,  {Hoyle} F.,  1944, \mn@doi [\mnras] {10.1093/mnras/104.5.273},
  \href {http://adsabs.harvard.edu/abs/1944MNRAS.104..273B} {104, 273}

\bibitem[\protect\citeauthoryear{{Booth} \& {Schaye}}{{Booth} \&
  {Schaye}}{2009}]{BoothBH2009}
{Booth} C.~M.,  {Schaye} J.,  2009, \mn@doi [\mnras]
  {10.1111/j.1365-2966.2009.15043.x}, \href
  {http://adsabs.harvard.edu/abs/2009MNRAS.398...53B} {398, 53}

\bibitem[\protect\citeauthoryear{{Borrow}, {Vogelsberger}, {O'Neil}, {McDonald}
   \& {Smith}}{{Borrow} et~al.}{2022}]{borrow22}
{Borrow} J.,  {Vogelsberger} M.,  {O'Neil} S.,  {McDonald} M.~A.,   {Smith} A.,
   2022, arXiv e-prints, \href
  {https://ui.adsabs.harvard.edu/abs/2022arXiv220510376B} {p. arXiv:2205.10376}

\bibitem[\protect\citeauthoryear{{Bortolas}, {Bonetti}, {Dotti}, {Lupi},
  {Capelo}, {Mayer}  \& {Sesana}}{{Bortolas} et~al.}{2022}]{bortolas22}
{Bortolas} E.,  {Bonetti} M.,  {Dotti} M.,  {Lupi} A.,  {Capelo} P.~R.,
  {Mayer} L.,   {Sesana} A.,  2022, \mn@doi [\mnras] {10.1093/mnras/stac645},
  \href {https://ui.adsabs.harvard.edu/abs/2022MNRAS.512.3365B} {512, 3365}

\bibitem[\protect\citeauthoryear{{Boylan-Kolchin}, {Springel}, {White},
  {Jenkins}  \& {Lemson}}{{Boylan-Kolchin} et~al.}{2009}]{bk09}
{Boylan-Kolchin} M.,  {Springel} V.,  {White} S. D.~M.,  {Jenkins} A.,
  {Lemson} G.,  2009, \mn@doi [\mnras] {10.1111/j.1365-2966.2009.15191.x},
  \href {https://ui.adsabs.harvard.edu/abs/2009MNRAS.398.1150B} {398, 1150}

\bibitem[\protect\citeauthoryear{{Bricman} \& {Gomboc}}{{Bricman} \&
  {Gomboc}}{2020}]{bricmanGomboc2020}
{Bricman} K.,  {Gomboc} A.,  2020, \mn@doi [\apj] {10.3847/1538-4357/ab6989},
  \href {https://ui.adsabs.harvard.edu/abs/2020ApJ...890...73B} {890, 73}

\bibitem[\protect\citeauthoryear{{Burke} et~al.,}{{Burke}
  et~al.}{2022}]{burke22}
{Burke} C.~J.,  et~al., 2022, \mn@doi [\mnras] {10.1093/mnras/stac2262}, \href
  {https://ui.adsabs.harvard.edu/abs/2022MNRAS.516.2736B} {516, 2736}

\bibitem[\protect\citeauthoryear{{Cameron}}{{Cameron}}{2011}]{cameron11}
{Cameron} E.,  2011, \mn@doi [Publications of the Astronomical Society of
  Australia] {10.1071/AS10046}, \href
  {https://ui.adsabs.harvard.edu/#abs/2011PASA...28..128C} {28, 128}

\bibitem[\protect\citeauthoryear{{Cann} et~al.,}{{Cann} et~al.}{2021}]{cann21}
{Cann} J.~M.,  et~al., 2021, \mn@doi [\apjl] {10.3847/2041-8213/abf56d}, \href
  {https://ui.adsabs.harvard.edu/abs/2021ApJ...912L...2C} {912, L2}

\bibitem[\protect\citeauthoryear{{Carleton}, {Guo}, {Munshi}, {Tremmel}  \&
  {Wright}}{{Carleton} et~al.}{2021}]{carleton21}
{Carleton} T.,  {Guo} Y.,  {Munshi} F.,  {Tremmel} M.,   {Wright} A.,  2021,
  \mn@doi [\mnras] {10.1093/mnras/stab031}, \href
  {https://ui.adsabs.harvard.edu/abs/2021MNRAS.502..398C} {502, 398}

\bibitem[\protect\citeauthoryear{{Carlsten}, {Greene}, {Beaton}  \&
  {Greco}}{{Carlsten} et~al.}{2022}]{carlsten22}
{Carlsten} S.~G.,  {Greene} J.~E.,  {Beaton} R.~L.,   {Greco} J.~P.,  2022,
  \mn@doi [\apj] {10.3847/1538-4357/ac457e}, \href
  {https://ui.adsabs.harvard.edu/abs/2022ApJ...927...44C} {927, 44}

\bibitem[\protect\citeauthoryear{{Carson}, {Barth}, {Seth}, {den Brok},
  {Cappellari}, {Greene}, {Ho}  \& {Neumayer}}{{Carson}
  et~al.}{2015}]{carson15}
{Carson} D.~J.,  {Barth} A.~J.,  {Seth} A.~C.,  {den Brok} M.,  {Cappellari}
  M.,  {Greene} J.~E.,  {Ho} L.~C.,   {Neumayer} N.,  2015, \mn@doi [\aj]
  {10.1088/0004-6256/149/5/170}, \href
  {https://ui.adsabs.harvard.edu/abs/2015AJ....149..170C} {149, 170}

\bibitem[\protect\citeauthoryear{{Chadayammuri}, {Tremmel}, {Nagai}, {Babul}
  \& {Quinn}}{{Chadayammuri} et~al.}{2021}]{chadayammuri21}
{Chadayammuri} U.,  {Tremmel} M.,  {Nagai} D.,  {Babul} A.,   {Quinn} T.,
  2021, \mn@doi [\mnras] {10.1093/mnras/stab1010}, \href
  {https://ui.adsabs.harvard.edu/abs/2021MNRAS.504.3922C} {504, 3922}

\bibitem[\protect\citeauthoryear{{Chandrasekhar}}{{Chandrasekhar}}{1943}]{chandrasekhar43}
{Chandrasekhar} S.,  1943, \mn@doi [\apj] {10.1086/144517}, \href
  {http://adsabs.harvard.edu/abs/1943ApJ....97..255C} {97, 255}

\bibitem[\protect\citeauthoryear{{Chaves-Montero}, {Angulo}, {Schaye},
  {Schaller}, {Crain}, {Furlong}  \& {Theuns}}{{Chaves-Montero}
  et~al.}{2016}]{chavesMontero16}
{Chaves-Montero} J.,  {Angulo} R.~E.,  {Schaye} J.,  {Schaller} M.,  {Crain}
  R.~A.,  {Furlong} M.,   {Theuns} T.,  2016, \mn@doi [\mnras]
  {10.1093/mnras/stw1225}, \href
  {https://ui.adsabs.harvard.edu/abs/2016MNRAS.460.3100C} {460, 3100}

\bibitem[\protect\citeauthoryear{{Chen}, {Ni}, {Tremmel}, {Di Matteo}, {Bird},
  {DeGraf}  \& {Feng}}{{Chen} et~al.}{2022}]{chen22}
{Chen} N.,  {Ni} Y.,  {Tremmel} M.,  {Di Matteo} T.,  {Bird} S.,  {DeGraf} C.,
   {Feng} Y.,  2022, \mn@doi [\mnras] {10.1093/mnras/stab3411}, \href
  {https://ui.adsabs.harvard.edu/abs/2022MNRAS.510..531C} {510, 531}

\bibitem[\protect\citeauthoryear{{Christensen}, {Governato}, {Quinn}, {Brooks},
  {Shen}, {McCleary}, {Fisher}  \& {Wadsley}}{{Christensen}
  et~al.}{2014}]{christensen14b}
{Christensen} C.~R.,  {Governato} F.,  {Quinn} T.,  {Brooks} A.~M.,  {Shen} S.,
   {McCleary} J.,  {Fisher} D.~B.,   {Wadsley} J.,  2014, \mn@doi [\mnras]
  {10.1093/mnras/stu399}, \href
  {http://adsabs.harvard.edu/abs/2014MNRAS.440.2843C} {440, 2843}

\bibitem[\protect\citeauthoryear{{Churazov}, {Sazonov}, {Sunyaev}, {Forman},
  {Jones}  \& {B{\"o}hringer}}{{Churazov} et~al.}{2005}]{churazov05}
{Churazov} E.,  {Sazonov} S.,  {Sunyaev} R.,  {Forman} W.,  {Jones} C.,
  {B{\"o}hringer} H.,  2005, \mn@doi [\mnras]
  {10.1111/j.1745-3933.2005.00093.x}, \href
  {https://ui.adsabs.harvard.edu/abs/2005MNRAS.363L..91C} {363, L91}

\bibitem[\protect\citeauthoryear{{Colpi} et~al.,}{{Colpi}
  et~al.}{2019}]{colpi19wp}
{Colpi} M.,  et~al., 2019, arXiv e-prints, \href
  {https://ui.adsabs.harvard.edu/abs/2019arXiv190306867C} {p. arXiv:1903.06867}

\bibitem[\protect\citeauthoryear{{Croton}, {Gao}  \& {White}}{{Croton}
  et~al.}{2007}]{croton07}
{Croton} D.~J.,  {Gao} L.,   {White} S. D.~M.,  2007, \mn@doi [\mnras]
  {10.1111/j.1365-2966.2006.11230.x}, \href
  {https://ui.adsabs.harvard.edu/abs/2007MNRAS.374.1303C} {374, 1303}

\bibitem[\protect\citeauthoryear{{Davies}, {Miller}  \& {Bellovary}}{{Davies}
  et~al.}{2011}]{davies11}
{Davies} M.~B.,  {Miller} M.~C.,   {Bellovary} J.~M.,  2011, \mn@doi [\apjl]
  {10.1088/2041-8205/740/2/L42}, \href
  {https://ui.adsabs.harvard.edu/abs/2011ApJ...740L..42D} {740, L42}

\bibitem[\protect\citeauthoryear{{Devecchi} \& {Volonteri}}{{Devecchi} \&
  {Volonteri}}{2009}]{devecchi2009}
{Devecchi} B.,  {Volonteri} M.,  2009, \mn@doi [\apj]
  {10.1088/0004-637X/694/1/302}, \href
  {http://adsabs.harvard.edu/abs/2009ApJ...694..302D} {694, 302}

\bibitem[\protect\citeauthoryear{{Di Matteo}, {Colberg}, {Springel},
  {Hernquist}  \& {Sijacki}}{{Di Matteo} et~al.}{2008}]{diMatteo2008}
{Di Matteo} T.,  {Colberg} J.,  {Springel} V.,  {Hernquist} L.,   {Sijacki} D.,
   2008, \mn@doi [\apj] {10.1086/524921}, \href
  {http://adsabs.harvard.edu/abs/2008ApJ...676...33D} {676, 33}

\bibitem[\protect\citeauthoryear{{Di Matteo}, {Croft}, {Feng}, {Waters}  \&
  {Wilkins}}{{Di Matteo} et~al.}{2017}]{dimatteo17}
{Di Matteo} T.,  {Croft} R. A.~C.,  {Feng} Y.,  {Waters} D.,   {Wilkins} S.,
  2017, \mn@doi [\mnras] {10.1093/mnras/stx319}, \href
  {https://ui.adsabs.harvard.edu/abs/2017MNRAS.467.4243D} {467, 4243}

\bibitem[\protect\citeauthoryear{{Drinkwater}, {Gregg}, {Hilker}, {Bekki},
  {Couch}, {Ferguson}, {Jones}  \& {Phillipps}}{{Drinkwater}
  et~al.}{2003}]{drinkwater03}
{Drinkwater} M.~J.,  {Gregg} M.~D.,  {Hilker} M.,  {Bekki} K.,  {Couch} W.~J.,
  {Ferguson} H.~C.,  {Jones} J.~B.,   {Phillipps} S.,  2003, \mn@doi [\nat]
  {10.1038/nature01666}, \href
  {https://ui.adsabs.harvard.edu/abs/2003Natur.423..519D} {423, 519}

\bibitem[\protect\citeauthoryear{{Dubois}, {Peirani}, {Pichon}, {Devriendt},
  {Gavazzi}, {Welker}  \& {Volonteri}}{{Dubois} et~al.}{2016}]{dubois16}
{Dubois} Y.,  {Peirani} S.,  {Pichon} C.,  {Devriendt} J.,  {Gavazzi} R.,
  {Welker} C.,   {Volonteri} M.,  2016, \mn@doi [\mnras]
  {10.1093/mnras/stw2265}, \href
  {http://adsabs.harvard.edu/abs/2016MNRAS.tmp.1372D} {}

\bibitem[\protect\citeauthoryear{{Dubois} et~al.,}{{Dubois}
  et~al.}{2021}]{dubois21}
{Dubois} Y.,  et~al., 2021, \mn@doi [\aap] {10.1051/0004-6361/202039429}, \href
  {https://ui.adsabs.harvard.edu/abs/2021A&A...651A.109D} {651, A109}

\bibitem[\protect\citeauthoryear{{Dunn}, {Bellovary}, {Holley-Bockelmann},
  {Christensen}  \& {Quinn}}{{Dunn} et~al.}{2018}]{dunn18}
{Dunn} G.,  {Bellovary} J.,  {Holley-Bockelmann} K.,  {Christensen} C.,
  {Quinn} T.,  2018, \mn@doi [\apj] {10.3847/1538-4357/aac7c2}, \href
  {https://ui.adsabs.harvard.edu/#abs/2018ApJ...861...39D} {861, 39}

\bibitem[\protect\citeauthoryear{{Fahrion} et~al.,}{{Fahrion}
  et~al.}{2020}]{fahrion20}
{Fahrion} K.,  et~al., 2020, \mn@doi [\aap] {10.1051/0004-6361/201937120},
  \href {https://ui.adsabs.harvard.edu/abs/2020A&A...634A..53F} {634, A53}

\bibitem[\protect\citeauthoryear{{Fahrion} et~al.,}{{Fahrion}
  et~al.}{2021}]{fahrion21}
{Fahrion} K.,  et~al., 2021, \mn@doi [\aap] {10.1051/0004-6361/202140644},
  \href {https://ui.adsabs.harvard.edu/abs/2021A&A...650A.137F} {650, A137}

\bibitem[\protect\citeauthoryear{{Fahrion}, {Leaman}, {Lyubenova}  \& {van de
  Ven}}{{Fahrion} et~al.}{2022a}]{fahrion22}
{Fahrion} K.,  {Leaman} R.,  {Lyubenova} M.,   {van de Ven} G.,  2022a, \mn@doi
  [\aap] {10.1051/0004-6361/202039778}, \href
  {https://ui.adsabs.harvard.edu/abs/2022A&A...658A.172F} {658, A172}

\bibitem[\protect\citeauthoryear{{Fahrion} et~al.,}{{Fahrion}
  et~al.}{2022b}]{fahrion22b}
{Fahrion} K.,  et~al., 2022b, \mn@doi [\aap] {10.1051/0004-6361/202244932},
  \href {https://ui.adsabs.harvard.edu/abs/2022A&A...667A.101F} {667, A101}

\bibitem[\protect\citeauthoryear{{Ferrarese} et~al.,}{{Ferrarese}
  et~al.}{2006}]{ferrarese06b}
{Ferrarese} L.,  et~al., 2006, \mn@doi [\apjl] {10.1086/505388}, \href
  {http://adsabs.harvard.edu/abs/2006ApJ...644L..21F} {644, L21}

\bibitem[\protect\citeauthoryear{{Gallo}, {Treu}, {Marshall}, {Woo}, {Leipski}
  \& {Antonucci}}{{Gallo} et~al.}{2010}]{gallo10}
{Gallo} E.,  {Treu} T.,  {Marshall} P.~J.,  {Woo} J.-H.,  {Leipski} C.,
  {Antonucci} R.,  2010, \mn@doi [\apj] {10.1088/0004-637X/714/1/25}, \href
  {https://ui.adsabs.harvard.edu/abs/2010ApJ...714...25G} {714, 25}

\bibitem[\protect\citeauthoryear{{Gao} \& {White}}{{Gao} \&
  {White}}{2007}]{gao07}
{Gao} L.,  {White} S. D.~M.,  2007, \mn@doi [\mnras]
  {10.1111/j.1745-3933.2007.00292.x}, \href
  {https://ui.adsabs.harvard.edu/abs/2007MNRAS.377L...5G} {377, L5}

\bibitem[\protect\citeauthoryear{{Gao}, {Springel}  \& {White}}{{Gao}
  et~al.}{2005}]{gao05}
{Gao} L.,  {Springel} V.,   {White} S. D.~M.,  2005, \mn@doi [\mnras]
  {10.1111/j.1745-3933.2005.00084.x}, \href
  {https://ui.adsabs.harvard.edu/abs/2005MNRAS.363L..66G} {363, L66}

\bibitem[\protect\citeauthoryear{{Geha}, {Blanton}, {Yan}  \& {Tinker}}{{Geha}
  et~al.}{2012}]{geha12}
{Geha} M.,  {Blanton} M.~R.,  {Yan} R.,   {Tinker} J.~L.,  2012, \mn@doi [\apj]
  {10.1088/0004-637X/757/1/85}, \href
  {http://adsabs.harvard.edu/abs/2012ApJ...757...85G} {757, 85}

\bibitem[\protect\citeauthoryear{{Georgiev}, {B{\"o}ker}, {Leigh},
  {L{\"u}tzgendorf}  \& {Neumayer}}{{Georgiev} et~al.}{2016}]{georgiev16}
{Georgiev} I.~Y.,  {B{\"o}ker} T.,  {Leigh} N.,  {L{\"u}tzgendorf} N.,
  {Neumayer} N.,  2016, \mn@doi [\mnras] {10.1093/mnras/stw093}, \href
  {https://ui.adsabs.harvard.edu/abs/2016MNRAS.457.2122G} {457, 2122}

\bibitem[\protect\citeauthoryear{{Governato} et~al.,}{{Governato}
  et~al.}{2015}]{governato15}
{Governato} F.,  et~al., 2015, \mn@doi [\mnras] {10.1093/mnras/stu2720}, \href
  {http://adsabs.harvard.edu/abs/2015MNRAS.448..792G} {448, 792}

\bibitem[\protect\citeauthoryear{{Greene}}{{Greene}}{2012}]{greene12}
{Greene} J.~E.,  2012, \mn@doi [Nature Communications] {10.1038/ncomms2314},
  \href {https://ui.adsabs.harvard.edu/abs/2012NatCo...3.1304G} {3, 1304}

\bibitem[\protect\citeauthoryear{{Guedes}, {Callegari}, {Madau}  \&
  {Mayer}}{{Guedes} et~al.}{2011}]{eris11}
{Guedes} J.,  {Callegari} S.,  {Madau} P.,   {Mayer} L.,  2011, \mn@doi [\apj]
  {10.1088/0004-637X/742/2/76}, \href
  {http://adsabs.harvard.edu/abs/2011ApJ...742...76G} {742, 76}

\bibitem[\protect\citeauthoryear{{Haardt} \& {Madau}}{{Haardt} \&
  {Madau}}{2012}]{HM12}
{Haardt} F.,  {Madau} P.,  2012, \mn@doi [\apj] {10.1088/0004-637X/746/2/125},
  \href {http://adsabs.harvard.edu/abs/2012ApJ...746..125H} {746, 125}

\bibitem[\protect\citeauthoryear{{Habouzit} et~al.,}{{Habouzit}
  et~al.}{2021}]{habouzit21}
{Habouzit} M.,  et~al., 2021, \mn@doi [\mnras] {10.1093/mnras/stab496}, \href
  {https://ui.adsabs.harvard.edu/abs/2021MNRAS.tmp..518H} {}

\bibitem[\protect\citeauthoryear{{Habouzit} et~al.,}{{Habouzit}
  et~al.}{2022}]{habouzit22}
{Habouzit} M.,  et~al., 2022, \mn@doi [\mnras] {10.1093/mnras/stab3147}, \href
  {https://ui.adsabs.harvard.edu/abs/2022MNRAS.509.3015H} {509, 3015}

\bibitem[\protect\citeauthoryear{{Haidar} et~al.,}{{Haidar}
  et~al.}{2022}]{haidaar22}
{Haidar} H.,  et~al., 2022, \mn@doi [\mnras] {10.1093/mnras/stac1659}, \href
  {https://ui.adsabs.harvard.edu/abs/2022MNRAS.514.4912H} {514, 4912}

\bibitem[\protect\citeauthoryear{{Henden}, {Puchwein}, {Shen}  \&
  {Sijacki}}{{Henden} et~al.}{2018}]{henden18}
{Henden} N.~A.,  {Puchwein} E.,  {Shen} S.,   {Sijacki} D.,  2018, \mn@doi
  [\mnras] {10.1093/mnras/sty1780}, \href
  {https://ui.adsabs.harvard.edu/abs/2018MNRAS.479.5385H} {479, 5385}

\bibitem[\protect\citeauthoryear{{Hirschmann}, {Dolag}, {Saro}, {Bachmann},
  {Borgani}  \& {Burkert}}{{Hirschmann} et~al.}{2014}]{Hirschmann14}
{Hirschmann} M.,  {Dolag} K.,  {Saro} A.,  {Bachmann} L.,  {Borgani} S.,
  {Burkert} A.,  2014, \mn@doi [\mnras] {10.1093/mnras/stu1023}, \href
  {http://adsabs.harvard.edu/abs/2014MNRAS.442.2304H} {442, 2304}

\bibitem[\protect\citeauthoryear{{Hoyer}, {Neumayer}, {Georgiev}, {Seth}  \&
  {Greene}}{{Hoyer} et~al.}{2021}]{hoyer2021}
{Hoyer} N.,  {Neumayer} N.,  {Georgiev} I.~Y.,  {Seth} A.~C.,   {Greene} J.~E.,
   2021, \mn@doi [\mnras] {10.1093/mnras/stab2277}, \href
  {https://ui.adsabs.harvard.edu/abs/2021MNRAS.507.3246H} {507, 3246}

\bibitem[\protect\citeauthoryear{{Inayoshi}, {Haiman}  \&
  {Ostriker}}{{Inayoshi} et~al.}{2016}]{inayoshi16}
{Inayoshi} K.,  {Haiman} Z.,   {Ostriker} J.~P.,  2016, \mn@doi [\mnras]
  {10.1093/mnras/stw836}, \href
  {https://ui.adsabs.harvard.edu/abs/2016MNRAS.459.3738I} {459, 3738}

\bibitem[\protect\citeauthoryear{{Jord{\'a}n} et~al.,}{{Jord{\'a}n}
  et~al.}{2007}]{jordan07}
{Jord{\'a}n} A.,  et~al., 2007, \mn@doi [\apjs] {10.1086/516840}, \href
  {https://ui.adsabs.harvard.edu/abs/2007ApJS..171..101J} {171, 101}

\bibitem[\protect\citeauthoryear{{Joshi}, {Parker}  \& {Wadsley}}{{Joshi}
  et~al.}{2016}]{joshi16}
{Joshi} G.~D.,  {Parker} L.~C.,   {Wadsley} J.,  2016, \mn@doi [\mnras]
  {10.1093/mnras/stw1699}, \href
  {https://ui.adsabs.harvard.edu/abs/2016MNRAS.462..761J} {462, 761}

\bibitem[\protect\citeauthoryear{{Jung} et~al.,}{{Jung} et~al.}{2022}]{jung22}
{Jung} S.~L.,  et~al., 2022, \mn@doi [\mnras] {10.1093/mnras/stac1622}, \href
  {https://ui.adsabs.harvard.edu/abs/2022MNRAS.515...22J} {515, 22}

\bibitem[\protect\citeauthoryear{{Kacharov}, {Neumayer}, {Seth}, {Cappellari},
  {McDermid}, {Walcher}  \& {B{\"o}ker}}{{Kacharov} et~al.}{2018}]{kacharov18}
{Kacharov} N.,  {Neumayer} N.,  {Seth} A.~C.,  {Cappellari} M.,  {McDermid} R.,
   {Walcher} C.~J.,   {B{\"o}ker} T.,  2018, \mn@doi [\mnras]
  {10.1093/mnras/sty1985}, \href
  {https://ui.adsabs.harvard.edu/abs/2018MNRAS.480.1973K} {480, 1973}

\bibitem[\protect\citeauthoryear{{Knebe} et~al.,}{{Knebe}
  et~al.}{2011}]{knebe11}
{Knebe} A.,  et~al., 2011, \mn@doi [\mnras] {10.1111/j.1365-2966.2011.18858.x},
  \href {https://ui.adsabs.harvard.edu/abs/2011MNRAS.415.2293K} {415, 2293}

\bibitem[\protect\citeauthoryear{{Knollmann} \& {Knebe}}{{Knollmann} \&
  {Knebe}}{2009}]{knollmann09}
{Knollmann} S.~R.,  {Knebe} A.,  2009, \mn@doi [\apjs]
  {10.1088/0067-0049/182/2/608}, \href
  {http://adsabs.harvard.edu/abs/2009ApJS..182..608K} {182, 608}

\bibitem[\protect\citeauthoryear{{Kormendy} \& {Ho}}{{Kormendy} \&
  {Ho}}{2013}]{kormendy2013}
{Kormendy} J.,  {Ho} L.~C.,  2013, \mn@doi [\araa]
  {10.1146/annurev-astro-082708-101811}, \href
  {http://adsabs.harvard.edu/abs/2013ARA%26A..51..511K} {51, 511}

\bibitem[\protect\citeauthoryear{{Kormendy} \& {Richstone}}{{Kormendy} \&
  {Richstone}}{1995}]{kormendy95}
{Kormendy} J.,  {Richstone} D.,  1995, \mn@doi [ARA\&A]
  {10.1146/annurev.aa.33.090195.003053}, \href
  {http://adsabs.harvard.edu/cgi-bin/nph-bib_query?bibcode=1995ARA%26A..33..581K&db_key=AST}
  {33, 581}

\bibitem[\protect\citeauthoryear{{Koudmani}, {Henden}  \& {Sijacki}}{{Koudmani}
  et~al.}{2021}]{koudmani21}
{Koudmani} S.,  {Henden} N.~A.,   {Sijacki} D.,  2021, \mn@doi [\mnras]
  {10.1093/mnras/stab677}, \href
  {https://ui.adsabs.harvard.edu/abs/2021MNRAS.503.3568K} {503, 3568}

\bibitem[\protect\citeauthoryear{{Koudmani}, {Sijacki}  \& {Smith}}{{Koudmani}
  et~al.}{2022}]{koudmani22}
{Koudmani} S.,  {Sijacki} D.,   {Smith} M.~C.,  2022, \mn@doi [\mnras]
  {10.1093/mnras/stac2252}, \href
  {https://ui.adsabs.harvard.edu/abs/2022MNRAS.516.2112K} {516, 2112}

\bibitem[\protect\citeauthoryear{{Kroupa}, {Subr}, {Jerabkova}  \&
  {Wang}}{{Kroupa} et~al.}{2020}]{kroupa20}
{Kroupa} P.,  {Subr} L.,  {Jerabkova} T.,   {Wang} L.,  2020, \mn@doi [\mnras]
  {10.1093/mnras/staa2276}, \href
  {https://ui.adsabs.harvard.edu/abs/2020MNRAS.498.5652K} {498, 5652}

\bibitem[\protect\citeauthoryear{{Latimer}, {Reines}, {Bogdan}  \&
  {Kraft}}{{Latimer} et~al.}{2021}]{latimer21}
{Latimer} L.~J.,  {Reines} A.~E.,  {Bogdan} A.,   {Kraft} R.,  2021, \mn@doi
  [\apjl] {10.3847/2041-8213/ac3af6}, \href
  {https://ui.adsabs.harvard.edu/abs/2021ApJ...922L..40L} {922, L40}

\bibitem[\protect\citeauthoryear{{Leja} et~al.,}{{Leja} et~al.}{2019}]{leja19}
{Leja} J.,  et~al., 2019, \mn@doi [\apj] {10.3847/1538-4357/ab1d5a}, \href
  {https://ui.adsabs.harvard.edu/abs/2019ApJ...877..140L} {877, 140}

\bibitem[\protect\citeauthoryear{{Lescaudron}, {Dubois}, {Beckmann}  \&
  {Volonteri}}{{Lescaudron} et~al.}{2022}]{lescaudron22}
{Lescaudron} S.,  {Dubois} Y.,  {Beckmann} R.~S.,   {Volonteri} M.,  2022,
  arXiv e-prints, \href {https://ui.adsabs.harvard.edu/abs/2022arXiv220913548L}
  {p. arXiv:2209.13548}

\bibitem[\protect\citeauthoryear{{Lodato} \& {Natarajan}}{{Lodato} \&
  {Natarajan}}{2006}]{LN2006BH}
{Lodato} G.,  {Natarajan} P.,  2006, \mn@doi [{MNRAS}]
  {10.1111/j.1365-2966.2006.10801.x}, \href
  {http://adsabs.harvard.edu/cgi-bin/nph-bib_query?bibcode=2006MNRAS.371.1813L&db_key=AST}
  {371, 1813}

\bibitem[\protect\citeauthoryear{{Lodato} \& {Natarajan}}{{Lodato} \&
  {Natarajan}}{2007}]{LN2007BH}
{Lodato} G.,  {Natarajan} P.,  2007, \mn@doi [\mnras]
  {10.1111/j.1745-3933.2007.00304.x}, \href
  {https://ui.adsabs.harvard.edu/abs/2007MNRAS.377L..64L} {377, L64}

\bibitem[\protect\citeauthoryear{{Mayer}, {Capelo}, {Zwick}  \& {Di
  Matteo}}{{Mayer} et~al.}{2023}]{mayer23}
{Mayer} L.,  {Capelo} P.~R.,  {Zwick} L.,   {Di Matteo} T.,  2023, \mn@doi
  [arXiv e-prints] {10.48550/arXiv.2304.02066}, \href
  {https://ui.adsabs.harvard.edu/abs/2023arXiv230402066M} {p. arXiv:2304.02066}

\bibitem[\protect\citeauthoryear{{Menon}, {Wesolowski}, {Zheng}, {Jetley},
  {Kale}, {Quinn}  \& {Governato}}{{Menon} et~al.}{2015}]{changa15}
{Menon} H.,  {Wesolowski} L.,  {Zheng} G.,  {Jetley} P.,  {Kale} L.,  {Quinn}
  T.,   {Governato} F.,  2015, \mn@doi [Comp. Astrophysics and Cosmology]
  {10.1186/s40668-015-0007-9}, \href
  {http://adsabs.harvard.edu/abs/2015ComAC...2....1M} {2, 1}

\bibitem[\protect\citeauthoryear{{Mezcua}, {Civano}, {Marchesi}, {Suh},
  {Fabbiano}  \& {Volonteri}}{{Mezcua} et~al.}{2018}]{mezcua18b}
{Mezcua} M.,  {Civano} F.,  {Marchesi} S.,  {Suh} H.,  {Fabbiano} G.,
  {Volonteri} M.,  2018, \mn@doi [\mnras] {10.1093/mnras/sty1163}, \href
  {https://ui.adsabs.harvard.edu/abs/2018MNRAS.478.2576M} {478, 2576}

\bibitem[\protect\citeauthoryear{{Miller} \& {Lotz}}{{Miller} \&
  {Lotz}}{2007}]{miller07}
{Miller} B.~W.,  {Lotz} J.~M.,  2007, \mn@doi [\apj] {10.1086/522323}, \href
  {https://ui.adsabs.harvard.edu/abs/2007ApJ...670.1074M} {670, 1074}

\bibitem[\protect\citeauthoryear{{Miller}, {Lotz}, {Ferguson}, {Stiavelli}  \&
  {Whitmore}}{{Miller} et~al.}{1998}]{miller98}
{Miller} B.~W.,  {Lotz} J.~M.,  {Ferguson} H.~C.,  {Stiavelli} M.,   {Whitmore}
  B.~C.,  1998, \mn@doi [\apjl] {10.1086/311739}, \href
  {https://ui.adsabs.harvard.edu/abs/1998ApJ...508L.133M} {508, L133}

\bibitem[\protect\citeauthoryear{{Miller}, {Gallo}, {Greene}, {Kelly}, {Treu},
  {Woo}  \& {Baldassare}}{{Miller} et~al.}{2015}]{miller15}
{Miller} B.~P.,  {Gallo} E.,  {Greene} J.~E.,  {Kelly} B.~C.,  {Treu} T.,
  {Woo} J.-H.,   {Baldassare} V.,  2015, \mn@doi [\apj]
  {10.1088/0004-637X/799/1/98}, \href
  {http://adsabs.harvard.edu/abs/2015ApJ...799...98M} {799, 98}

\bibitem[\protect\citeauthoryear{{Mistani} et~al.,}{{Mistani}
  et~al.}{2016}]{mistani16}
{Mistani} P.~A.,  et~al., 2016, \mn@doi [\mnras] {10.1093/mnras/stv2435}, \href
  {https://ui.adsabs.harvard.edu/#abs/2016MNRAS.455.2323M} {455, 2323}

\bibitem[\protect\citeauthoryear{{Molina}, {Reines}, {Latimer}, {Baldassare}
  \& {Salehirad}}{{Molina} et~al.}{2021}]{molina21}
{Molina} M.,  {Reines} A.~E.,  {Latimer} L.~J.,  {Baldassare} V.,   {Salehirad}
  S.,  2021, \mn@doi [\apj] {10.3847/1538-4357/ac1ffa}, \href
  {https://ui.adsabs.harvard.edu/abs/2021ApJ...922..155M} {922, 155}

\bibitem[\protect\citeauthoryear{{Mu{\~n}oz} et~al.,}{{Mu{\~n}oz}
  et~al.}{2015}]{munos15}
{Mu{\~n}oz} R.~P.,  et~al., 2015, \mn@doi [\apj] {10.1088/2041-8205/813/1/L15},
  \href {https://ui.adsabs.harvard.edu/abs/2015ApJ...813L..15M} {813, L15}

\bibitem[\protect\citeauthoryear{{Munshi} et~al.,}{{Munshi}
  et~al.}{2013}]{munshi13}
{Munshi} F.,  et~al., 2013, \mn@doi [\apj] {10.1088/0004-637X/766/1/56}, \href
  {http://adsabs.harvard.edu/abs/2013ApJ...766...56M} {766, 56}

\bibitem[\protect\citeauthoryear{{Munshi}, {Brooks}, {Applebaum},
  {Christensen}, {Quinn}  \& {Sligh}}{{Munshi} et~al.}{2021}]{munshi21}
{Munshi} F.,  {Brooks} A.~M.,  {Applebaum} E.,  {Christensen} C.~R.,  {Quinn}
  T.,   {Sligh} S.,  2021, \mn@doi [\apj] {10.3847/1538-4357/ac0db6}, \href
  {https://ui.adsabs.harvard.edu/abs/2021ApJ...923...35M} {923, 35}

\bibitem[\protect\citeauthoryear{{Natarajan}}{{Natarajan}}{2011}]{natarajan11}
{Natarajan} P.,  2011, Bulletin of the Astronomical Society of India, \href
  {https://ui.adsabs.harvard.edu/abs/2011BASI...39..145N} {39, 145}

\bibitem[\protect\citeauthoryear{{Natarajan}}{{Natarajan}}{2014}]{natarajan14}
{Natarajan} P.,  2014, \mn@doi [General Relativity and Gravitation]
  {10.1007/s10714-014-1702-6}, \href
  {https://ui.adsabs.harvard.edu/abs/2014GReGr..46.1702N} {46, 1702}

\bibitem[\protect\citeauthoryear{{Natarajan}}{{Natarajan}}{2021}]{natarajan21}
{Natarajan} P.,  2021, \mn@doi [\mnras] {10.1093/mnras/staa3724}, \href
  {https://ui.adsabs.harvard.edu/abs/2021MNRAS.501.1413N} {501, 1413}

\bibitem[\protect\citeauthoryear{{Natarajan}, {Pacucci}, {Ferrara}, {Agarwal},
  {Ricarte}, {Zackrisson}  \& {Cappelluti}}{{Natarajan}
  et~al.}{2017}]{natarajan17}
{Natarajan} P.,  {Pacucci} F.,  {Ferrara} A.,  {Agarwal} B.,  {Ricarte} A.,
  {Zackrisson} E.,   {Cappelluti} N.,  2017, \mn@doi [\apj]
  {10.3847/1538-4357/aa6330}, \href
  {http://adsabs.harvard.edu/abs/2017ApJ...838..117N} {838, 117}

\bibitem[\protect\citeauthoryear{{Nelson} et~al.,}{{Nelson}
  et~al.}{2019}]{nelson19}
{Nelson} D.,  et~al., 2019, \mn@doi [\mnras] {10.1093/mnras/stz2306}, \href
  {https://ui.adsabs.harvard.edu/abs/2019MNRAS.490.3234N} {490, 3234}

\bibitem[\protect\citeauthoryear{{Nguyen} et~al.,}{{Nguyen}
  et~al.}{2018}]{nguyen18}
{Nguyen} D.~D.,  et~al., 2018, \mn@doi [\apj] {10.3847/1538-4357/aabe28}, \href
  {https://ui.adsabs.harvard.edu/abs/2018ApJ...858..118N} {858, 118}

\bibitem[\protect\citeauthoryear{{Nguyen} et~al.,}{{Nguyen}
  et~al.}{2019}]{nguyen19}
{Nguyen} D.~D.,  et~al., 2019, \mn@doi [\apj] {10.3847/1538-4357/aafe7a}, \href
  {https://ui.adsabs.harvard.edu/abs/2019ApJ...872..104N} {872, 104}

\bibitem[\protect\citeauthoryear{{Ni} et~al.,}{{Ni} et~al.}{2022}]{ni22}
{Ni} Y.,  et~al., 2022, \mn@doi [\mnras] {10.1093/mnras/stac351}, \href
  {https://ui.adsabs.harvard.edu/abs/2022MNRAS.513..670N} {513, 670}

\bibitem[\protect\citeauthoryear{{Okamoto}, {Nemmen}  \& {Bower}}{{Okamoto}
  et~al.}{2008}]{okamoto2008_BHFB}
{Okamoto} T.,  {Nemmen} R.~S.,   {Bower} R.~G.,  2008, \mn@doi [\mnras]
  {10.1111/j.1365-2966.2008.12883.x}, \href
  {http://adsabs.harvard.edu/abs/2008MNRAS.385..161O} {385, 161}

\bibitem[\protect\citeauthoryear{{Onions} et~al.,}{{Onions}
  et~al.}{2012}]{onions12}
{Onions} J.,  et~al., 2012, \mn@doi [\mnras]
  {10.1111/j.1365-2966.2012.20947.x}, \href
  {https://ui.adsabs.harvard.edu/abs/2012MNRAS.423.1200O} {423, 1200}

\bibitem[\protect\citeauthoryear{{Ostriker}}{{Ostriker}}{1999}]{ostriker99}
{Ostriker} E.~C.,  1999, \mn@doi [\apj] {10.1086/306858}, \href
  {http://adsabs.harvard.edu/abs/1999ApJ...513..252O} {513, 252}

\bibitem[\protect\citeauthoryear{{Pacucci} et~al.,}{{Pacucci}
  et~al.}{2019}]{pacucci19}
{Pacucci} F.,  et~al., 2019, \baas, \href
  {https://ui.adsabs.harvard.edu/abs/2019BAAS...51c.117P} {51, 117}

\bibitem[\protect\citeauthoryear{{Pacucci}, {Mezcua}  \& {Regan}}{{Pacucci}
  et~al.}{2021}]{pacucci21}
{Pacucci} F.,  {Mezcua} M.,   {Regan} J.~A.,  2021, \mn@doi [\apj]
  {10.3847/1538-4357/ac1595}, \href
  {https://ui.adsabs.harvard.edu/abs/2021ApJ...920..134P} {920, 134}

\bibitem[\protect\citeauthoryear{{Peng} et~al.,}{{Peng} et~al.}{2008}]{peng08}
{Peng} E.~W.,  et~al., 2008, \mn@doi [\apj] {10.1086/587951}, \href
  {https://ui.adsabs.harvard.edu/abs/2008ApJ...681..197P} {681, 197}

\bibitem[\protect\citeauthoryear{{Pfister}, {Volonteri}, {Dubois}, {Dotti}  \&
  {Colpi}}{{Pfister} et~al.}{2019}]{pfister19}
{Pfister} H.,  {Volonteri} M.,  {Dubois} Y.,  {Dotti} M.,   {Colpi} M.,  2019,
  \mn@doi [\mnras] {10.1093/mnras/stz822}, \href
  {https://ui.adsabs.harvard.edu/abs/2019MNRAS.486..101P} {486, 101}

\bibitem[\protect\citeauthoryear{{Planck Collaboration} et~al.,}{{Planck
  Collaboration} et~al.}{2016}]{planck16}
{Planck Collaboration} et~al., 2016, \mn@doi [\aap]
  {10.1051/0004-6361/201525830}, \href
  {http://adsabs.harvard.edu/abs/2016A%26A...594A..13P} {594, A13}

\bibitem[\protect\citeauthoryear{{Poggianti} et~al.,}{{Poggianti}
  et~al.}{2017}]{poggianti17}
{Poggianti} B.~M.,  et~al., 2017, \mn@doi [\nat] {10.1038/nature23462}, \href
  {https://ui.adsabs.harvard.edu/#abs/2017Natur.548..304P} {548, 304}

\bibitem[\protect\citeauthoryear{{Pontzen} \& {Tremmel}}{{Pontzen} \&
  {Tremmel}}{2018}]{tangos}
{Pontzen} A.,  {Tremmel} M.,  2018, \mn@doi [\apjs] {10.3847/1538-4365/aac832},
  \href {https://ui.adsabs.harvard.edu/abs/2018ApJS..237...23P} {237, 23}

\bibitem[\protect\citeauthoryear{{Pontzen} et~al.,}{{Pontzen}
  et~al.}{2008}]{pontzen08}
{Pontzen} A.,  et~al., 2008, \mn@doi [\mnras]
  {10.1111/j.1365-2966.2008.13782.x}, \href
  {http://adsabs.harvard.edu/abs/2008MNRAS.390.1349P} {390, 1349}

\bibitem[\protect\citeauthoryear{{Pontzen}, {Ro{\v s}kar}, {Stinson}  \&
  {Woods}}{{Pontzen} et~al.}{2013}]{pynbody}
{Pontzen} A.,  {Ro{\v s}kar} R.,  {Stinson} G.,   {Woods} R.,  2013, {pynbody:
  N-Body/SPH analysis for python}, Astrophysics Source Code Library, record
  ascl 1305.002 (\mn@eprint {ascl} {1305.002})

\bibitem[\protect\citeauthoryear{{Power}, {Navarro}, {Jenkins}, {Frenk},
  {White}, {Springel}, {Stadel}  \& {Quinn}}{{Power} et~al.}{2003}]{power03}
{Power} C.,  {Navarro} J.~F.,  {Jenkins} A.,  {Frenk} C.~S.,  {White} S.~D.~M.,
   {Springel} V.,  {Stadel} J.,   {Quinn} T.,  2003, \mn@doi [\mnras]
  {10.1046/j.1365-8711.2003.05925.x}, \href
  {http://adsabs.harvard.edu/abs/2003MNRAS.338...14P} {338, 14}

\bibitem[\protect\citeauthoryear{{Regan}}{{Regan}}{2023}]{regan23}
{Regan} J.,  2023, \mn@doi [The Open Journal of Astrophysics]
  {10.21105/astro.2210.04899}, \href
  {https://ui.adsabs.harvard.edu/abs/2023OJAp....6E..12R} {6, 12}

\bibitem[\protect\citeauthoryear{{Regan}, {Visbal}, {Wise}, {Haiman},
  {Johansson}  \& {Bryan}}{{Regan} et~al.}{2017}]{regan17}
{Regan} J.~A.,  {Visbal} E.,  {Wise} J.~H.,  {Haiman} Z.,  {Johansson} P.~H.,
  {Bryan} G.~L.,  2017, \mn@doi [Nature Astronomy] {10.1038/s41550-017-0075},
  \href {https://ui.adsabs.harvard.edu/abs/2017NatAs...1E..75R} {1, 0075}

\bibitem[\protect\citeauthoryear{{Regan}, {Haiman}, {Wise}, {O'Shea}  \&
  {Norman}}{{Regan} et~al.}{2020a}]{regan20b}
{Regan} J.~A.,  {Haiman} Z.,  {Wise} J.~H.,  {O'Shea} B.~W.,   {Norman} M.~L.,
  2020a, \mn@doi [The Open Journal of Astrophysics]
  {10.21105/astro.2006.14625}, \href
  {https://ui.adsabs.harvard.edu/abs/2020OJAp....3E...9R} {3, E9}

\bibitem[\protect\citeauthoryear{{Regan}, {Wise}, {Woods}, {Downes}, {O'Shea}
  \& {Norman}}{{Regan} et~al.}{2020b}]{regan20}
{Regan} J.~A.,  {Wise} J.~H.,  {Woods} T.~E.,  {Downes} T.~P.,  {O'Shea} B.~W.,
    {Norman} M.~L.,  2020b, \mn@doi [The Open Journal of Astrophysics]
  {10.21105/astro.2008.08090}, \href
  {https://ui.adsabs.harvard.edu/abs/2020OJAp....3E..15R} {3, 15}

\bibitem[\protect\citeauthoryear{{Regan}, {Pacucci}  \&
  {Bustamante-Rosell}}{{Regan} et~al.}{2023}]{regan23b}
{Regan} J.~A.,  {Pacucci} F.,   {Bustamante-Rosell} M.~J.,  2023, \mn@doi
  [\mnras] {10.1093/mnras/stac3463}, \href
  {https://ui.adsabs.harvard.edu/abs/2023MNRAS.518.5997R} {518, 5997}

\bibitem[\protect\citeauthoryear{{Reines} \& {Deller}}{{Reines} \&
  {Deller}}{2012}]{reines12}
{Reines} A.~E.,  {Deller} A.~T.,  2012, \mn@doi [\apjl]
  {10.1088/2041-8205/750/1/L24}, \href
  {http://adsabs.harvard.edu/abs/2012ApJ...750L..24R} {750, L24}

\bibitem[\protect\citeauthoryear{{Reines}, {Sivakoff}, {Johnson}  \&
  {Brogan}}{{Reines} et~al.}{2011}]{reines11}
{Reines} A.~E.,  {Sivakoff} G.~R.,  {Johnson} K.~E.,   {Brogan} C.~L.,  2011,
  \mn@doi [\nat] {10.1038/nature09724}, \href
  {http://adsabs.harvard.edu/abs/2011Natur.470...66R} {470, 66}

\bibitem[\protect\citeauthoryear{{Reines}, {Condon}, {Darling}  \&
  {Greene}}{{Reines} et~al.}{2020}]{reines20}
{Reines} A.~E.,  {Condon} J.~J.,  {Darling} J.,   {Greene} J.~E.,  2020,
  \mn@doi [\apj] {10.3847/1538-4357/ab4999}, \href
  {https://ui.adsabs.harvard.edu/abs/2020ApJ...888...36R} {888, 36}

\bibitem[\protect\citeauthoryear{{Ricarte} \& {Natarajan}}{{Ricarte} \&
  {Natarajan}}{2018}]{ricarte18}
{Ricarte} A.,  {Natarajan} P.,  2018, \mn@doi [\mnras] {10.1093/mnras/sty2448},
  \href {https://ui.adsabs.harvard.edu/abs/2018MNRAS.481.3278R} {481, 3278}

\bibitem[\protect\citeauthoryear{{Ricarte}, {Tremmel}, {Natarajan}  \&
  {Quinn}}{{Ricarte} et~al.}{2019}]{ricarte19}
{Ricarte} A.,  {Tremmel} M.,  {Natarajan} P.,   {Quinn} T.,  2019, \mn@doi
  [\mnras] {10.1093/mnras/stz2161}, \href
  {https://ui.adsabs.harvard.edu/abs/2019MNRAS.489..802R} {489, 802}

\bibitem[\protect\citeauthoryear{{Ricarte}, {Tremmel}, {Natarajan}  \&
  {Quinn}}{{Ricarte} et~al.}{2020}]{ricarte20}
{Ricarte} A.,  {Tremmel} M.,  {Natarajan} P.,   {Quinn} T.,  2020, \mn@doi
  [\apjl] {10.3847/2041-8213/ab9022}, \href
  {https://ui.adsabs.harvard.edu/abs/2020ApJ...895L...8R} {895, L8}

\bibitem[\protect\citeauthoryear{{Ricarte}, {Tremmel}, {Natarajan}, {Zimmer}
  \& {Quinn}}{{Ricarte} et~al.}{2021a}]{ricarte21}
{Ricarte} A.,  {Tremmel} M.,  {Natarajan} P.,  {Zimmer} C.,   {Quinn} T.,
  2021a, \mn@doi [\mnras] {10.1093/mnras/stab866}, \href
  {https://ui.adsabs.harvard.edu/abs/2021MNRAS.503.6098R} {503, 6098}

\bibitem[\protect\citeauthoryear{{Ricarte}, {Tremmel}, {Natarajan}  \&
  {Quinn}}{{Ricarte} et~al.}{2021b}]{ricarte21b}
{Ricarte} A.,  {Tremmel} M.,  {Natarajan} P.,   {Quinn} T.,  2021b, \mn@doi
  [\apjl] {10.3847/2041-8213/ac1170}, \href
  {https://ui.adsabs.harvard.edu/abs/2021ApJ...916L..18R} {916, L18}

\bibitem[\protect\citeauthoryear{{Ritchie} \& {Thomas}}{{Ritchie} \&
  {Thomas}}{2001}]{ritchie01}
{Ritchie} B.~W.,  {Thomas} P.~A.,  2001, \mn@doi [\mnras]
  {10.1046/j.1365-8711.2001.04268.x}, \href
  {http://adsabs.harvard.edu/abs/2001MNRAS.323..743R} {323, 743}

\bibitem[\protect\citeauthoryear{{Rosas-Guevara}, {Bower}, {Schaye},
  {McAlpine}, {Dalla Vecchia}, {Frenk}, {Schaller}  \&
  {Theuns}}{{Rosas-Guevara} et~al.}{2016}]{rosasGuevara16}
{Rosas-Guevara} Y.,  {Bower} R.~G.,  {Schaye} J.,  {McAlpine} S.,  {Dalla
  Vecchia} C.,  {Frenk} C.~S.,  {Schaller} M.,   {Theuns} T.,  2016, \mn@doi
  [\mnras] {10.1093/mnras/stw1679}, \href
  {https://ui.adsabs.harvard.edu/abs/2016MNRAS.462..190R} {462, 190}

\bibitem[\protect\citeauthoryear{{Ro{\v s}kar}, {Fiacconi}, {Mayer},
  {Kazantzidis}, {Quinn}  \& {Wadsley}}{{Ro{\v s}kar} et~al.}{2015}]{roskar15}
{Ro{\v s}kar} R.,  {Fiacconi} D.,  {Mayer} L.,  {Kazantzidis} S.,  {Quinn}
  T.~R.,   {Wadsley} J.,  2015, \mn@doi [\mnras] {10.1093/mnras/stv312}, \href
  {http://adsabs.harvard.edu/abs/2015MNRAS.449..494R} {449, 494}

\bibitem[\protect\citeauthoryear{{Saitoh} \& {Makino}}{{Saitoh} \&
  {Makino}}{2009}]{saitoh09}
{Saitoh} T.~R.,  {Makino} J.,  2009, \mn@doi [\apjl]
  {10.1088/0004-637X/697/2/L99}, \href
  {https://ui.adsabs.harvard.edu/abs/2009ApJ...697L..99S} {697, L99}

\bibitem[\protect\citeauthoryear{{S{\'a}nchez-Janssen} \&
  {Aguerri}}{{S{\'a}nchez-Janssen} \& {Aguerri}}{2012}]{sanchezjanssen12}
{S{\'a}nchez-Janssen} R.,  {Aguerri} J.~A.~L.,  2012, \mn@doi [\mnras]
  {10.1111/j.1365-2966.2012.21301.x}, \href
  {https://ui.adsabs.harvard.edu/abs/2012MNRAS.424.2614S} {424, 2614}

\bibitem[\protect\citeauthoryear{{S{\'a}nchez-Janssen}
  et~al.,}{{S{\'a}nchez-Janssen} et~al.}{2019}]{sanchezJanssen19}
{S{\'a}nchez-Janssen} R.,  et~al., 2019, \mn@doi [\apj]
  {10.3847/1538-4357/aaf4fd}, \href
  {https://ui.adsabs.harvard.edu/abs/2019ApJ...878...18S} {878, 18}

\bibitem[\protect\citeauthoryear{{Sassano}, {Capelo}, {Mayer}, {Schneider}  \&
  {Valiante}}{{Sassano} et~al.}{2023}]{sessano23}
{Sassano} F.,  {Capelo} P.~R.,  {Mayer} L.,  {Schneider} R.,   {Valiante} R.,
  2023, \mn@doi [\mnras] {10.1093/mnras/stac3608}, \href
  {https://ui.adsabs.harvard.edu/abs/2023MNRAS.519.1837S} {519, 1837}

\bibitem[\protect\citeauthoryear{{Scott} \& {Graham}}{{Scott} \&
  {Graham}}{2013}]{scott13}
{Scott} N.,  {Graham} A.~W.,  2013, \mn@doi [\apj]
  {10.1088/0004-637X/763/2/76}, \href
  {https://ui.adsabs.harvard.edu/abs/2013ApJ...763...76S} {763, 76}

\bibitem[\protect\citeauthoryear{{Sesana}}{{Sesana}}{2013}]{sesana13}
{Sesana} A.,  2013, \mn@doi [\mnras] {10.1093/mnrasl/slt034}, \href
  {http://adsabs.harvard.edu/abs/2013MNRAS.433L...1S} {433, L1}

\bibitem[\protect\citeauthoryear{{Sesana} \& {Khan}}{{Sesana} \&
  {Khan}}{2015}]{sesana15}
{Sesana} A.,  {Khan} F.~M.,  2015, \mn@doi [\mnras] {10.1093/mnrasl/slv131},
  \href {http://adsabs.harvard.edu/abs/2015MNRAS.454L..66S} {454, L66}

\bibitem[\protect\citeauthoryear{{Seth}, {Dalcanton}, {Hodge}  \&
  {Debattista}}{{Seth} et~al.}{2006}]{seth06}
{Seth} A.~C.,  {Dalcanton} J.~J.,  {Hodge} P.~W.,   {Debattista} V.~P.,  2006,
  \mn@doi [\aj] {10.1086/508994}, \href
  {https://ui.adsabs.harvard.edu/abs/2006AJ....132.2539S} {132, 2539}

\bibitem[\protect\citeauthoryear{{Seth}, {Ag{\"u}eros}, {Lee}  \&
  {Basu-Zych}}{{Seth} et~al.}{2008}]{seth08}
{Seth} A.,  {Ag{\"u}eros} M.,  {Lee} D.,   {Basu-Zych} A.,  2008, \mn@doi
  [\apj] {10.1086/528955}, \href
  {https://ui.adsabs.harvard.edu/abs/2008ApJ...678..116S} {678, 116}

\bibitem[\protect\citeauthoryear{{Seth} et~al.,}{{Seth} et~al.}{2014}]{seth14}
{Seth} A.~C.,  et~al., 2014, \mn@doi [\nat] {10.1038/nature13762}, \href
  {https://ui.adsabs.harvard.edu/abs/2014Natur.513..398S} {513, 398}

\bibitem[\protect\citeauthoryear{{Sharma}, {Brooks}, {Somerville}, {Tremmel},
  {Bellovary}, {Wright}  \& {Quinn}}{{Sharma} et~al.}{2020}]{sharma20}
{Sharma} R.~S.,  {Brooks} A.~M.,  {Somerville} R.~S.,  {Tremmel} M.,
  {Bellovary} J.,  {Wright} A.~C.,   {Quinn} T.~R.,  2020, \mn@doi [\apj]
  {10.3847/1538-4357/ab960e}, \href
  {https://ui.adsabs.harvard.edu/abs/2020ApJ...897..103S} {897, 103}

\bibitem[\protect\citeauthoryear{{Sharma}, {Brooks}, {Tremmel}, {Bellovary},
  {Ricarte}  \& {Quinn}}{{Sharma} et~al.}{2022a}]{sharma22}
{Sharma} R.~S.,  {Brooks} A.~M.,  {Tremmel} M.,  {Bellovary} J.,  {Ricarte} A.,
    {Quinn} T.~R.,  2022a, arXiv e-prints, \href
  {https://ui.adsabs.harvard.edu/abs/2022arXiv220305580S} {p. arXiv:2203.05580}

\bibitem[\protect\citeauthoryear{{Sharma}, {Brooks}, {Tremmel}, {Bellovary}  \&
  {Quinn}}{{Sharma} et~al.}{2022b}]{sharma22b}
{Sharma} R.~S.,  {Brooks} A.~M.,  {Tremmel} M.,  {Bellovary} J.,   {Quinn}
  T.~R.,  2022b, arXiv e-prints, \href
  {https://ui.adsabs.harvard.edu/abs/2022arXiv221105275S} {p. arXiv:2211.05275}

\bibitem[\protect\citeauthoryear{{Shen}, {Wadsley}  \& {Stinson}}{{Shen}
  et~al.}{2010}]{shen10}
{Shen} S.,  {Wadsley} J.,   {Stinson} G.,  2010, \mn@doi [\mnras]
  {10.1111/j.1365-2966.2010.17047.x}, \href
  {http://adsabs.harvard.edu/abs/2010MNRAS.407.1581S} {407, 1581}

\bibitem[\protect\citeauthoryear{{Shen}, {Hopkins}, {Faucher-Gigu{\`e}re},
  {Alexander}, {Richards}, {Ross}  \& {Hickox}}{{Shen} et~al.}{2020}]{shen20}
{Shen} X.,  {Hopkins} P.~F.,  {Faucher-Gigu{\`e}re} C.-A.,  {Alexander} D.~M.,
  {Richards} G.~T.,  {Ross} N.~P.,   {Hickox} R.~C.,  2020, \mn@doi [\mnras]
  {10.1093/mnras/staa1381}, \href
  {https://ui.adsabs.harvard.edu/abs/2020MNRAS.495.3252S} {495, 3252}

\bibitem[\protect\citeauthoryear{{Sijacki}, {Vogelsberger}, {Genel},
  {Springel}, {Torrey}, {Snyder}, {Nelson}  \& {Hernquist}}{{Sijacki}
  et~al.}{2015}]{IllustrisBH15}
{Sijacki} D.,  {Vogelsberger} M.,  {Genel} S.,  {Springel} V.,  {Torrey} P.,
  {Snyder} G.~F.,  {Nelson} D.,   {Hernquist} L.,  2015, \mn@doi [\mnras]
  {10.1093/mnras/stv1340}, \href
  {http://adsabs.harvard.edu/abs/2015MNRAS.452..575S} {452, 575}

\bibitem[\protect\citeauthoryear{{Stinson}, {Seth}, {Katz}, {Wadsley},
  {Governato}  \& {Quinn}}{{Stinson} et~al.}{2006}]{Stinson06}
{Stinson} G.,  {Seth} A.,  {Katz} N.,  {Wadsley} J.,  {Governato} F.,   {Quinn}
  T.,  2006, \mn@doi [\mnras] {10.1111/j.1365-2966.2006.11097.x}, \href
  {http://adsabs.harvard.edu/abs/2006MNRAS.373.1074S} {373, 1074}

\bibitem[\protect\citeauthoryear{{Trebitsch} et~al.,}{{Trebitsch}
  et~al.}{2021}]{obelisk21}
{Trebitsch} M.,  et~al., 2021, \mn@doi [\aap] {10.1051/0004-6361/202037698},
  \href {https://ui.adsabs.harvard.edu/abs/2021A&A...653A.154T} {653, A154}

\bibitem[\protect\citeauthoryear{{Tremaine} et~al.,}{{Tremaine}
  et~al.}{2002}]{tremaine_etal02}
{Tremaine} S.,  et~al., 2002, \mn@doi [\apj] {10.1086/341002}, \href
  {http://adsabs.harvard.edu/cgi-bin/nph-bib_query?bibcode=2002ApJ...574..740T&db_key=AST}
  {574, 740}

\bibitem[\protect\citeauthoryear{{Tremmel}, {Governato}, {Volonteri}  \&
  {Quinn}}{{Tremmel} et~al.}{2015}]{tremmel15}
{Tremmel} M.,  {Governato} F.,  {Volonteri} M.,   {Quinn} T.~R.,  2015, \mn@doi
  [\mnras] {10.1093/mnras/stv1060}, \href
  {http://adsabs.harvard.edu/abs/2015MNRAS.451.1868T} {451, 1868}

\bibitem[\protect\citeauthoryear{{Tremmel}, {Karcher}, {Governato},
  {Volonteri}, {Quinn}, {Pontzen}, {Anderson}  \& {Bellovary}}{{Tremmel}
  et~al.}{2017}]{tremmel17}
{Tremmel} M.,  {Karcher} M.,  {Governato} F.,  {Volonteri} M.,  {Quinn} T.~R.,
  {Pontzen} A.,  {Anderson} L.,   {Bellovary} J.,  2017, \mn@doi [\mnras]
  {10.1093/mnras/stx1160}, \href
  {http://adsabs.harvard.edu/abs/2017MNRAS.470.1121T} {470, 1121}

\bibitem[\protect\citeauthoryear{{Tremmel}, {Governato}, {Volonteri}, {Quinn}
  \& {Pontzen}}{{Tremmel} et~al.}{2018a}]{tremmel18}
{Tremmel} M.,  {Governato} F.,  {Volonteri} M.,  {Quinn} T.~R.,   {Pontzen} A.,
   2018a, \mn@doi [\mnras] {10.1093/mnras/sty139}, \href
  {https://ui.adsabs.harvard.edu/#abs/2018MNRAS.475.4967T} {475, 4967}

\bibitem[\protect\citeauthoryear{Tremmel, Governato, Volonteri, Pontzen  \&
  Quinn}{Tremmel et~al.}{2018b}]{tremmel18b}
Tremmel M.,  Governato F.,  Volonteri M.,  Pontzen A.,   Quinn T.~R.,  2018b,
  The Astrophysical Journal Letters, 857, L22

\bibitem[\protect\citeauthoryear{{Tremmel} et~al.,}{{Tremmel}
  et~al.}{2019}]{tremmel19}
{Tremmel} M.,  et~al., 2019, \mn@doi [\mnras] {10.1093/mnras/sty3336}, \href
  {https://ui.adsabs.harvard.edu/abs/2019MNRAS.483.3336T} {483, 3336}

\bibitem[\protect\citeauthoryear{{Tremmel}, {Wright}, {Brooks}, {Munshi},
  {Nagai}  \& {Quinn}}{{Tremmel} et~al.}{2020}]{tremmel20}
{Tremmel} M.,  {Wright} A.~C.,  {Brooks} A.~M.,  {Munshi} F.,  {Nagai} D.,
  {Quinn} T.~R.,  2020, \mn@doi [\mnras] {10.1093/mnras/staa2015}, \href
  {https://ui.adsabs.harvard.edu/abs/2020MNRAS.497.2786T} {497, 2786}

\bibitem[\protect\citeauthoryear{{Van Nest}, {Munshi}, {Wright}, {Tremmel},
  {Brooks}, {Nagai}  \& {Quinn}}{{Van Nest} et~al.}{2022}]{jvn22}
{Van Nest} J.~D.,  {Munshi} F.,  {Wright} A.~C.,  {Tremmel} M.,  {Brooks}
  A.~M.,  {Nagai} D.,   {Quinn} T.,  2022, \mn@doi [\apj]
  {10.3847/1538-4357/ac43b7}, \href
  {https://ui.adsabs.harvard.edu/abs/2022ApJ...926...92V} {926, 92}

\bibitem[\protect\citeauthoryear{{Voggel}, {Hilker}  \& {Richtler}}{{Voggel}
  et~al.}{2016}]{voggel16}
{Voggel} K.,  {Hilker} M.,   {Richtler} T.,  2016, \mn@doi [\aap]
  {10.1051/0004-6361/201527070}, \href
  {https://ui.adsabs.harvard.edu/abs/2016A&A...586A.102V} {586, A102}

\bibitem[\protect\citeauthoryear{{Voggel}, {Seth}, {Baumgardt}, {Mieske},
  {Pfeffer}  \& {Rasskazov}}{{Voggel} et~al.}{2019}]{voggel19}
{Voggel} K.~T.,  {Seth} A.~C.,  {Baumgardt} H.,  {Mieske} S.,  {Pfeffer} J.,
  {Rasskazov} A.,  2019, \mn@doi [\apj] {10.3847/1538-4357/aaf735}, \href
  {https://ui.adsabs.harvard.edu/abs/2019ApJ...871..159V} {871, 159}

\bibitem[\protect\citeauthoryear{{Volonteri}}{{Volonteri}}{2010}]{volonteri10}
{Volonteri} M.,  2010, \mn@doi [\aapr] {10.1007/s00159-010-0029-x}, \href
  {http://adsabs.harvard.edu/abs/2010A%26ARv..18..279V} {18, 279}

\bibitem[\protect\citeauthoryear{{Volonteri}}{{Volonteri}}{2012}]{volonteri12}
{Volonteri} M.,  2012, \mn@doi [Science] {10.1126/science.1220843}, \href
  {http://adsabs.harvard.edu/abs/2012Sci...337..544V} {337, 544}

\bibitem[\protect\citeauthoryear{{Volonteri} \& {Gnedin}}{{Volonteri} \&
  {Gnedin}}{2009}]{volonteri09}
{Volonteri} M.,  {Gnedin} N.~Y.,  2009, \mn@doi [\apj]
  {10.1088/0004-637X/703/2/2113}, \href
  {https://ui.adsabs.harvard.edu/abs/2009ApJ...703.2113V} {703, 2113}

\bibitem[\protect\citeauthoryear{{Volonteri} \& {Natarajan}}{{Volonteri} \&
  {Natarajan}}{2009}]{volonteriMSIGMA2009}
{Volonteri} M.,  {Natarajan} P.,  2009, \mn@doi [\mnras]
  {10.1111/j.1365-2966.2009.15577.x}, \href
  {http://adsabs.harvard.edu/abs/2009MNRAS.400.1911V} {400, 1911}

\bibitem[\protect\citeauthoryear{{Volonteri} \& {Rees}}{{Volonteri} \&
  {Rees}}{2005}]{volonteriRees2005}
{Volonteri} M.,  {Rees} M.~J.,  2005, \mn@doi [\apj] {10.1086/466521}, \href
  {https://ui.adsabs.harvard.edu/abs/2005ApJ...633..624V} {633, 624}

\bibitem[\protect\citeauthoryear{{Volonteri}, {Lodato}  \&
  {Natarajan}}{{Volonteri} et~al.}{2008}]{volonteri08}
{Volonteri} M.,  {Lodato} G.,   {Natarajan} P.,  2008, \mn@doi [\mnras]
  {10.1111/j.1365-2966.2007.12589.x}, \href
  {http://adsabs.harvard.edu/abs/2008MNRAS.383.1079V} {383, 1079}

\bibitem[\protect\citeauthoryear{{Volonteri} et~al.,}{{Volonteri}
  et~al.}{2020}]{volonteri20}
{Volonteri} M.,  et~al., 2020, \mn@doi [\mnras] {10.1093/mnras/staa2384}, \href
  {https://ui.adsabs.harvard.edu/abs/2020MNRAS.498.2219V} {498, 2219}

\bibitem[\protect\citeauthoryear{{Wadsley}, {Stadel}  \& {Quinn}}{{Wadsley}
  et~al.}{2004}]{wadsley04}
{Wadsley} J.~W.,  {Stadel} J.,   {Quinn} T.,  2004, New Astronomy, \href
  {http://adsabs.harvard.edu/cgi-bin/nph-bib_query?bibcode=2004NewA....9..137W&db_key=AST}
  {9, 137}

\bibitem[\protect\citeauthoryear{{Wadsley}, {Veeravalli}  \&
  {Couchman}}{{Wadsley} et~al.}{2008}]{wadsley08}
{Wadsley} J.~W.,  {Veeravalli} G.,   {Couchman} H.~M.~P.,  2008, \mn@doi
  [\mnras] {10.1111/j.1365-2966.2008.13260.x}, \href
  {http://adsabs.harvard.edu/abs/2008MNRAS.387..427W} {387, 427}

\bibitem[\protect\citeauthoryear{{Wadsley}, {Keller}  \& {Quinn}}{{Wadsley}
  et~al.}{2017}]{wadsley17}
{Wadsley} J.~W.,  {Keller} B.~W.,   {Quinn} T.~R.,  2017, \mn@doi [\mnras]
  {10.1093/mnras/stx1643}, \href
  {http://adsabs.harvard.edu/abs/2017MNRAS.471.2357W} {471, 2357}

\bibitem[\protect\citeauthoryear{{Wehner} \& {Harris}}{{Wehner} \&
  {Harris}}{2006}]{wehner06}
{Wehner} E.~H.,  {Harris} W.~E.,  2006, \mn@doi [\apjl] {10.1086/505387}, \href
  {http://adsabs.harvard.edu/abs/2006ApJ...644L..17W} {644, L17}

\bibitem[\protect\citeauthoryear{{Wise}, {Regan}, {O'Shea}, {Norman}, {Downes}
  \& {Xu}}{{Wise} et~al.}{2019}]{wise19}
{Wise} J.~H.,  {Regan} J.~A.,  {O'Shea} B.~W.,  {Norman} M.~L.,  {Downes}
  T.~P.,   {Xu} H.,  2019, \mn@doi [\nat] {10.1038/s41586-019-0873-4}, \href
  {https://ui.adsabs.harvard.edu/abs/2019Natur.566...85W} {566, 85}

\bibitem[\protect\citeauthoryear{{Woo}, {Cho}, {Gallo}, {Hodges-Kluck}, {Le},
  {Shin}, {Son}  \& {Horst}}{{Woo} et~al.}{2019}]{woo19}
{Woo} J.-H.,  {Cho} H.,  {Gallo} E.,  {Hodges-Kluck} E.,  {Le} H. A.~N.,
  {Shin} J.,  {Son} D.,   {Horst} J.~C.,  2019, \mn@doi [Nature Astronomy]
  {10.1038/s41550-019-0790-3}, \href
  {https://ui.adsabs.harvard.edu/abs/2019NatAs...3..755W} {3, 755}

\bibitem[\protect\citeauthoryear{{Wright}, {Tremmel}, {Brooks}, {Munshi},
  {Nagai}, {Sharma}  \& {Quinn}}{{Wright} et~al.}{2021}]{wright21}
{Wright} A.~C.,  {Tremmel} M.,  {Brooks} A.~M.,  {Munshi} F.,  {Nagai} D.,
  {Sharma} R.~S.,   {Quinn} T.~R.,  2021, \mn@doi [\mnras]
  {10.1093/mnras/stab081}, \href
  {https://ui.adsabs.harvard.edu/abs/2021MNRAS.502.5370W} {502, 5370}

\bibitem[\protect\citeauthoryear{{van Wassenhove}, {Volonteri}, {Walker}  \&
  {Gair}}{{van Wassenhove} et~al.}{2010}]{vanwassenhove10}
{van Wassenhove} S.,  {Volonteri} M.,  {Walker} M.~G.,   {Gair} J.~R.,  2010,
  \mn@doi [\mnras] {10.1111/j.1365-2966.2010.17189.x}, \href
  {https://ui.adsabs.harvard.edu/abs/2010MNRAS.408.1139V} {408, 1139}

\bibitem[\protect\citeauthoryear{{van de Voort}, {Schaye}, {Booth}, {Haas}  \&
  {Dalla Vecchia}}{{van de Voort} et~al.}{2011}]{vandevoort11}
{van de Voort} F.,  {Schaye} J.,  {Booth} C.~M.,  {Haas} M.~R.,   {Dalla
  Vecchia} C.,  2011, \mn@doi [\mnras] {10.1111/j.1365-2966.2011.18565.x},
  \href {https://ui.adsabs.harvard.edu/#abs/2011MNRAS.414.2458V} {414, 2458}

\bibitem[\protect\citeauthoryear{{van den Bosch} \& {Ogiya}}{{van den Bosch} \&
  {Ogiya}}{2018}]{vdBosch18b}
{van den Bosch} F.~C.,  {Ogiya} G.,  2018, \mn@doi [\mnras]
  {10.1093/mnras/sty084}, \href
  {https://ui.adsabs.harvard.edu/abs/2018MNRAS.475.4066V} {475, 4066}

\makeatother
\end{thebibliography}

%\begin{thebibliography}{99}
%\end{thebibliography}

%%%%%%%%%%%%%%%%%%%%%%%%%%%%%%%%%%%%%%%%%%%%%%%%%%

%%%%%%%%%%%%%%%%% APPENDICES %%%%%%%%%%%%%%%%%%%%%

%\appendix



%%%%%%%%%%%%%%%%%%%%%%%%%%%%%%%%%%%%%%%%%%%%%%%%%%

% Don't change these lines
%\bsp	% typesetting comment
%\label{lastpage}
\end{document}